\documentclass[prb,preprint]{revtex4-1} 
\usepackage{amsmath}  % needed for \tfrac, \bmatrix, etc.
\usepackage{amsfonts} % needed for bold Greek, Fraktur, and blackboard bold
\usepackage{graphicx} % needed for figures
\usepackage{xfrac}
\usepackage{listings}   % Mathematica code
\usepackage{lipsum}
\newcommand{\be}[1]{\begin{equation}\label{#1}}
\newcommand{\ee}{\end{equation}}
\newcommand{\bea}[1]{\begin{eqnarray}\label{#1}}
\newcommand{\eea}{\end{eqnarray}}
\newcommand{\no}{\nonumber \\}
\newcommand{\Fig}[1]{Fig.(\ref{#1})}
\newcommand{\Tbl}[1]{Table~\ref{#1}}
\newcommand{\Eq}[1]{Eq.(\ref{#1})}
\newcommand{\App}[1]{Appendix~\ref{#1}}
\newcommand{\Sec}[1]{Section~\ref{#1}}
\newcommand{\Lst}[1]{Listing~\ref{#1}}
\newcommand{\bsub}{\begin{subequations}}
\newcommand{\esub}{\end{subequations}}
\newcommand{\bwt}{\begin{widetext}}
\newcommand{\ewt}{\end{widetext}}

\usepackage{color}

\def\trm#1{\textrm{#1}}
\def\tit#1{\textit{#1}}

\def\ttt#1{\texttt{#1}}
\def\bs#1{\boldsymbol{#1}}
\def\hbs#1{\hat{\boldsymbol{#1}}}
\def\a0{{\alpha_0}}

\def\da0{{\dot{\alpha}_0}}

\def\myoverDefn#1#2{\hbox{\space \raise-2mm\hbox{$\textstyle{#1} \atop \scriptstyle{#2}$} }}
\def\k{{\kappa}}

\def\G{{\Gamma}}

\def\g{{\gamma}}

\def\a{{\alpha}}

\def\dag{\dagger}

\def\F{\mathcal{F}}

\def\rp{r_{P}}
\def\rp2{r_{p}^{2}}

\def\eps{\epsilon}
\def\defeq{\overset{\textrm{def}}{=}}
\def\u{\textrm{\bf u}}
\def\p{\textrm{\bf p}}
\def\e{\textrm{\bf e}}

\def\phat{\hat{\textrm{\textbf{p}}}}
\def\qhat{\hat{\textrm{\textbf{q}}}}
\def\sigmahat{\hat{\boldsymbol{\sigma}}}
\def\F{{\mathcal{F}}}
\def\Qhat{{\hat{\mathcal{Q}}}}
\def\rootF{{\sqrt{\mathcal{F}}}}
\def\Ohat{{\hat{\mathcal{O}}}}

\newcommand{\half}{\frac{1}{2}}
\newcommand{\thalf}{\tfrac{1}{2}}

\newcommand{\pd}{\partial}

\newcommand{\IP}[2]{\langle {#1} | {#2} \rangle}  % inner product
\begin{document}
%=================================================
%======== fancyhdr: puts the page number back in on the [R]ight vs the [L]eft =====
%%\fancyhead[R]{\ifnum\value{page}<2\relax\else\thepage\fi}
%================================================================
% Be sure to use the \title, \author, \affiliation, and \abstract macros
% to format your title page.  Don't use lower-level macros to  manually
% adjust the fonts and centering.

\title{Bound states of the Dirac equation in Schwarzschild spacetime: \\ an exploration of intuition for the curious student}
 %For AJP review - remove name and affiliations, and acknowledgements
\author{Paul M. Alsing}
\email{paul.alsing@us.af.mil; paul.alsing@afrl.af.mil}
\affiliation{Air Force Research Laboratory, Information Directorate, 525 Brooks Rd, Rome, NY, 13441}

% In a long title you can use \\ to force a line break at a certain location.

%When submitting the manuscript for review, do not include the author's name or institution
%\author{Daniel V. Schroeder}
%\email{dschroeder@weber.edu} % optional
%\altaffiliation[permanent address: ]{101 Main Street, Anytown, USA} % optional second address
% If there were a second author at the same address, we would put another 
% \author{} statement here.  Don't combine multiple authors in a single
% \author statement.
%\affiliation{Department of Physics, Weber State University, Ogden, UT 84408-2508}
% Please provide a full mailing address here.

%\author{David P. Jackson}
%\email{ajp@dickinson.edu}
%\affiliation{Department of Physics, Dickinson College, Carlisle, PA 17013}

% See the REVTeX documentation for more examples of author and affiliation lists.

\date{\today}

\begin{abstract}
In this work we explore the possibility of quantum bound states in a Schwarzschild gravitational field leveraging the analogy of the elementary derivation of bound states in the Coulomb potential as taught in an undergraduate course in Quantum Mechanics. For this we will also need to go beyond non-relativistic quantum mechanics and utilize the relativistic Dirac equation for a central potential as taught in an advanced undergraduate or first year graduate (special) relativistic quantum mechanics course. Finally, the special relativistic Dirac equation must be extended to the general relativistic version for curved spacetime. All these disparate component pieces exist in excellent, very readable textbooks written for the student reader, with sufficient detail for a curious student to learn and explore. We pull all these threads together in order to explore a very natural question that a student might ask: ``If the effective $1/r$ radial potential of the Schwarzschild metric (with angular momentum barrier), as taught in elementary GR courses for undergraduates, appears Newtonian-like (with a $1/r^3$ correction), then is it possible to derive quantum bound states in the Schwarzschild spacetime by simply changing the radial potential $V(r)$ from 
$V_C(r)=-e^2/r$ to $V_{Schw}=-G M m/r$?" 
\end{abstract}

\maketitle % title page is now complete
%================================================
%=================
% needed for fancyhdr
%=================
%%\thispagestyle{fancy}
%=================

%=========================================================
\section{Introduction}\label{sec:Intro}
In the course of an undergraduate to first year graduate student's physcis education, the student will learn quantum mechanics (QM) encompassing the solution of bound states of the hydrogen atom, and the application and manipulation of spin. In specialized physics courses the student may have the opportunity to take an introduction to general relativity (GR), typically encompassing particle and photon orbits in the Schwarzschild spacetime (SST). Finally, in an advanced undergraduate (or first year graduate) QM course, the student will be exposed to the special relativistic (SR) Dirac equation (DE), and its solutions in special cases.

Given the above archetypal physics education,  a curious student may make the following observations, leading to the the subsequently posed question to a professor, or better yet, to themselves. ``In my introductory GR class, we saw that in the Schwarzschild metric, the radial equation for particles and photons orbits has a form reminiscent of a classical energy equation
(see Hartle\cite{Hartle:2009}, p195), which for particles ($m>0$) takes the form
\bea{Hartle:9.32} 
E_{Newt} &=& \frac{m}{2}\,\left( \frac{dr}{d\tau}\right)^2 +V_{eff}(r), \\
\textrm{where}\quad V_{eff}(r) &=& -\frac{G M m}{r}+\frac{L^2}{2 m r^2}  - \frac{G M L^2}{c^2 m r^3},
\eea
where $M$ is the mass of the gravitating body, $m$ is the test mass, 
$e = (E_{Newt} + m c^2)/ (m c^2)$ and $\ell = L/m$
are the conserved energy and angular momentum per unit rest mass, respectively (constants of the motion along the trajectory). Now if we ignore the $1/r^3$ term, this looks just like the central potential with the angular momentum  barrier term included for an attractive $1/r$ potential, like the Coulomb problem.  In my (non-relativistic) quantum mechanics (NRQM) class, we solved for the wave functions and eigenvalues of the 
bound quantum states of a hydrogen-like atom in a central potential $V_{C} = -(Ze)e/r$. 
Thus, can one expect to find bound states of $V_{Schw} \equiv V_{eff}$ if one were to simply replace $M\to Z e$ and $m\to e$? 
Could one at least find s-orbital solutions if one sets $L=0$?
Would this make any sense? The scale of $M$, say for a solar mass black hole (BH), and $Ze$ are so vastly different? How would one go about answering this (intuitive) question?" 

The purpose of this article is to assemble the disparate, requisite components for the students in order to answer this question in the affirmative, at a level within the students undergraduate, to first year graduate physics background, aided by detailed, readable textbook material that would enable the student to answer this question for themselves.
\Tbl{tbl:outline} gives a description of the sections of this paper, the author recommended associated reference books that the student could use for self- or guided- study, along with an estimate of the level of physics maturity required. \Tbl{tbl:outline} also serves as an entry point guide for readers that are already familiar with the results of some of the sections.

%=======================================
% Outline and entry points into Sections
% see p 181 and 117 of Latex for Everyone
%=======================================
% p181 of Latex for Everyone
\begin{table}[ht]
\caption{Section Number and Titles, Reference Book, and typical physics prerequisite}\label{tbl:outline}
\renewcommand{\arraystretch}{0.5} % this reduces the vertical spacing between rows
\linespread{0.5}\selectfont\centering
% p117 Latex for Everyone
\begin{tabular}{|c|p{3.0in}|p{3.0in}|} \hline
\multicolumn{3}{|c|}{\bf Section Number, Titles and Reference Book}\\ [0.5em] \hline
\multicolumn{1}{|c|}{\; \bf Section \; } & 
\multicolumn{1}{|c|}{\bf Description}  &
\multicolumn{1}{|c|}{\; \bf Ref. Book\; } 
\\  [0.25em]\hline\hline
\Sec{NRQM:Coulomb:solns} & Wave functions and eigenenergies of the Non-relativistic QM Coulomb problem: a refresher & Cohen-Tannoudji \textit{et al.}, {QM Vol 1}\cite{Cohen-Tannoudji:1977} 
\newline \hspace*{0.15in}(undergraduate quantum mechanics)  \\ [0.5em] \hline
\Sec{sec:SRDE:vector:scalar} &The special relativistic Dirac equation for spin $1/2$ particles and its solutions for vector (Coulomb) and scalar (mass) coupling & Greiner, \tit{Relativistic QM: Wave Eqns}\cite{Greiner:1990} 
\newline (advance undergraduate, first year graduate)
\\ [0.5em] \hline
\Sec{sec:DE:CST} & The Dirac equation in curved spacetime & Ryder, \tit{Introduction to General Relativity} \cite{Ryder:2009} 
\newline (advance undergraduate, first year graduate)
\\ [0.5em] \hline
\Sec{sec:bound:states:DE:SST} & Bound states of the Dirac equation in SST & \tit{Mathematica} (undergraduate) 
\\ [0.5em]\hline
\Sec{sec:DE:SST:inside:horizon} & Solutions of the Dirac equation in SST inside the horizon? & \tit{Mathematica}  (undergraduate)\\  [0.5em] \hline\hline
\end{tabular}
\end{table}
%=======================================
To save time repeatedly writing out common terms,  \Tbl{tbl:abbreviations} lists the acronyms used throughout this work.
%=======================================
% Table of Abbreiviations
%=======================================
\begin{table}[ht]
\caption{Acronyms used throughout this work}\label{tbl:abbreviations}
\renewcommand{\arraystretch}{0.5} % this reduces the vertical spacing between rows
\linespread{0.5}\selectfont\centering
% p117 Latex for Everyone
%\hspace{-0.5in}
\begin{tabular}{|c|p{1.5in}|c|p{1.55in}|c|p{1.75in}|} \hline
\multicolumn{6}{|c|}{\bf Abbreviations}\\ [0.25em] \hline\hline
\multicolumn{1}{|c|}{\bf Acronym}   & 
\multicolumn{1}{|c|}{\bf Term} &
\multicolumn{1}{|c|}{\bf Acroynm}     &
\multicolumn{1}{|c|}{\bf Term} &
\multicolumn{1}{|c|}{\bf Acroynm}     &
\multicolumn{1}{|c|}{\bf Term} 
\\ \hline
SR & Special Relativity & GR & General Relativity& NR & Non-Relativisitc  \\ [0.5em] \hline
SE & Schr\"{o}dinger Equation & DE & Dirac Equation & QM & Quantum Mechanics \\ [0.5em] \hline
SST/CST & Schwarzschild/Curved Spacetime & CTQMv1 & Cohen-Tannoudji \textit{et al.}, {QM Vol 1}\cite{Cohen-Tannoudji:1977} 
& RQMWE & Greiner, \tit{Relativistic QM: Wave Eqns}\cite{Greiner:1990} \\ [0.5em] \hline
LT & Lorentz Transformation & BH & Black Hole & SM & Supplemental Material  \\ [0.5em] \hline\hline
\end{tabular}
\end{table}
%=======================================

Let us now ask, what component physics, in order of difficulty, is required by the student to answer this question, and point out (from the author's personal perspective) those reference books containing enough explicit worked out material/derivations that would enable the student solve their question.

%\blue{This whole motivating section  in the Intro (indicated in \red{red}) could be summarized more succinctly and reduced in size. A discussion of the helpful reference books for the students used could be moved to the SM.}
%\medskip

% red: 369-374
{\color{black} 
First, one has to be able to solve for the bound states of hydrogen atom. This is standard fare in almost all introductory QM textbooks, but is lucidly well described and worked out in detail in one of the classic undergraduate QM texts, ``Quantum Mechanics: Volume 1," by Cohen-Tannoudgi, Diu and Laloe (CTQMv1) \cite{Cohen-Tannoudji:1977}. The eigenenergies of the hydrogen atom ($Z=1$) are known to be given by
$E_n = -E_I/n^2$ where the ionizations energy is given by (CTQMv1, Chapter VII) $E_I = \frac{\alpha^2}{2} \mu c^2 = 13.6$ eV, where $\mu = (m_e\,m_p)/(m_e+m_p)\sim m_e$ is the reduced mass of the electron-proton hydrogen atom system, $\mu c^2\sim m_e c^2$ is the rest mass of the electron, $e>0$ is the charge of the proton, and 
$\alpha = e^2/(\hbar c) \sim 1/137$ is the fine structure constant.
This is the subject of \Sec{NRQM:Coulomb:solns}, which the reader can skip if already familiar.
}% red from 369

However, since we are inquiring about possible quantum bound states in a classical background GR spacetime, it is not enough to have a NRQM solution. One could attempt to invoke the special relativity (SR) energy equation and posit that by including the rest mass, the SRQM energy  $\mathcal{E}$ should be given by 
$E \sim m c^2 + E_n$ (where from now on we just denote $m_e\to m$). Working backwards, one might surmise that this might the small energy approximation to either
\bea{SRQM:eval:guess}
\frac{E}{m c^2}  &=& \mathcal{E}_s   \overset{?}{=} \sqrt{1 - \frac{2\,E_I/m\,c^2}{n^2}}\approx 1 - \frac{\alpha^2}{2\,n^2}, \label{E_s}\\
\frac{E}{m c^2} &=& \mathcal{E}_v \overset{?}{=} \frac{1}{\sqrt{1 + \frac{2\,E_I/m\,c^2}{n^2}}}\;\; \approx 1 - \frac{\alpha^2}{2\,n^2}. \label{E_v}
\eea
(The meaning of the $s$ and $v$ subscripts on $\mathcal{E}$ above, denoting scalar and vector coupling, respectively, will be explain in due time).
But, which formula is more appropriate, and how would one physically decide between the two?
It appears that a much more complete description of the Coulomb problem is required - i.e. the solution of the Coulomb problem in a SR setting using the Dirac equation (DE), which is the subject of \Sec{sec:SRDE:vector:scalar}.

% red: 389-412
{\color{black} 
The DE and its solution in a central potential are treated in the extremely informative worked out example problems in one of the extraordinary set of textbooks by Walther Greiner, namely ``Relativistic Quantum Mechanics: Wave Equations (RQMWE)."\cite{Greiner:1990} This textbook series, including volumes on QM, SRQM, Electrodynamics, Nuclear Physics, Thermodynamics, Quantum Field Theory and Statistical and Thermal Physics, is essentially (in the author's opinion) a ``reworking" of the classic Landau and Lipfshitz (informative, but terse) set of graduate level physics, now geared for advanced undergraduates and/or first year graduate students (the prefaces just state that it is for ``students"). In addition to its exceptional lucidity, each book involves typically over 100 solved problems, worked out in full detail, showing all intermediate steps, which greatly aids the students proficiency in solving problems. 
Chapter 2 introduces the DE and Chapter 9 solves the Coulomb problem in a central potential for both (electromagnetic) vector and scalar (mass) coupling.
Here, vector coupling means that the potential $V_v(r)=e\phi(r) /c$ arises from the
 timelike component of a SR 4-vector potential $A^\mu=(\phi, \bs{A})$ such that the relativistic energy $E$ is modified to $E\to E-V_v$ in the DE.
 Scalar coupling means that the potential $V_s(r)$ couples to the mass, so that
 $m c^2 \to mc^2 + V_s(r)$ in the DE. So intuition would lead one to suspect favoring the scalar coupling formula \Eq{E_s}. 
 We shall explore the validity of this intuition.
 The most involved part of the solution of the SR DE is the separation of the angular from the radial degrees of freedom. 
 Readers familiar with the solution the SR DE for the Coulomb potential, may wish to skim \Sec{sec:SRDE:vector:scalar} in the main text and the greater details in {sec:DE:central:potential}, since complications (at a much greater level) arise when we discuss the DE in SST.
 
 If one is able to master the solution of the DE wave equation in SR in a central potential (aided by RQMWE), the student still has to tackle the thorny question of how to incorporate QM into GR. Of course, this is an outstanding unsolved problem in physics, but such a full theory is not required here. What one requires is the solution of a quantum mechanical wave function in a classical background GR spacetime. In effect, one needs to solve a wave equation in a medium with a spatially (here radially) dependent index of refraction. Thus, what is required is the DE in classical GR (curved spacetime) background. But how does one obtain this equation, since in RQM we solve the DE in a flat (Minkowski) background spacetime, incorporating only SR?
 This is the subject of \Sec{sec:DE:CST}.
 
 The solution is well-known (to the QM/GR Illuminati), but the key concept is easy to grasp - Einstein's Equivalence Principle. Namely, the global inertial frames of SR no longer hold in GR, and are only locally valid at each point (actually within the observer's local laboratory\cite{Hartle:2009}, if small on the length scale of changes of the spacetime curvature). Thus, one needs to ``project" the SR DE into the observer's instantaneous local laboratory, and solve the resulting GR DE in this local (moving) frame. This involves the use of \tit{tetrads}, 
describing the four axes of the observer's frame $\mathbf{e}_a$, with $a\in\{0,1,2,3\}$, moving in the surrounding spacetime, with world vector components 
$\{e_a^{\;\;\mu}(x)\}=(\mathbf{e}_a)^\mu(x)$, for $\mu=\{0,1,2,3\}$. In the local observer's frame the GR DE takes on a first order form analogous to the SR DE, with all the complication going into the \tit{connection} describing how the spinor (2-vector with complex entries) components change (parallel transport and all that) as one moves from one point to another in a curved spacetime.

There are now numerous introductory GR textbooks aimed at the upper level undergraduate student. The only one that tackles the GR DE, and in lucid, informative detail (but only setting up the equation, not solving it) is Lewis Ryder's (a particle physicist) wonderful ``Introduction to General Relativity, Chapter 11."\cite{Ryder:2009}. This is the formulation of the GR DE that we will use in this work.
It is assumed that readers are in general less familiar with this material so that
\Sec{sec:DE:CST} aims to provided helpful guidance for the key concepts and the relevant derivations.

\Sec{sec:bound:states:DE:SST} is the main focus of this work where we will explore the eigenvalues of the DE in SST with a gravitational potential, and compare and contrast with the eigenvalues of the SR DE and NR SE for the Coulomb potential (all of $1/r$ form).
\Sec{sec:DE:SST:inside:horizon} explores a numerical analytic extension of the previous section to develop ``eigenvalues" for inside the black hole (BH) horizon, with a discussion of the reasonableness of such solutions.

In the \App{app:Units} we provide a discussion of the dimensionless units in which we will write our equations.
In the appendices of this SM\cite{SM} we provide further details on \App{app:commutators:SST}: the coordinate and orthonormal bases used in GR (used in the computation of the spinor connection required for the DE in SST), \App{app:other:spinor:solns}: a discussion of other particular spinor solutions to the DE in SST, and finally \App{app:NumericalMethods}: a discussion of the numerical methods (with simple illustrative examples) used in this work to derive the eigenvalues of both the SR DE (as a validation of the numerical method) and DE in SST \Sec{sec:bound:states:DE:SST} and \Sec{sec:DE:SST:inside:horizon}.

As one can see, there is much the student must master in order to answer their own intuitive question.
And it may seem too daunting a task to tackle; a mountain too high to climb even for the motivated student. 
In NRQM, the solution of the Schr\"{o}dinger equation (SE) in a central potential brings in the additional complication of orbitial angular momentum operators, spherical harmonics and power series solutions of the SE, which are all standard fare in an undergraduate QM class (and clearly elucidated in many textbook, including CTQMv1).
Mastering just the SR DE involves understanding an appreciation for covariant formalism, spinor transformations under Lorentz transformations (LT), and clever manipulation of the Dirac matrices to bring out the orbital and spin angular momentum terms (as explained in detail in RQMWE).
Finally, GR itself presents conceptual and mathematical manipulation hurdles itself to the average student, even before one attempts tackling the GR DE. The latter, then involves additional concepts and mathematical manipulations of ``gauging" Lorentz symmetry, and constructing the appropriate tetrads for the local observer.
} % red from 389

In spite of this seeming mountain of required background, the claim here in this work is that the above mentioned textbooks provide more than sufficient background to  educate the motivated student step-by-step, and allow them to learn the necessary prerequisites required to answer their self-posed question. The purpose of this article is to summarize the highlights of the relevant, requisite material (rather than reproduce it in full) and guide the student to the those portions of the above mentioned textbooks that give vastly more comprehensive and worked out details. The end result is a solution of the eigenvalues of bound states in the background Schwarzschild spacetime (SST) and a comparison to the SR DE formulas and their simplified Schr\"{o}dinger plus special relativity (put in ``by hand") forms \Eq{E_s} and \Eq{E_v}, answering the student's intuitively posed question. To keep the problem as simple as possible, we focus on spherically symmetric $s$-orbital states with $L=0$, so that the gravitational $V_{eff}$ is directly analogous to Coulomb $V_C$ (i.e. we drop the angular momentum barrier terms in the former). We include Appendices with enough background material to make this as work self-contained as possible when supplemented by the aforementioned textbooks. We will also write down equations in dimensionless form using the length, energy, etc. scales appropriate to the Coulomb and gravitational problems (see \App{app:Units}), so that both equations have the same form for purposes of comparison.

So let us now embark on our investigative journey of the student's intuitive question.
%=========================================================

\section{Wave functions and eigenenergies of the NRQM Coulomb problem: a refresher}\label{NRQM:Coulomb:solns}
As any student exposed to a NRQM course is well award, the Schr\"{o}dinger equation (SE)
$i \hbar\,\partial \psi(\bs{r},t)/\partial t 
= \left( \bs{p}^{\,2}/2m + V_C(r) \right)\, \psi(\bs{r},t)$
for stationary states 
$\psi(\bs{r},t) = e^{-i E t/\hbar}\,\phi(\mathbf{r})$ for $E<0$ constant, with $E\to i \hbar \frac{\partial}{\partial t}$ and
$\mathbf{p}\to -i \hbar\,\mathbf{\nabla}$ yields 
\be{SchrEqn}
\mathcal{H}\, \phi(\mathbf{r})\equiv
\left[-\frac{\hbar^2}{2 m} \nabla^2 - \frac{e^2}{r} \right]\,\phi(\mathbf{r}) = E\,\phi(\bs{r}).
\ee
Using the differential operator identities
(see the inside cover of CTQMv1 and Arfken\cite{Arfken:2012}, Chapter 3)
\bea{Del:Opr}
\boldsymbol{\nabla} &=& \mathbf{e}_r \frac{\partial}{\partial r} - \frac{i}{\hbar r^2}\,\mathbf{r}\times\mathbf{L}, \\
\mathbf{L} &=& \frac{\hbar}{i}\,\mathbf{r}\times\boldsymbol{\nabla} = \mathbf{r}\times\mathbf{p},
\eea
where $\mathbf{p} = -i\,\hbar\,\boldsymbol{\nabla}$ is the QM momentum operator, and 
$\mathbf{e}_r = \bs{r}/r= (\sin\theta\cos\phi, \sin\theta\sin\phi, \cos\theta)$ is the orthonormal unit vector in the radial direction.

Following CTQMv1  (p662 and p778), we can write the well known formula for the Laplacian $\boldsymbol{\nabla}^2$ in spherical coordinates as
\bea{Laplacian}
\boldsymbol{\nabla}^2 &=& -\frac{\hbar^2}{2 m}\,\frac{1}{r}\, \frac{\partial^2}{\partial r^2}\, r + 
\frac{\mathbf{L}^2}{2 m r^2}, \\
\ell^2 = \frac{\mathbf{L}^2}{\hbar^2} &=& 
 -\left(
 \frac{\partial^2}{\partial\theta^2} + \frac{1}{\tan\theta}\,\frac{\partial}{\partial \theta}
 + \frac{1}{\sin^2\theta}\,\frac{\partial^2}{\partial\varphi^2 }
 \right)
\eea
Since the Coulomb force is radially directed, there is no torque, so that the  angular momentum $\mathbf{L}$  is constant of the motion. As such, we can consider the classical motion of particles as taking place in a plane, typically taken to be the equator $\theta=\pi/2$.
Thus, we seek solutions $\phi(\mathbf{r})$ involving a complete set of commuting observables
$\{ \mathcal{H}, \mathbf{L}^2, L_z\}$ such that
\be{NRQM:sep:vars}
\mathcal{H}\,\phi(\mathbf{r}) = E\,\phi(\mathbf{r}),\quad 
\mathbf{L}^2\,\phi(\mathbf{r}) = \hbar^2\,\ell\,(\ell+1)\,\phi(\mathbf{r}),\quad 
L_z\,\phi(\mathbf{r}) = m\hbar\,\phi(\mathbf{r}).
\ee
Using the well known properties of the spherical harmonics\cite{Cohen-Tannoudji:1977,Arfken:2012} 
$Y_{\ell, m}(\theta,\varphi)$
as the eigenfunctions of the angular momentum operator,
\be{Ylm}
\ell^2\, Y_{\ell, m}(\theta,\varphi) = \ell\,(\ell + 1)Y_{\ell, m}(\theta,\varphi),\quad
-i \frac{\partial}{\partial \varphi}Y_{\ell, m}(\theta,\varphi) = m\,Y_{\ell, m}(\theta,\varphi),
\ee
for $m\in\{-\ell, -\ell+1,\ldots, \ell-1, \ell\}$,
we choose the separation of variables as
$\phi(\mathbf{r}) = R(r)\,Y_{\ell, m}(\theta,\varphi)$ to obtain 
\be{radial:SE:R}
\left[
 -\frac{\hbar^2}{2 m}\,\frac{1}{r}\, \frac{\partial^2}{\partial r^2}\, r + \frac{\ell\,(\ell + 1)\,\hbar^2}{2 m r^2} + V_C(r)
\right]\,R(r) = E\,R(r),
\ee
where the spherical harmonics have been canceled from both sides of the equation, leading to a purely radial Schr\"{o}dinger equation.
In addition to \Eq{radial:SE:R} we impose, on physical grounds, the boundary condition that the radial wave function $R(r)$ decay to zero as $r\to\infty$, It is convenient to make one last substitution $R(r) = u_{k,\ell}(r)/r$, and based on physical ground argue that one needs to impose the boundary condition $u_{k,\ell}(0)=0$, at the singularity that exists at the origin for the point particle electron. This leads to the final, simplified final form of the radial equation
\be{radial:SE:u}
\left[
 -\frac{\hbar^2}{2 m}\, \frac{\partial^2}{\partial r^2}\,  + \frac{\ell\,(\ell + 1)\,\hbar^2}{2 m r^2} + V_C(r) 
\right]\,u_{k,\ell}(r) = E\,u_{k,\ell}(r), \quad V_C(r) = -\frac{e^2}{r}.
\ee
Here, the $\ell$ subscripts on $u_{k,\ell}(r)$ indicates the dependence on the angular momentum, and the index $k$ represents all other non-angular momentum dependencies of the eigensolutions.

We can now write \Eq{radial:SE:u}  in two dimensionless forms:
\bea{radial:SE:u:dimless}
\left[
  \frac{\partial^2}{\partial \rho^2}\,  - \frac{\ell\,(\ell + 1)}{\rho^2} + \frac{2}{\rho} - \lambda^2_{k,\ell}
\right]\,u_{k,\ell}(\rho) &=& 0,  \quad \rho = \frac{r}{a_0}, \quad 
\lambda_{k,\ell} = \sqrt{\frac{-E_{k,\ell}}{E_I}}\ge 0, 
\label{radial:SE:u:dimless:atomic:units}
\\ 
&{}& \no
\left[
  \frac{\partial^2}{\partial \rho^2}\,  - \frac{\ell\,(\ell + 1)}{\rho^2} + \frac{2 \alpha_C}{\rho} - \eps_{k,\ell}
\right]\,u_{k,\ell}(\rho) &=& 0, \quad \rho = \frac{r}{\lambda_c}, \quad \eps_{k,\ell} = \frac{E_{k,\ell}}{m c^2}.
\label{radial:SE:u:dimless:fundamental:units}
\eea
\Eq{radial:SE:u:dimless:atomic:units} is expressed in atomic units: i.e. lengths and energy in terms of the Bohr radius $a_0$ and ionization energy $E_I$ of the hydrogen atom (see \App{app:Units}), and is the form that is typically employed (see CTQMv1, p794).
The second form \Eq{radial:SE:u:dimless:fundamental:units} is expressed in terms of fundamental units of the electron: i.e. lengths and energy in terms of the Compton wavelength $\lambda_c=\hbar/(m c)$ and rest mass  $m c^2$ of the electron (see \App{app:Units}). While this latter form is not typically employed, we list it here for later reference when we want to consistently compare to the DE, especially in the case of SST.
This latter form explicitly brings out the role of fine structure constant $\alpha_C = e^2/(\hbar c) \approx 1/137$.

%\blue{Move the following series solution of the SE to the SM.}

% red: 505-535
{\color{black} 
Forgoing the use of (non-intuitive) special (confluent hypergeometric) functions, we will follow CTQMv1 and outline the solution to 
\Eq{radial:SE:u:dimless:atomic:units} using a power series expansion, the typical approach taught in an undergraduate QM course. Following CTQMv1, pp794-797, we outline the essentials of the solution method.
First one looks at the asymptotic form of the equation for $\rho\to\infty$, keeping only the dominant terms 
(i.e. dropping terms involving $\rho^{-1}$ and $\rho^{-2}$), which yields $[\partial^2_\rho - \lambda^2_{k,\ell}]\, u_{k,\ell}(\rho)=0$ with solutions $e^{\pm\lambda_{k,\ell}\,\rho}$. Since we want the wavefunction to decay to zero at spatial infinity (so that they are square integrable for finite total probability) we reject the positive exponent, and look for solutions of the form $u_{k,\ell}(\rho) =e^{-\lambda_{k,\ell}\,\rho} \,y_{k,\ell}(\rho)$, where
we take $y_{k,\ell}(\rho)$ of the power series form $y_{k,\ell}(\rho) = \rho^s\,\sum_{q=0}^\infty c_q\,\rho^q$ with 
$c_0\ne 0$.
The equation satisfied by $y_{k,\ell}(\rho)$ is
\be{y:eqn}
\left[
\frac{d^2}{d\rho^2} - 2\,\lambda_{k,\ell}\,\frac{d}{d\rho} +
 \left( 
 \frac{2}{\rho}-\frac{\ell(\ell+1)}{\rho^2}
 \right)\,
\right]\,y_{k,\ell}(\rho)=0,
\ee
which must be satisfied for each coefficient $c_q$ term by term. 
In particular, the $c_0$ term yields the simple equation
$[ s(s+1) - \ell(\ell+1) ]\,c_0=0$ yielding the solutions $s=\{\ell+1, -\ell\}$.
Since we have the boundary condition $y_{k,\ell}(0)=0$ we must reject $s=-\ell$ (since $\ell\ge0$) and retain
$s=\ell+1$. It is then straight forward to develop the following one term recursion 
$q (q+ 2\,\ell +1)\,c_q = 2\,[(q+\ell)\,\lambda_{k,\ell} -1]\,c_{q-1}$, for $q\ge 1$.
The key quantization concept is that if series is infinite then 
$\frac{c_q}{c_{q-1}}\underset{q\to\infty}{\sim}\frac{2\,\lambda_{k,\ell}}{q}$
which is the same ratio of coefficients in the power series expansion of
the function $e^{2\,\lambda_{k,\ell}\,\rho}$, which then violates our requirement for solutions decaying as we approach infinity. The only way to avoid this, is for the above one term recursion relation to terminate.
By examining the coefficient of $c_{q-1}$ above, this is easily satisfied if $\lambda_{k,\ell} = \frac{1}{k+\ell}$.
Thus, one obtains that the bound energies are given by $E_{k,\ell}= -E_I/(k+\ell)^2$ for a given $\ell$ for 
$k\in\{1, 2, 3, \ldots\}$ and $\ell=\{0, 1, 2, \ldots\}$.
We therefore rename $k+\ell\to n$ and recover the Bohr energies in terms of the principle quantum number $n$. 
} % red from 505
The well known radial wavefunctions, with eigenenergies $E_n = -E_I/n^2 = -\half\,\alpha_C^2\,m c^2/n^2$, are then given by (CTQMv1, p797) 
\bea{R:wavefunctions}
n=1, \ell=0:\quad R_{k=1,\ell=0}(r) &=& 2\, (a_0)^{-3/2}\, e^{-r/a_0}, \\
n=2, \ell=0:\quad R_{k=2,\ell=0}(r) &=& 2\, (2\,a_0)^{-3/2}\,\left(1- \frac{r}{2\,a_0} \right) \,e^{-r/2\,a_0}, \\
n=2, \ell=1:\quad R_{k=2,\ell=0}(r) &=& (2\,a_0)^{-3/2}\,\frac{1}{\sqrt{3}}\,\frac{r}{a_0}  \,e^{-r/2\,a_0}
\eea
Finally, since $E_I\ll m\,c^2$, one is justified in using the NRQM SE.
%=========================================================

%=========================================================
\section{The SR Dirac Equation for spin $\mathbf{1/2}$ particles and its solutions for vector (Coulomb) and scalar (mass) coupling}\label{sec:SRDE:vector:scalar}
%=========================================================
Having traversed familiar ground, we now steadfastly venture in territory that may be less familiar to some students.

\subsection{Free DE: no potentials}\label{sec:DE:no:V}
%\blue{(This is most likely less familiar material for many students, so I would suggest leaving it in the main text).\;}
As is historically well-known, Dirac was bothered by the inconsistency of the SE being first order in time, yet second order in space. Since SR treats time and space equivalently he was led to seek an equation that was first order in both time and space. For the following we will closely follow the exposition of Chapter 2 of Greiner\cite{Greiner:1990} which we denote as RQMWE. Dirac postulated a wave equation given by 
$i\,\hbar\,\frac{\pd \psi}{\pd t} = \hat{H}\,\psi$ of the form (RQMWE, p75, p82)
\be{DE:form}
i\,\hbar\,\frac{\pd \psi}{\pd t} =\left[c\,\hat{\boldsymbol{\alpha}}\cdot\hat{\mathbf{p}} 
+ \hat{\beta}\,m\,c^2\right]\,\psi \equiv \hat{H}_f\,\psi.
\ee
Here  $\hat{\mathbf{p}} = -i\,\hbar\boldsymbol{\nabla}$ is the usual QM momentum operator and again 
we have taken $\hat{E}\to i\,\hbar\,\pd/\pd t$ as the energy operator.
The quantities $\hat{\boldsymbol{\alpha}}= \{\hat{\alpha}_1, \hat{\alpha}_2,\hat{\alpha}_3\}$
and  $\hat{\beta}$ cannot be numbers if one wishes to preserve the form invariance of the equation under simple spatial rotations, and these are considered as $N\times N$ matrices.
Therefore, $\psi$ itself must be a column vector of $N$ complex functions of space and time $x=(t,\mathbf{x})$
\be{psi:DE}
\psi = 
\left(\begin{array}{c}
\psi_1(t,\mathbf{x}) \\
\psi_2(t,\mathbf{x})   \\
\vdots \\
\psi_N(t,\mathbf{x})  
\end{array}\right)
\quad\Rightarrow\quad
\rho(x) = \psi^\dag\,\psi(x) = \sum_{i=1}^N\, \psi^*(x)\,\psi(x)\ge 0,
\ee
where $\rho(x)$ is the positive probability density of the particle.

One also wishes to preserve the SR energy equation $E^2 = \mathbf{p}^2\,c^2 + (m\,c^2)^2$ where $\mathbf{p}$ without the operator circumflex $\verb+^+$ denotes the ordinary classical momentum vector.
Under the usual NR substitution (quantization) $E\to i\,\hbar\,\pd/\pd t$ and $\mathbf{p}\to -i\,\boldsymbol{\nabla}$ the SR energy equations becomes the Klein-Gordon wave equation 
$-\hbar^2\,\frac{\pd^2}{\pd t^2}\psi_\sigma = \left(-\hbar^2\,c^2\nabla^2 + m^2\,c^4 \right)\psi_\sigma$
for each component of the wave function $\psi_\sigma$ (where one typically denotes the components of the wave function with Greek indices). Carrying out this straightforward algebra (RQMWE, p76-77) the net result is that the matrices $\hat{\alpha}_i$ and $\hat{\beta}$ must satisfy
\bea{alpha:beta:DE}
\{ \hat{\alpha}_i, \hat{\alpha}_j\} &\equiv& \hat{\alpha}_i\,\hat{\alpha}_j + \hat{\alpha}_j\, \hat{\alpha}_i = 2\,\delta_{ij}\,\mathbb{I} , \label{alpha:beta:DE:line1} \\
\{ \hat{\alpha}_i, \hat{\beta}\} &=& 0, \label{alpha:beta:DE:line2} \\
\hat{\alpha}^2_i &=& \hat{\beta}^2= \mathbb{I}, \label{alpha:beta:DE:line3} 
\eea
where $\mathbb{I}$ is the $N\times N$ identity matrix.
It is not hard to show that no $N=2$ solution exists, and the first non-trivial solution that exists is for $N=4$.
This is called the Dirac 4-spinor, denoted by
\be{Dirac:4spinor}
\psi(x) = 
\left(\begin{array}{c}
\varphi(t,\mathbf{x}) \\
\chi(t,\mathbf{x})
\end{array}\right)
\ee
where $\varphi$ and $\chi$ are each themselves 2-spinors (a 2-vector with complex entries), 
familiar from NRQM.
The physical interpretation is that upper two components $\varphi$ of the 4-spinor $\psi$ represents particles of spin-$1/2$, while
the lower two components $\chi$  represents anti-particles of spin-$1/2$ (same mass, but opposite charge).

A conventional set of $4\times 4$ matrices satisfying \Eq{alpha:beta:DE:line1} -\Eq{alpha:beta:DE:line3} are given by
\be{DE:alpha:beta}
\hat{\alpha}_i = 
\left(\begin{array}{cc}
0 & \hat{\sigma}_i \\
 \hat{\sigma}_i  & 0
 \end{array}\right),
 \qquad
 \hat{\beta} = 
\left(\begin{array}{cc}
 \mathbb{I} & 0 \\
 0& -\mathbb{I} 
 \end{array}\right),
\ee
where $\sigma_i$ are the standard $2\times 2$ Pauli matrices, and 
$\mathbb{I}$ is the $2\times 2$ identity matrix,
\be{Pauli:matrices}
\sigma_1 = 
\left(\begin{array}{cc}
 0 & 1 \\
 1 & 0
 \end{array}\right),\quad
 \sigma_2 = 
\left(\begin{array}{cc}
 0 & -i \\
 i & 0
 \end{array}\right),\quad
 \sigma_3 = 
\left(\begin{array}{cc}
 1 & 0 \\
 0& -1
 \end{array}\right), \quad
  \mathbb{I} = 
\left(\begin{array}{cc}
 1 & 0 \\
 0& 1
 \end{array}\right).
\ee

Another commonly employed set of $4\times 4$ matrices $\gamma^\mu$ 
satisfying $\{\gamma^\mu, \gamma^\nu\} = 2\,\eta^{\mu\nu}\mathbb{I}$
when using the particle physicist convention for the SR metric (RQMWE\cite{Greiner:1990})
$\eta_{\mu\nu}=\eta^{\mu\nu} = \trm{diagonal}\{1,-1,-1,-1\}$,
(note: SR and GR books, e.g. see Chapter 11.3, Ryder\cite{Ryder:2009}, most commonly use the alternative convention, $\eta_{\mu\nu}=\eta^{\mu\nu} = \trm{diagonal}\{-1,1,1,1\}$ - the SR and GR metric sign convention) 
called the Dirac gamma matrices, are given by 
\be{DE:gamma:matrices}
\gamma^0=
\left(\begin{array}{cc}
 \mathbb{I} & 0 \\
 0& -\mathbb{I} 
 \end{array}\right),
 \quad
\gamma^i =
\left(\begin{array}{cc}
0 & \hat{\sigma}_i \\
 -\hat{\sigma}_i  & 0
 \end{array}\right).
\ee
These are employed so that the DE can be written in covariant form
\be{DE:covariant:form}
i\,\hbar\,\gamma^\mu\,\pd_\mu\,\psi = m\,c\,\psi, \qquad \eta_{\mu\nu} =\eta^{\mu\nu} =  \trm{diagonal}\{1,-1,-1,-1\},
\ee
where the $4\times 4$ identity matrix $\mathbb{I}_{4\times 4}$ is implied (and often omitted) on the righthand side.
In the particle physicist convention $\hat{p}^\mu = i\,\hbar\, \{\frac{1}{c}\frac{\pd}{\pd t}, -\boldsymbol{\nabla}\}$
so the covariant DE is given by $\gamma^\mu\,\hat{p}_\mu\,\psi~=~m\,c\,\psi$,
where  $\hat{p}_\mu = \eta_{\mu\nu}\,\hat{p}^\nu = i\,\hbar\, \{\frac{1}{c}\frac{\pd}{\pd t}, \boldsymbol{\nabla}\}$.

Lastly, the operator $\hat{\mathbf{S}}$ 
\be{DE:spin:opr} 
\hat{\mathbf{S}} = \frac{\hbar}{2}\,\hat{\mathbf{\Sigma}}\equiv  \frac{\hbar}{2}\,
\left(\begin{array}{cc}
 \hat{\boldsymbol{\sigma}} & 0\\
0 &   \hat{\boldsymbol{\sigma}}  
 \end{array}\right), 
 \qquad
 \hat{\mathbf{J}} =  \hat{\mathbf{L}}+\hat{\mathbf{S}}.
\ee
which commutes with the Dirac Hamiltonian $\hat{H}_f$ is called the spin vector operator, and represents the intrinsic spin of the electron, in addition to its orbital angular momentum. The total angular momentum $\hat{\mathbf{J}}$ is the sum of the orbital and intrinsic angular momentum, and is a constant (of the motion).

For stationary states (which is our primary interest) we let $\psi(t,\mathbf{x}) = e^{-i\,E/\hbar}\psi(\mathbf{x})$ for $E$ constant, such that $ i\,\hbar\,\pd/\pd t \,\psi(t,\mathbf{x}) = E\,\psi(t,\mathbf{x})$, and the DE simply becomes
 of the form (RQMWE, p82)
\be{DE:form:E}
E\,\psi =\left[c\,\hat{\boldsymbol{\alpha}}\cdot\hat{\mathbf{p}} 
+ \hat{\beta}\,m\,c^2\right]\,\psi \equiv \hat{H}_f\,\psi.
\ee

%=========================================================
\subsection{DE with potentials}\label{sec:DE:with:V}
%=========================================================
If the electron interacts with an electromagnetic potential $A^\mu = \{A^0, \mathbf{A}\}$ one obtains the 
usual NRQM substitutions (RQMWE, p94)
 $E\to i\,\hbar\,\pd/\pd t - e\,A_0$ and 
$\hat{\mathbf{p}} \to \hat{\mathbf{p}} - \frac{e}{c}\, \mathbf{A}$ 
so that the DE for stationary states becomes
\be{DE:form:EM:general}
i\,\hbar\,\frac{\pd \psi}{\pd t} =
\left[c\,\hat{\boldsymbol{\alpha}}\cdot\left(\hat{\mathbf{p}} -  \frac{e}{c}\, \mathbf{A}\right) 
+ e\,A_0
+ \hat{\beta}\,m\,c^2\right]\,\psi,  \\
\ee
If we define say the Coulomb potential in the Coulomb gauge ($\mathbf{\nabla}\cdot\mathbf{A}(t, \mathbf{x})=0$) via 
$V_v \overset{\trm{def}}{=} e\,A_0$ and $\mathbf{A}=0$ 
the DE for stationary states becomes
\be{DE:form:EM:vector:coupling}
  \left[c\,\hat{\boldsymbol{\alpha}}\cdot\hat{\mathbf{p}} 
+ \hat{\beta}\,m\,c^2\right]\,\psi =(E-V_v)\,\psi.
\ee
One can also introduce a scalar potential $V_s$ which couples to the mass so that
 $m\,c^2~\to~m\,c^2+ V_s$. Thus, the DE for stationary states, with both vector and scalar coupling, is given by 
 (RQMWE, section 9.8, p184-187)
\be{DE:form:EM:scalar:coupling}
  \left[c\,\hat{\boldsymbol{\alpha}}\cdot\hat{\mathbf{p}} 
+ \hat{\beta}\,(m\,c^2 + V_s)\right]\,\psi =(E-V_v)\,\psi.
\ee

%=========================================================
\subsection{DE in a central potential}\label{sec:DE:central:potential}
The DE in a central potential (vector coupling), in this section denoted by
$V_v \to V(r)$, presents some involved matrix and vector identity manipulations which are
explicitly worked out in RQMWE (see section 9.3, p169-172). 
The goal of this exercise is to (i) explicitly reveal the terms involving the angular momentum 
$\hat{\mathbf{L}}$ (as in the NRQM SE), as well as the intrinsic spin $\hat{\mathbf{S}}$, 
and (ii) ultimately remove the spinor dependency of the equations to produce a purely radial equation 
(actually, a pair of first order radial equations, vs a single second order radial equation as in the NRQM SE).
Rather than reproducing all these algebraic machinations in detail here, we simply highlight the essential points 
which lead to our objectives (i) and (ii).

%\blue{The important, but lengthy bit here, and hardest for the novice student, is the eventual removal of the spinors from the DE in order to end up with two coupled ordinary differential equations. However, the following could be motivated, but the details moved to the SM.}

Using the representation of 
$\hat{\boldsymbol{\alpha}}=
\tiny{\left(\begin{array}{cc} 0 & \hat{\boldsymbol{\sigma}} \\ \hat{\boldsymbol{\sigma}} & 0 \end{array}\right)}$ 
and 
$\hat{\beta}=
\tiny{\left(\begin{array}{cc}   \mathbb{I} &0 \\0& -\mathbb{I}  \end{array}\right)}$ 
given previously, and the representation of the Dirac 4-spinor 
in terms of a pair of 2-spinors 
$\psi = \tiny{\left(\begin{array}{c}\varphi \\ \chi \end{array}\right)}$ the DE becomes the pair of coupled ordinary differential equations
\bea{DE:RQMWE:p170:5}
c(\hat{\boldsymbol{\sigma}}\cdot\hat{\mathbf{p}})\chi &=& [(E-V) - m\,c^2 ]\,\varphi, 
\label{DE:RQMWE:p170:5:line:1} \\
c(\hat{\boldsymbol{\sigma}}\cdot\hat{\mathbf{p}})\varphi &=& [(E-V) + m\,c^2 ]\,\chi. \label{DE:RQMWE:p170:5:line:2} 
\eea

One must now make an ansatz about the form of the 2-spinors $\varphi$ and $\chi$.
From the NRQM SE, we suspect that they must contain the spherical harmonics 
$Y_{l,m}(\theta,\phi)$ in some form. This is the trickiest part of the separation of variables problem of the SR DE in a central potential, and involves much detailed algebra (explicitly worked out in RQMWE, section 9.3 pp169-172. For an intricate symmetry proof also see (10.54) pp212-214). These results are required in order that the angular portion of the 2-spinors drop out the of DE (analogous to the NRQM SE), 
leaving only two coupled, scalar radial equations - which is the end goal of all these algebraic manipulations.
As mentioned above, we will state some of these important results, without proof, but with explanation, %providing some additional details in \App{app:SRDE:sep:var}, 
primarily directing the reader to the detailed derivations in RQMWE.\cite{Greiner:1990}

%\blue{Move the following spinor details to the SM}

%red 765-990
{\color{black}
The ansatz for the pair of 2-spinors is
\be{2:spinor:ansatx}
\varphi\to \varphi_{j\ell m} = i\,g(r)\,\Omega_{j\ell m}(\mathbf{r}/r), \qquad
\chi\to \chi_{j\ell' m} = -f(r)\,\Omega_{j\ell' m}(\mathbf{r}/r),
\ee
where $\Omega_{j\ell m}$ is called a spherical spinor and
$\mathbf{r}/r = \mathbf{e}_r=(\sin\theta\,\cos\phi, \sin\theta\,\sin\phi, \cos\theta)$ so that the former only depend on the angles $(\theta,\phi)$ and not on $r$.
Here the indices $\{j\ell m\}$ indicate the 
total angular moment $j$ 
(the eigenvalues of $\mathbf{J}^2$ are $\hbar^2\,j(j+1)$), 
orbital angular momentum $\ell$
(the eigenvalues of $\mathbf{L}^2$ are $\hbar^2\,\ell(\ell+1)$),
and the magnetic quantum number $m$ takes values from $\{-j, -j+1,\ldots,j\}$.
Here, the value of the intrinsic spin is $s=\thalf$
(the eigenvalues of $\mathbf{S}^2$ are $\hbar^2\,s(s+1)$).
Note that the expression for $\chi$ carries an index $\ell'$ given by
\be{ell:prime}
\ell' = 
\begin{cases}
  \ell+1, \quad \trm{if}\; j=\ell+\half,\\
  \ell-1,       \quad \trm{if}\; j=\ell-\half.
\end{cases}
\ee
This is the subtlety that warrants much algebra to prove.
The physics of it lies with the fact that solutions of the SR DE (as well as the NRQM SE) must also be eigenstate of the parity operator, which simply changes 
$\mathbf{r}\to-\mathbf{r}$, and is equivalent to $\{\theta, \phi\}\to\{\pi-\theta, \pi+\phi\}$. 
This is manifested in the choice of spherical harmonic functions
$Y_{\ell m}(\theta,\phi)$
that appear in $\Omega_{j\ell m}$ and $\Omega_{j\ell' m}$.
Under the operation of parity we have have
 $Y_{\ell m}(\theta,\phi)\to Y_{\ell m}(\pi-\theta, \pi+\phi) 
 = (-1)^\ell\,Y_{\ell m}(\theta,\phi)$ so that the parity of the spherical harmonics depends solely on $\ell$.
 The implication of \Eq{ell:prime}, and the primary physical point, is that the parity of 
 $\varphi_{j\ell m}$ is the negative (i.e. opposite) of that of $\chi_{j\ell' m}$.
 Thus, under parity, the orbital orbital angular momentum value $\ell'$ of $\chi_{j\ell' m}$ can only change by one unit from that $\ell$ of $\varphi_{j\ell m}$.
 Formally, this statement is expressed as
 \be{Greiner:p171:12}
 \left(
 \hat{\boldsymbol{\sigma}}\cdot \frac{\mathbf{r}}{r} 
 \right)\,\Omega_{j\ell m}
 = -\Omega_{j\ell' m}.
 \ee
 Note the use of $\ell$ on the lefthand side of \Eq{Greiner:p171:12}, but 
 $\ell'$ on the righthand side.
 Here $\hat{\boldsymbol{\sigma}}\cdot \mathbf{r}/r$ is a scalar operator that changes sign under parity
 ($\mathbf{r}\to-\mathbf{r}$).
  \Eq{Greiner:p171:12} is a non-trivial statement which should be proved
  (explicitly, see RQMWE, (10.54) pp212-214), but for which here, we will simply utilize.

Ultimately,  the above additional complexity arises from the addition of the orbital and spin angular momentum to yield a fixed total angular momentum. 
Addition of angular momentum, and the manipulation of Clebsch-Gordon (CG) coefficients are standard topics covered in most undergraduate QM textbooks, but also quite explicitly and lucidly (with many worked examples) in the second volume (CTQMv2) of the two volume classic QM text by
Cohen-Tannoudji, Diu and  Laloe.\cite{Cohen-Tannoudji:1977:2}

The key concept is that for a given orbital angular momentum $\ell$, the addition of an intrinsic spin-1/2 ($s=\half$) leads to the two possible values for the angular momentum
$j\in\{\ell-\half, \ell+\half\}$. This is often expressed as 
$\ell\otimes\half = (\ell-\half)\oplus(\ell+\half)$ to indicate how the composition (addition, tensor product $\otimes$) of two angular momentum leads to the (direct sum, $\oplus$) of subspaces of total angular momentum
$j=\ell\pm\half$.
Thus, if $\varphi_{j\ell m}$ has orbital angular momentum $\ell$ and total angular momentum, say, $j=\ell + \half$, then since $\chi_{j\ell' m}$ has opposite parity with orbital angular momentum differing by one unit, then it can only have $\ell'=\ell+1$. 
This is the case, since by addition of its angular momenta, $\chi_{j\ell m}$ can only possibly have total the angular momentum values
$\ell'=(\ell+1)\otimes\half = (\ell+\half\equiv j) \oplus (\ell+\frac{3}{2})$.
The other potential choice $\ell'=\ell-1$ must be rejected
since it can only result in total angular momentum values of 
$\ell'\overset{?}{=}(\ell-1)\otimes\half = (\ell-\frac{3}{2}) \oplus (\ell-\half)$, neither which are $j=\ell+\half$.
(Similar arguments hold if we take $j=\ell-\half$ implying that $\ell'=\ell-1$).
The explicit expressions for the spherical 2-spinors are then given by (RQMWE, (11a) and (11b), p170 - which are not really needed - just their symmetry property \Eq{Greiner:p171:12})
\be{Greiner:p170:11a:11b}
\Omega_{j=(\ell+\half),\ell,m} = 
\left(\begin{array}{c}
\sqrt{\frac{j+m}{2j}}\,Y_{\ell,m-\half} \\
\sqrt{\frac{j-m}{2j}}\,Y_{\ell,m+\half}\end{array}\right), \quad
\Omega_{j=(\ell-\half),\ell,m} = 
\left(\begin{array}{c}
-\sqrt{\frac{j-m+1}{2(j+1)}}\,Y_{\ell,m-\half} \\
\sqrt{\frac{j+m+1}{2(j+1)}}\,Y_{\ell,m+\half}\end{array}\right).
\ee

With this (important but involved) preliminary out of the way, one can now turn to the desired goal of extracting the pair of radial equations from our coupled 
\Eq{DE:RQMWE:p170:5:line:1} and  \Eq{DE:RQMWE:p170:5:line:2}.
The solution method is straightforward ``in theory," namely we want to apply
$\hat{\boldsymbol{\sigma}}\cdot\hat{\mathbf{p}}$ to \Eq{DE:RQMWE:p170:5:line:1} so that it can act on 
$\varphi$ on the righthand side, for which we can then
substitute in  \Eq{DE:RQMWE:p170:5:line:2}. The ``rub" comes in that
$\hat{\mathbf{p}} = -i\,\hbar\,\boldsymbol{\nabla}$, and so it acts on both the radial and angular parts of $\varphi$ and $\chi$. 
We provide the relevants steps here (see the more complete details in RQMWE, p171) so that the student can see where the orbital angular momentum operator $\hat{\mathbf{L}}$ arises.

Let us first consider the following calculation:
\bea{Greiner:p171:10}
(\hbs{\sigma}\cdot\hbs{p})\,\varphi_{j\ell m} &=& 
(\hbs{\sigma}\cdot\hbs{p})\Big(i\,g(r)\,\Omega(\bs{r}/r)_{j\ell m}\Big), \no
&=& 
(\hbs{\sigma}\cdot\hbs{p})\big(i\,g(r)\big)\,\Omega_{j\ell m} + 
i\,g(r)\,(\hbs{\sigma}\cdot\hbs{p})\,\Omega_{j\ell m}, \no
&=& 
\hbar\,\frac{d g(r)}{dr} \left(\hbs{\sigma}\cdot\frac{\bs{r}}{r} \right) \,\Omega_{j\ell m}
+ i\,g(r)\,(\hbs{\sigma}\cdot\hbs{p})\,\Omega_{j\ell m},
\eea
where the parity term of \Eq{Greiner:p171:12}
$\left(\hbs{\sigma}\cdot\frac{\bs{r}}{r} \right) \Omega_{j\ell m}$
is clearly exhibited.

Let us now apply $\hat{\boldsymbol{\sigma}}\cdot\hat{\mathbf{p}}$ to $\Omega_{j\ell m}$
and use \Eq{Greiner:p171:12} 
(with $\left(\hbs{\sigma}\cdot\frac{\bs{r}}{r} \right)^2=\mathbb{I}$) 
to obtain
\be{Greiner:p171:13}
-\left(\hat{\boldsymbol{\sigma}}\cdot\hat{\mathbf{p}}\right)\,
\Omega_{j\ell m} =
\left(
\hat{\boldsymbol{\sigma}}\cdot\hat{\mathbf{p}}
\right)\,
\left(
\hat{\boldsymbol{\sigma}}\cdot\frac{\mathbf{r}}{r}
\right)\,
\Omega_{j\ell' m}.
\ee
We can now use the well known product formula for Pauli matrices
$(\hat{\boldsymbol{\sigma}}\cdot\mathbf{A})
 (\hat{\boldsymbol{\sigma}}\cdot\mathbf{B}) = 
 \mathbf{A}\cdot\mathbf{B} + i\,\hat{\boldsymbol{\sigma}}\cdot\mathbf{A}\times\mathbf{B}$, a relationship which is straightforwardly proved by expanding the vector operations in terms of components and using the well known Pauli matrix relations 
 $\sigma_i\,\sigma_j = \delta_{ij} + i\,\eps_{ijk}\,\sigma_j\,\sigma_k$.
 Applying this to the righthand side of \Eq{Greiner:p171:13} one obtains
 \be{Greiner:p171:15}
-\left(\hat{\boldsymbol{\sigma}}\cdot\hat{\mathbf{p}}\right)\,
\Omega_{j\ell m} =
\left(
\hat{\mathbf{p}}\cdot\frac{\mathbf{r}}{r}
+ i\, \hat{\boldsymbol{\sigma}}\cdot
  \left(
  \hat{\mathbf{p}}\times\frac{\hat{\mathbf{r}}}{r}
  \right)
\right)
\,\Omega_{j\ell' m},
\ee
from which one can identify the angular momentum operator
$\hat{\mathbf{L}}= \mathbf{r}\times\hat{\mathbf{p}}$.
However, one must be careful of operator ordering since 
$\hat{\mathbf{p}} = -i\hbar\,\boldsymbol{\nabla}$ is a differential operator that also
acts on $\mathbf{r}$. Thus, we write the above term in the parentheses 
on the righthand side as
\bea{Greiner:p171:16}
\big(
\hat{\mathbf{p}}\cdot\mathbf{r}
&+& 
i\, \hat{\boldsymbol{\sigma}}\cdot
  \left(
  \hat{\mathbf{p}}\times\mathbf{r}
  \right)
\big)
\,\frac{1}{r}\,\Omega_{j\ell' m}, \no
&=& 
\big(
-i\hbar\,(\boldsymbol{\nabla}\cdot\mathbf{r})
-i\,\hbar\,\mathbf{r}\cdot\boldsymbol{\nabla}
-i\, \hat{\boldsymbol{\sigma}}\cdot(\mathbf{r}\times\hat{\mathbf{p}})
\big)
\,\frac{1}{r}\,\Omega_{j\ell' m}, \no
&=& 
\left(
-i\,\hbar\,\frac{3}{r} -i\,\hbar\,r\,\left(-\frac{1}{r^2}\right)
-i\,\frac{\hbs{\sigma}\cdot\hbs{L}}{r}
\right)
\,\Omega_{j\ell' m}, \no
&=& -i\,\frac{1}{r}\,
\left(
2\,\hbar + \hbs{L}\cdot\hbs{\sigma}
\right)
\,\Omega_{j\ell' m}.
\eea
One then performs the standard calculation
$\hbs{J}^2 = \left(\hbs{L}+\half\,\hbar\,\hbs{\sigma}\right)^2 = 
\hbs{L}^2+\left(\half\,\hbar\,\hbs{\sigma}\right)^2+\hbar\,\hbs{\sigma}\cdot\hbs{L}$ 
in order to write
\bea{Greiner:p171:18}
\hbar\,\hbs{\sigma}\cdot\hbs{L}\,\Omega_{j\ell' m} &=&
\left(
\hbs{J}^2 -\hbs{L}^2-\left(\half\,\hbar\,\hbs{\sigma}\right)^2
\right)
\,\Omega_{j\ell' m}, \no
&=& 
\Big(
j(j+1) - \ell(\ell+1) - s(s+1)
\Big)
\,\hbar^2\,\Omega_{j\ell' m},  \quad \trm{for}\; s=\half.
\eea

Further, one can now cleverly define $\kappa$ via
\bea{kappa:defn}
\kappa = \mp\,(j+\half)=
\begin{cases}
-(\ell+1), \quad\trm{for}\quad j=\ell+\half, \\
  \quad\ell,       \hspace{0.55in}\trm{for}\quad j=\ell-\half,
\end{cases}
\eea
with $|\kappa| = j+\half$ or $j=|\kappa|-\half$.
Then using $\ell' \equiv 2\,j-1$ and \Eq{Greiner:p171:18}
one has (after some clever, but simple algebra, RQMWE, p171)
that the term in parentheses on the righthand side of \Eq{Greiner:p171:16} can be written as
$(2\hbar + \hbs{L}\cdot\hbs{\sigma})\,\Omega_{j\ell' m} = (1+\kappa)\,\hbar\Omega_{j\ell' m}$.
If one had instead begun this whole calculation with $\Omega_{j\ell m}$ one would similarly
arrive at 
$(2\hbar + \hbs{L}\cdot\hbs{\sigma})\,\Omega_{j\ell m} = (1-\kappa)\,\hbar\,\Omega_{j\ell m}$.
These can be rearranged (by simply subtracting $\hbar\,\Omega_{j\ell' m}$ from both sides of the equality, and similarly with $\hbar\,\Omega_{j\ell m}$) 
in the form of and eigenvalue equation for the operator
$\hat{\kappa}\overset{\trm{def}}{=} \hbar + \hbs{L}\cdot\hbs{\sigma}$ yielding
$\hat{\kappa}\,\Omega_{j\ell m}= -\hbar\,\kappa\,\Omega_{j\ell m}$ and
$\hat{\kappa}\,\Omega_{j\ell' m}= \hbar\,\kappa\,\Omega_{j\ell' m}$. 
Thus, one often defines the spinors
$\chi_{\k,\mu}\equiv\Omega_{j\ell m}$ and
$\chi_{-\k,\mu}\equiv\Omega_{j\ell' m}$
where $\mu=\pm\half$  is the magnetic quantum number indexing spin up and spin down, respectively, so that these eigenvalue equations read  equivalently as
$\hat{\k}\,\chi_{\k,\mu}= -\hbar\,\k\,\chi_{\k,\mu}$ and
$\hat{\k}\,\chi_{-\k,\mu}= \hbar\,\k\,\chi_{-\k,\mu}$. 
} % red from 765

We can therefore write the (spatial portion) of the 4-spinor in a central potential 
for stationary states as
\bea{Greiner:172:4:spinor}
\psi_{j\ell m}(\bs{r}) &=& 
\left(\begin{array}{c}\varphi_{j\ell m}(\bs{r}) \\ \chi_{j\ell' m}(\bs{r})\end{array}\right) = 
\left(\begin{array}{c}i\,g(r)\,\Omega_{j\ell m}(\bs{r}/r) \\ -f(r)\,\Omega_{j\ell' m}(\bs{r}/r)\end{array}\right), \\
&=& 
\left(\begin{array}{c}i\,g(r)\,\chi_{\k, m} \\ -f(r)\,\chi_{-\k, m}\end{array}\right) = 
i\,\left(\begin{array}{c}\,g(r)\,\chi_{\k, m} \\ i\,f(r)\,\chi_{-\k, m}\end{array}\right).
\eea
Finally, using \Eq{Greiner:p171:16} and \Eq{Greiner:p171:12}, 
\Eq{Greiner:p171:10} takes the sought after form
\be{Greiner:p172:22}
(\hbs{\sigma}\cdot\hbs{p})\,\varphi_{j\ell m} = 
-\Omega_{j\ell' m}
\left(
\hbar\,\frac{d g(r)}{dr} + \hbar\,\frac{\k+1}{r}\,g(r)
\right).
\ee
Notice the parity flip from $\ell$ to $\ell'$ in going from the 
righthand to the lefthand side.
Similarly, one derives
\be{Greiner:p172:23}
(\hbs{\sigma}\cdot\hbs{p})\,\chi_{j\ell' m} = 
-\Omega_{j\ell m}
\left(
\hbar\,\frac{d f(r)}{dr} - \hbar\,\frac{\k-1}{r}\,f(r)
\right).
\ee
We can now insert these expression into our coupled SR DE \Eq{DE:RQMWE:p170:5:line:1} and \Eq{DE:RQMWE:p170:5:line:2} with the spherical spinors cancelling from both sides of both equations (the fruition of all the above involved labor)
to obtain one of our final forms
\bea{Greiner:p172:24}
\hbar c\, \frac{dg(r)}{dr} + (1+\k)\,\hbar c\,\frac{g(r)}{r} - [(E-V(r)) + m\,c^2]\,f(r) &=& 0, \\
\hbar c\, \frac{df(r)}{dr} + (1-\k)\,\hbar c\,\frac{f(r)}{r} + [(E-V(r)) - m\,c^2]\,g(r) &=& 0.
\eea
With the standard substitution $G= r\,g$ and $F= r\,f$ 
ensuring $G(0)=F(0)=0$ at the origin singularity,
the above equations further simplify to our final form
\bea{Greiner:p172:24}
\hbar c\, \frac{dG(r)}{dr} + \hbar c\,\frac{\k}{r}\,G(r) - [(E-V(r)) + m\,c^2]\,F(r) &=& 0, \\
\hbar c\, \frac{dF(r)}{dr} - \hbar c\,\frac{\k}{r}\,F(r)+ [(E-V(r)) - m\,c^2]\,G(r) &=& 0.
\eea
%

%\blue{Insert the scalar coupling of (55-56) directly into the above equations: (53-54).}

% red 1042-1049
{\color{black}
Again, if we wish to include a scalar coupling $V_s(r)$ to the mass, in addition to the vector coupling $V\to V_v(r)$ (as computed above), these equations become
\bea{Greiner:with:vector:scalar:coupling}
\hbar c\, \frac{dG(r)}{dr} + \hbar c\,\frac{\k}{r}\,G(r) - [(E-V_v(r)) + \big(m\,c^2+V_s(r)\big)]\,F(r) &=& 0, \label{Greiner:with:vector:scalar:coupling:line:1} \\
\hbar c\, \frac{dF(r)}{dr} - \hbar c\,\frac{\k}{r}\,F(r)+ [(E-V_v(r)) - \big(m\,c^2+V_s(r)\big)]\,G(r) &=& 0. \label{Greiner:with:vector:scalar:coupling:line:2}
\eea
} % red from 1042
The coupled first order equations
\Eq{Greiner:with:vector:scalar:coupling:line:1} and 
\Eq{Greiner:with:vector:scalar:coupling:line:2}
are the SR extension of the second order radial equations in the NRQM SE.

%=========================================================
\subsection{Eigenenergies of the DE in a central potential with vector and scalar coupling}\label{DE:Solns:vector:scalar:coupling}
%=========================================================
We now wish to outline the solution for the eigenenergies of the  stationary states of the DE containing both vector (Coulomb-like) and scalar (gravitational-like) coupling. Typically, for scalar coupling, the mass acts like a position dependent mass term, and is often associated with very massive particles such as the $\sigma$ meson so that the range is very short (inversely proportional to the mass). In quantum field theory (QFT) one describes this interaction in terms of the exchange of massless scalar mesons mediating the force, in analogy with the exchange of massless photons mediating the Coulomb force. Since we are dealing with SR particles, bound states mean that the magnitude of the energy of the stationary state is less than the rest mass, i.e.
$-m\,c^2<E< m\,c^2$, where positive energies are associated with bound particles and negative energies with bound anti particles. Energies outside this range are associated with continuum states for both particles and anti-particles. 

The strategy to solve \Eq{Greiner:with:vector:scalar:coupling:line:1} and 
\Eq{Greiner:with:vector:scalar:coupling:line:2} is analogous to that of the solution of the NRQM SE, except that now one must contend with two coupled first order radial equations for the DE, vs a single second order equation for the SE. This means there will ultimately be two coupled power series expansions. Once again the quantization condition will involve the truncations of these infinite series to polynomials to ensure that the boundary condition at spatial infinite (i.e. the wave function approaches zero) is satisfied, which in turn implies the functions are square integrable, thus ensuring the finiteness of the associate probabilities over all space.
Again, one begins by examining the solution near the origin to develop a leading non-negative exponent to $r$ so that the wave function is also well behaved (i.e. zero) at the origin (the location of the point source).

%\blue{Motivate the following solution of the DE that somewhat mimics that of the solution of the previous SE, but move all the details to the SM}

% red 1067-1128
{\color{black}
The details of such a calculation for the case of the DE with both vector and scalar coupling is 
explicitly worked out in RQMWE (section 9.8, pp184-187, see also section 9.6 p178-182  for the Coulomb solution only). Here, we just outline the highlights, but strongly encourage the reader to go through the detailed worked problems in RQMWE.\cite{Greiner:1990}. 

Using the convention of RQMWE we write $V_v(r) = -\alpha/r$ for the vector coupling 
(e.g. $\alpha = e^2/\hbar c = \alpha_C$) and 
 $V_s(r) = -\hbar c \alpha'/r$ for the scalar coupling (e.g. $\alpha' = GMm/\hbar c = \alpha_G$).
 From  \Eq{Greiner:with:vector:scalar:coupling:line:1} and 
\Eq{Greiner:with:vector:scalar:coupling:line:2}, the radial equations take the form
\bea{Greiner:p184:2}
\frac{dG(r)}{dr} &=& -\frac{\k}{r}\,G(r) + 
\left[ \frac{E + m\,c^2}{\hbar c} + \frac{\alpha-\alpha'}{r}\right]\,F(r), \label{Greiner:p184:2:line1} \\
\frac{dF(r)}{dr} &=& \frac{\k}{r}\,F(r) - 
\left[ \frac{E - m\,c^2}{\hbar c} + \frac{\alpha+\alpha'}{r}\right]\,G(r). \label{Greiner:p184:2:line2}
\eea

We first consider the region $r\sim 0$, where we can then keep only the $1/r$ terms and drop the constant terms. With the ansatz $G=a\,r^\g$ and $G=b\,r^\g$ were are lead to a homogeneous set of linear equations 
$a(\g+\k) - b(\alpha-\alpha')=0$ and
$a(\alpha-\alpha')+b(\g-\k) =0$, which upon setting the determinant of the coefficients equal to zero yields $\g = \pm \sqrt{\k^2 -\alpha^2+\alpha^{'2}}$. 
To allow for normalization of the wave functions, one must select the positive sign for $\g$.
One can now define $\lambda \defeq \frac{\sqrt{m^2 c^4-E^2}}{\hbar c}$ and $\rho = 2\lambda r$ so that the radial equations become 
\bea{Greiner:p185:11}
\frac{dG}{d\rho} &=& -\frac{\k}{\rho}\,G + 
\left[ \frac{E + m\,c^2}{2\hbar c\,\lambda} + \frac{\alpha-\alpha'}{\rho}\right]\,F, \\
\frac{dF}{d\rho} &=& \frac{\k}{\rho}\,F  
-\left[ \frac{E - m\,c^2}{2\hbar c\,\lambda} + \frac{\alpha+\alpha'}{\rho}\right]\,G.
\eea

It turns out to be convenient to define the functions $\phi_1(\rho)$ and $\phi_2(\rho)$ via
$G = \sqrt{mc^2+E}\, e^{-\rho/2}\,$ $(\phi_1+\phi_2)$ and
$F = \sqrt{mc^2+E}\, e^{-\rho/2}\,(\phi_1-\phi_2)$ 
and expand each in a power series given by
$\phi_1=\rho^{\gamma}\,\sum_{m=0}^\infty \alpha_m\,\rho^m$ and
$\phi_2=\rho^{\gamma}\,\sum_{m=0}^\infty \beta_m\,\rho^m$.
The justification for introducing $\phi_1$ and $\phi_2$ is that resulting equation for $\phi_2$ allows one to write the coefficients $\beta_m$ in terms of the $\alpha_m$, via (see RQMWE, p186)
\bea{Greiner:p186:17}
\frac{\beta_m}{\alpha_m} &=& 
\frac
{
-\k + \alpha m c^2/\hbar c\lambda + \alpha' E/\hbar c\lambda
}
{
m+\g - \alpha m c^2/\hbar c\lambda - \alpha'E/\hbar c\lambda
}, \no
&=&
\frac
{
\k - \alpha m c^2/\hbar c\lambda - \alpha' E/\hbar c\lambda
}
{n'-m}, \no
\trm{with}\;\; n' &\defeq& \frac{\alpha E}{\hbar c \lambda} + \frac{\alpha' m c^2}{\hbar c \lambda}-\g.
\eea
Inserting this into the series relationship from $\phi_1$, which involves
$\alpha_m$ expressed in terms of $\alpha_{m-1}$ and $\beta_m$,
one can develop an expression for $\alpha_m/\alpha_{0}$ whose numerator
involves the product $(n'-1)(n'-2)\ldots(n'-m)$.
} % red from 1067
Once again, if the series do not terminate, they lead to a sum that scales as $e^{\rho}$ so that the wave function does not converge at spatial infinity. Thus, as in the SE quantization case, requiring the series to terminate, demands that $n'=\{0,1,2,\ldots\}$.
Inserting this result back into \Eq{Greiner:p186:17} this equation produces a quadratic equation for the energy  
$E$ (since $\lambda\sim E$), which can be easily solved to finally give
\bea{Greiner:p186:20}
\frac{E}{mc^2} &=& 
\left\{ \frac{-\alpha\alpha'}{\alpha^2 + (n-(j+\half)+\gamma)^2} 
\pm 
\left[
\left(
\frac{\alpha\alpha'}{\alpha^2 + (n-(j+\half)+\gamma)^2}
\right)^2
\right. 
\right.
\no
&-&
\left.
\left.
\frac
{
\alpha^{'2} - (n-(j+\half)+\gamma)^2
}
{
\alpha^{2} + (n-(j+\half)+\gamma)^2
}
\right]^{1/2}
\right\}.
\eea
In the above, the principal quantum number $n$ is defined as $n=n'+|\k| = n' + j + \half$ with
$n\in\{1, 2, \ldots\}$.
Special cases of the above are both interesting and relevant.

\subsubsection{Vector Coupling only}
Here $\alpha'=0$ and $\g = \sqrt{\k^2-\alpha^2}$ apropos for the Coulomb field and one obtains
\be{Greiner:p186:case:2}
\frac{E}{mc^2} = 
\left[
1+ \frac{\alpha^2}{(n-(j+\half)+\g)^2}
\right]^{-1/2} \approx 1 - \half\frac{\alpha^2}{n^2} + \ldots,
\ee
recovering the SE result of \Eq{E_v} since $\alpha=\alpha_C \approx 1/137\ll 1$.
(Note: the negative energies do not fulfill the original quadratic equation 
\Eq{Greiner:p186:17} for $E$ in the case of vector coupling  
since both $n'+\g$ and $\frac{\alpha\,E}{\hbar c \lambda} + \frac{\alpha' m c^2}{\hbar c \lambda}$ are positive,
and so  that the $E<0$ solutions must be excluded).
Note that $\gamma^2=\kappa^2-\alpha^2\ge0$ implies $|\alpha|\le|\kappa| = (j+1/2)$ from \Eq{kappa:defn}.
Further, note that if we change $\alpha\to Z\alpha$ for a nucleus of charge $Z$, and take $j=1/2$ and $n=1$ \Eq{Greiner:p186:case:2} reduces to (recall $|\kappa| = j +1/2$)
\be{Greiner:p182:Exercise:9.7}
\frac{E}{mc^2} = 
\left[
1+ \frac{(Z\,\alpha)^2}{1-(Z\,\alpha)^2}
\right]^{-1/2} 
= \sqrt{1 - (Z\,\alpha)^2}, \quad\Rightarrow\quad Z<\frac{1}{\alpha} = 137.
\ee
The subsequent prediction that no bound states exist for nuclei with $Z>137$ can be understood in connection with the  supercritical phenomena of the \tit{collapse of the vacuum} (see RQMWE\cite{Greiner:1990} p182) in which the upper and lower bounds 
$\pm m\,c^2$ for the region of bound states are ``pulled" towards each other with increasing $Z$, and collapse at $E=0$ at $Z=137$.

\subsubsection{Scalar Coupling only}
In the case of scalar coupling $\alpha=0$ and $\g = \sqrt{\k^2+\alpha^{'2}}$ (apropos for a gravitational-like field),  one obtains instead the expression 
analogous to the SE result \Eq{E_s}
\be{Greiner:p186:case:1}
\frac{E}{mc^2} = 
\pm\left[
1- \frac{\alpha^{'2}}{(n-(j+\half)+\g)^2}
\right]^{1/2} \approx \pm\left(1 - \half\frac{\alpha'^2}{n^2} + \ldots\right), \quad\trm{if}\quad \alpha'\ll 1.
\ee
Further, the term in the square parentheses in \Eq{Greiner:p186:case:1} is positive for all values of $\alpha'>0$, and in fact  $\tfrac{E}{mc^2}\approx \sqrt{\tfrac{2\,(n-(j+1/2))}{\alpha'}}$ for $\alpha'\gg 1$.

Thus, the above two formulas \Eq{Greiner:p186:case:2} and \Eq{Greiner:p186:case:1}
almost answer the student's original question when we note the similarity of the placement of the square roots when comparing 
\Eq{E_v} with \Eq{Greiner:p186:case:2},
and 
\Eq{E_s} with \Eq{Greiner:p186:case:1}.
We see that in the small coupling limit 
both formulas have the form of $1-\frac{\alpha^2}{ 2 n^2}$, but this requires that 
$\alpha'=GMm/\hbar c = \frac{M}{M_p} \frac{m}{M_p}\ll 1$ (see \Eq{alpha:G}).
 While this is true say for $\sigma$ mesons and other elementary particles, for the originally posed question of a solar mass BH,  $\frac{M}{M_p}$ is astronomically huge, so that
 $\alpha'\to\alpha_G\approx 4\times 10^{15}\gg 1$ (to say the least!). Still, one's intuition is somewhat borne out by this analysis, but not fully yet answered to the student's satisfaction. 
Thus, in order to fully answer the student's question, we need to turn now to the DE in curved spacetime where the role of the metric encoding the gravitational field enters in a fundamentally new way. Further, the strange new concept of a horizon, dividing spacetime into two distinct regions: outside and inside, with its conceptually unusual (and confounding) ``one-way membrane" property (i.e. particles can cross the horizon from outside to in, but not the reverse!) rears it strange and complicated head.
%=========================================================

%=========================================================
\section{The DE in curved spacetime}\label{sec:DE:CST}
%=========================================================
%\blue{Keep the spirit of this motivating Section Intro (which introduces concepts that are probably new to the novice student), but make it more succinct. }

General Relativity (GR) presents its own unique challenges, both conceptually and mathematically, even to the motivated student. However, most students are familiar at least with some basic concepts, including the curved spacetime metric, black holes (BH) and particle orbits about them, and the bending of light rays around a massive object, even if they might not be quite capable of deriving them. Therefore, while there is some degree of heavy lifting involved when learning GR (the fundamentally new concept of spacetime vs space and time, tensors, covariant derivatives, parallel transport, etc...) there currently exists enough excellent GR textbooks now geared to the upper level undergraduate that attempts to ease these burdens. One excellent text that we will use here is that by James Hartle\cite{Hartle:2009}, which emphasizes ``physical concepts first" before mathematical manipulation. For most students (of all ages!) it's the concepts that both SR and GR introduce that are at first glance difficult to wrap one's head around. But, 100 years on now, there has been enough well explained exposition (and textbooks) that physics such as BHs and the bending of light are now part of general (and popular) knowledge. 

Therefore, the goal here is not review all of GR, but just those salient points that will get us to our goal most expeditiously. 
For those familiar with GR at the level of particle orbits in SST one may wish to simply peruse \Sec{sec:Veff:SST}.
Those with already familiar with covariant derivatives may wish to skim through \Sec{sec:DE:CST:preliminaries} and \Sec{sec:DE:CST:covar:deriv:GR}.
On top of these topics, we also want to include the non-standard (for the novice) topic of the DE in curved spacetime (CST). To that end, there does exist an excellent undergraduate GR text by Lewis Ryder\cite{Ryder:2009} that clearly and lucidly deals with this topic, leading to the DE in CST (see Ryder, Chapter 11.3, pp409-416). Again, it is not our goal to reproduce all of Ryder's illuminating discussion.  In the spirit of wanting to drive our car first before knowing how the engine under the hood was built (but allowing ourselves to read the service manual now and again), we again will point out the key concepts involved and refer the reader to Ryder's excellent GR textbook.

In the following we will first draw from Hartle\cite{Hartle:2009} (see section 9.3, p191-204) in order to get to a form of the classical GR radial equation (no quantum here!) that looks like an ordinary NR energy equation. This involves being given the Schwarzschild metric (sans derivation), and manipulating it (the 4-velocity) in order to produce the desired equation. The goal is to identify the GR radial potential for comparison to the DE in CST.

We then next switch gears, and focus on the derivation of the DE in CST (Ryder, Chapter 11.3, pp409-416). The key concept from GR is the covariant derivative, required to explain how to take derivatives in the surrounding CST as one moves from point to point. As mentioned earlier, the global inertial frames of SR are now ``collapsed" to local regions around each point which (following Hartle) we refer to as the observer's local laboratory. As long as the extent of the observer's spatial axes, and the duration of time  measured, are in a sense ``small" (with respect to changes in the curvature of the CST), the Equivalence Principle holds, and hence physics appears special relativistic. Thus, one must be able to describe the observer's local (frame) laboratory, and hence we are led down the road of introducing a tetrad of 4-vectors describing the local laboratory. Measurements and description of particles (e.g. their 4-momentum) passing through the observer's local laboratory are then made by ``projecting" then onto the observer's local axes (3 space and 1 time axes). 

The introduction of tetrads is the key concept required then to similarly ``project" the DE in the surrounding CST  ``into" the observer's local frame, which is locally (Minkowski) flat (i.e where SR holds). The relevant concept could be stated as such: ``Since we already know how to write down and solve the DE in SR (see the previous section), then by use of the Equivalence Principle, the strategy is to ``simply" project the CST DE into the observer's locally flat frame at each spacetime point $x$, where we already know how to solve it." This is what Ryder does in Chapter 11.3. Again, our objective is to get to the DE in CST as quickly as possible, so again we point out the major highlights of this derivation (with an emphasis more on the why and how we get there, which can be daunting to the un-initiated, rather than on the detailed proofs - which are shown in Ryder). 

So let us begin.
%=========================================================

%=========================================================
\subsection{The effective radial potential for particles in the Schwarzschild metric}\label{sec:Veff:SST}
%=========================================================
As is well known, the Schwarzschild metric for a central symmetric gravitational source of mass M is (in full units, see Hartle, p186)
\bea{Schw:metric}
\hspace{-0.5in}
ds^2 &=&-\F(r)\,c^2\,dt^2 + \frac{1}{\F(r)}\,dr^2 + r^2\,d\Omega^2, \;\;
\F(r) \defeq 1-\frac{2GM}{c^2 r}, \;\; 
d\Omega^2 = \left(d\theta^2 + \sin^2\theta\,d\phi^2 \right), \label{Schw:metric:line:1} \\
\hspace{-0.5in}
&=& g_{\mu\nu}\,dx^\mu\,dx^\nu, \qquad \mu,\nu\in\{0,1,2,3\}, \label{Schw:metric:line:2}\\
\hspace{-0.5in}
&\approx& -\left(1+\frac{2 V_G(r)}{c^2}\right)\,c^2\,dt^2 + \left(1-\frac{2 V_G(r)}{c^2}\right)\,dr^2 + r^2\,d\Omega^2, 
\qquad 
V_G(r) = -\frac{G M}{r},\label{Schw:metric:line:3}
\eea
where we have defined the Schwarzschild factor $\F(r) \defeq 1-\frac{2GM}{c^2 r}$ that will figure prominently throughout the discussions of Schwarzschild spacetime (SST). (Note that we have switched here to the SR and GR metric sign convention of $\{-1,1,1,1\}$).
\Eq{Schw:metric:line:3} is the Schwarzschild metric in the weak field approximation (apropos for, say about the Sun or the Earth), in which the Newtonian potential $V_G(r) = -\frac{G M}{r}$ explicitly appears. The relevant scale length (see \App{app:Units}) is the well known Schwarzschild radius $r_s\defeq 2GM/c^2$ 
(so that $2 V_G/c^2 = -\left(\frac{r_s}{r}\right)$ is unitless). 
For a stationary observer (i.e. one that sits at a fixed coordinate position) at fixed $r$, the metric yields
$ds^2 \defeq -d\tau^2 = -\F\,c^2\,dt^2 = g_{00} (dx^0)^2$ (taking $x^0= c t$) where $\tau$ is then seen as the \tit{proper time}, i.e. the time as measured on a clock carried with the observer.

\begin{figure}[h]
%\begin{tabular}{cc}
%\includegraphics[width=4.5in,height=3.0in]{fig_observers_local_laboratory_alsing_spin_half_paper_fig1} 
\includegraphics[width=4.5in,height=3.0in]{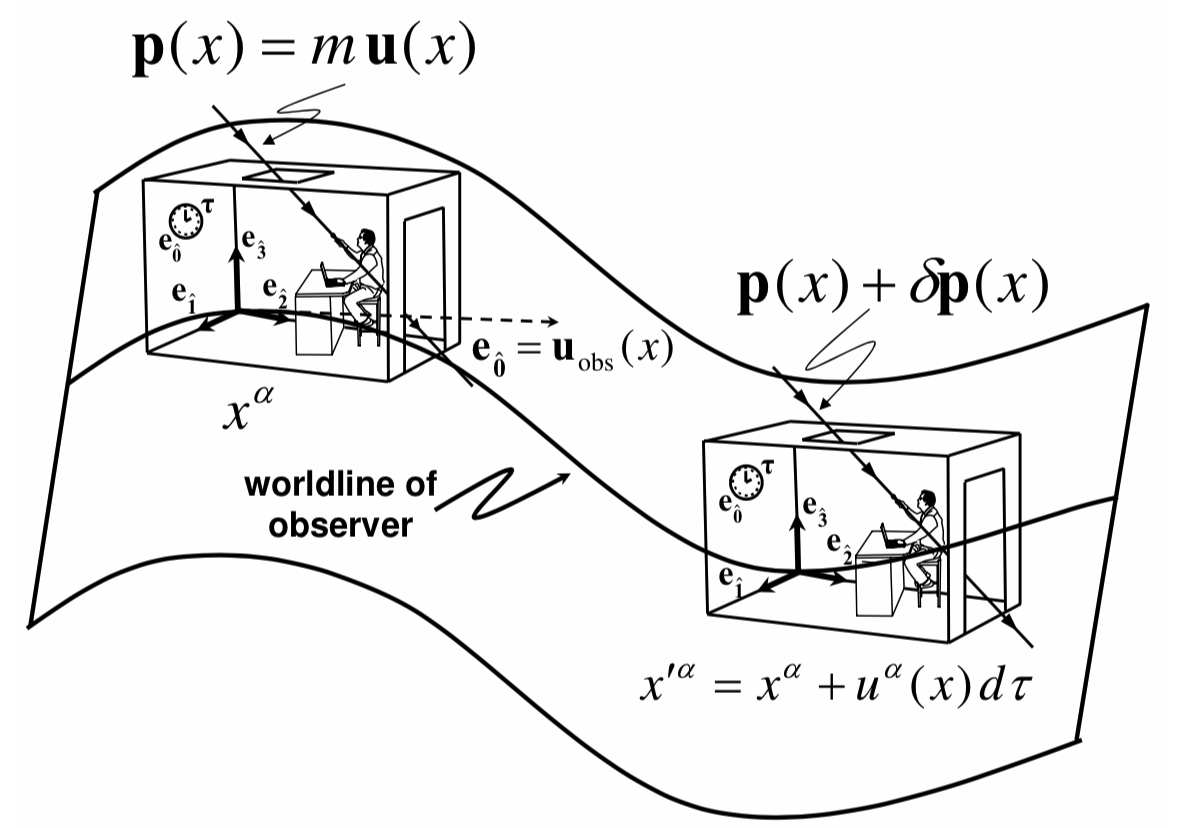}
%&
%\includegraphics[width=3.0in,height=1.5in]{alphaplus_wfdivw0_2_beta_0p1}
%\end{tabular}
\caption{The observer's \tit{local laboratory} (small room with physicist) at the curved spacetime (CST) point $x$, defined by the orthonormal tetrad $\e_a(x)$, $a=(0,1,2,3)$. The three spatial axes $\e_i(x)$, $i=(1,2,3)$ are located at the origin of the observer's laboratory (the universal coordinates system ``in the corner of the room"), while  $\e_0(x)=\u_{obs}(x)$ is the temporal axis, defined as the observer's 4-velocity, which is the tangent to the physicist's geodesic trajectory. A particle of 4-momentum $\p(x) = m \u(x)$ and world components $p^\alpha(x)$ passes through the observer's local laboratory. The observer measures the \tit{local} laboratory components $p^a(x) = e^a_{\;\;\alpha}(x)\,p^\alpha(x)$. At a small proper time later $d\tau$, the particle has moved from 
$x^\alpha \to x^{'\alpha} = x^\alpha + u^\alpha(x)\,d\tau$, which is measured by the observer in their local laboratory at the spacetime point $x^{'\alpha}$.
}
\label{fig:observers:local:laboratory} 
\end{figure}

Writing the metric as 
$-d\tau^2 =ds^2 = g_{\mu\nu}\,\left(\frac{dx^\mu}{d\tau^2}\right)\,
                                                                                   \left(\frac{dx^\nu}{d\tau^2}\right)\, d\tau^2 
                                                                                   \defeq \u^2_{obs}\, d\tau^2
$
we have $\u^2_{obs}~=~-~1$, where we have defined the observer's 4-velocity
$\u^\mu = dx^\mu/d\tau$, i.e. the velocity of the observer (rate of change of their coordinates) with respect their proper time $\tau$ (vs their coordinate time $t=t(\tau)$).
The observer's 4-velocity will be in fact the ``time axis" of the local local laboratory, which we denote
as $\e_0(x) \defeq \u_{obs}(x)$ a timelike unit vector (i.e. has magnitude $-1$) and is tangent to the  (freely falling geodesic) trajectory of the observer's motion (worldline) through the surrounding CST.
The other three spatial axes of the observer's local laboratory, denoted as
$\{\e_1(x), \e_2(x), \e_3(x)\}$ are chosen orthogonal to $\e_0(x)$ 
(see \Fig{fig:observers:local:laboratory}), and will be discussed later.
For now, in order to extract radial orbits of the Schwarzschild metric, we only require $\u_{obs}(x)$.
Note that if a particle of 4-momentum $\p$ passes through the observer's local laboratory, 
then the observer will measure its energy as $E/c= -\p\cdot\u = -\p\cdot\e_0$, and its 3-momentum components as
$p_a=\p\cdot\e_a$ for $a\in\{1,2,3\}$ yielding a local description of the particle with 
a SR-like local laboratory 4-vector $(E/c, \bs{p})$ (as an instantiation of the Equivalence Principle). 

%\blue{Keep the visual representation of observer's local laboratory and tetrads, but make the figure smaller, and reduce the size of the figure caption.}
%============================
%============================

%\blue{The next few paragraphs introduce the concept of Killing vectors and conserved quantities along the geodesic. Motivate, but move details to the SM}

% red 1275-1291
{\color{black} 
Particles (here both massive and massless until otherwise specified) under no other external forces (i.e. accelerations) travel on \tit{geodesics}, the analogue of straight line motion in Euclidean space, extended to CST. Such observers are called freely falling (FF). Now without delving into the geodesic equation per say (since we will not need to solve this equation directly in the subsequent discussion), an important concept is that of conserved quantities along the geodesic motion. If the metric is independent of a particular coordinate, then a conserved quantity exist along this motion (called an isometry; this is essentially Noether's theorem applied to geodesics, see Hartle, section 8.2, pp175-178). For example, for the Schwarzschild metric, we see that it is independent of the coordinates $t$ (stationarity) and azimuthal angle $\phi$ (rotationally invariant about the $z$-axis). 
 Let $\bs{\xi}$ be a coordinate vector along the the \tit{isometry} (i.e. motion along which the metric does not change). 
$\bs{\xi}$ is called a \tit{Killiing vector} (after Wilhelm Killing (1847-1923), see Hartle, Chapter 8.2, pp175-178) and there exists an equation named after him, that allows one to derive the isometries systematically for any metric. However, quite often one can deduce the Killing vectors by inspection
of the symmetry of the metric.
 The key result is that $\bs{\xi}\cdot\u_{obs}$ is a conserved quantity all along the geodesic.
 For the Schwarzschild metric, the metric does not change for translations in time $t$, so that 
 $\bs{\xi}_t = (1,0,0,0)$, and therefore we define the quantity $e\defeq -\bs{\xi}_t\cdot\u_{obs}$, which can be physically interpreted as the particle's rest energy per unit mass (at large $r$).
In the following we will follow the convention of Hartle (and most GR textbooks) and work in units of $G=c=1$. Physical units can be restored by resorting to dimensional analysis.
A second conserved quantity for particles on Schwarzschild geodesics is the 
$\ell = \bs{\xi}_\phi\cdot\u_{obs}$ the orbital angular momentum per unit mass (at large $r$), with 
$\bs{\xi}_\phi = (0,0,0,1)$ (in spherical polar coordinates $(t, r, \theta, \phi)$) indicating the independence of the Schwarzschild metric in the azimuthal coordinate $\phi$. Using these two conserved quantities and the timelike unit magnitude of the observer's 4-velocity, there exists enough symmetry to deduce the particle orbits directly. Note that since the orbital angular momentum is conserved, the geodesic motion occurs in a plane, which is conventionally (for convenience) take to be the equator, $\theta=\pi/2$.
} % red from 1275

In the following we will work in a \tit{coordinate basis} which means that 
$\e_a = \delta^\mu_a\,\pd_\mu$ are just coordinate derivatives 
in the direction indicated and their inner products
gives the metric components. This means that for two arbitrary 4-vectors $\bs{a}$ and $\bs{b}$ 
with coordinate components $a^\mu$ and $b^\nu$, respectively, we 
have $\bs{a}\cdot\bs{b}= g_{\mu\nu}a^\mu\,b^\nu$.
Thus, for the Schwarzschild metric we define the conserved quantities as
(following Hartle and using units of $G=c=1$)
\bea{Hartle:p193:9.21:9.22}
e &=& -\bs{\xi}_t\cdot\u_{obs} = \left(1-\frac{2 M}{r}\right)\,\frac{dt}{d\tau}=\F(r)\,\frac{dt}{d\tau}, \label{Hartle:p193:9.21} \\
\ell &=& \bs{\xi}_\phi\cdot\u_{obs} = r^2\sin^2\theta\,\frac{d\phi}{d\tau}. \label{Hartle:p193:9.22} 
\eea 
With the above conserved quantities in hand, and 
$\u_{obs} = (u^t, u^r, u^\theta, u^\phi)$ we have from its normalization $\u_{obs}\cdot\u_{obs}=-1$,
\bea{Hartle:p194:9.25}
-1 &=&-\left(1-\frac{2M}{r} \right)\,\left(u^t\right)^2
+ \left(1-\frac{2M}{r} \right)^{-1}\,\left(u^r\right)^2
+ r^2\,\left(u^\phi\right)^2, \\
\trm{or}\quad  -1 &=& 
-\left(1-\frac{2M}{r} \right)^{-1}\,e^2
+ \left(1-\frac{2M}{r} \right)^{-1}\,\left(\frac{dr}{d\tau}\right)^2
+ \frac{\ell^2}{r^2}, \\
\Rightarrow\quad
\mathcal{E} \defeq \frac{e^2-1}{2} &=& \half\,\left(\frac{dr}{d\tau}\right)^2
+ 
\left[
\left(1-\frac{2M}{r} \right)\,\left(1+\frac{\ell^2}{r^2} \right)-1
\right], \\
\trm{or}\quad
\mathcal{E}&=& \half\,\left(\frac{dr}{d\tau}\right)^2 + V_{eff}(r), \label{Netwon:E:eqn}
\eea
where we have defined the effective potential (see Hartle, p194)
\bea{V:eff}
V_{eff}(r)\equiv
\left[
\left(1-\frac{2M}{r} \right)\,\left(1+\frac{\ell^2}{r^2} \right)-1
\right]
&=&
-\frac{M}{r} + \frac{\ell^2}{2\,r^2} - \frac{M\,\ell^2}{r^3}, \label{V:eff:line1} \\
&=& 
\frac{1}{c^2}\,
\left(
-\frac{GM}{r} + \frac{\ell^2}{2\,r^2} - \frac{G M\,\ell^2}{c^2\,r^3} \label{V:eff:line2}
\right),
\eea
where in the last line we have restored all the physical constants.
\Eq{Netwon:E:eqn} has the form of a NR energy equation with a Newtonian-like potential with orbital angular momentum barrier (first two terms of \Eq{V:eff:line2}), but with an additional attractive GR correction (last term of \Eq{V:eff:line2}) that scales as $1/r^3$. This last term dominates for small $r$ and is responsible for the characteristic GR effects such as the precession of the perihelion, bending of light, etc.
We can make the Newtonian analogy even stronger by defining $E_{Newt}$ via
$e^2 = (mc^2 + E_{Newt})/mc^2$, i.e. as the small correction to the particle's rest mass, 
in strong analogy  to SR.
%to the bound energies of an electron in Coulomb (vector, or scalar) field.
With this substitution, the radial equation becomes (see Hartle p195, restoring full units)
\be{Hartle:p195:9.32}
E_{Newt} \approx \frac{m}{2}\,\left(\frac{dr}{d\tau}\right)^2 + \frac{L^2}{2 m r^2} 
-\frac{G M m}{r} - \frac{G M L^2}{c^2 m r^3}, \qquad L\defeq m\,\ell,
\ee
where we have approximated $\mathcal{E}=(e+1)(e-1)/2\approx (e-1)$.
\Eq{Hartle:p195:9.32} now has the exact same form as the energy integral in Newtonian gravity with
an additional relativistic correction to the potential proportional to $1/r^3$.
Note that
\be{Hartle:p195:9.33}
V_{eff}(r) \underset{r\to\infty}{\longrightarrow} -\frac{G M}{c^2\,r} = -\frac{r_s/2}{r}, \qquad
V_{eff}(r_s) = -\half. 
\ee
Further, at the Schwarzschild radius $r=r_s$ the first term of $V_{eff}(r_s)=-\half$, and the second and third terms exactly cancel each other.

 A detailed investigation for both radial plunge orbits ($\ell=0$), circular and hyperbolic orbits ($\ell\ne 0$)
 are well explored in Hartle Chapter 9 (and many other GR textbooks), and will not be covered here.
 For our goals, we are primarily interested in radial plunge orbits, since in this case 
 $V_{eff}(r) = -\frac{G M}{c^2\,r}=-\frac{r_s/2}{r}$ has the exact form of the scalar coupling potential, discussed previously for the SR DE.
 
 The question now, is how to incorporate the GR effects into the DE. We turn to this next, leveraging the discussion in Ryder\cite{Ryder:2009} (Chapter 11.3).
 
 %=========================================================
 \subsection{The DE in CST: GR preliminaries}\label{sec:DE:CST:preliminaries}
 %=========================================================
% \blue{I believe this section should be retained in the main text, since we will eventually write the DE in an orthonormal basis, so distinguishing it from a coordinate basis can be confusing to the novice student.}

In a seminal paper in 1976 Chandrasekhar \cite{Chandrasekhar:1976} showed that the Dirac equation could be separated in the Kerr metric for a rotating (uncharged) black hole, a surprising and unanticipated result at the time.
The separation of the angular momentum from the radial equations was a major accomplishment, and here can be appreciated from the machinations that were discussed to perform such a separation in the SR DE discussed in \Sec{sec:SRDE:vector:scalar}.
The method involved a sophisticated Newman-Penrose spinor formalism  which is fully explained in his subsequent famous book \tit{The Mathematical Theory of Black Holes} \cite{Chandrasekhar:1992}. Very shortly after, Page extended this result to the  (Kerr-Newman) charged, rotating black hole. The prime motivation of these works was to study scattering of massive particles and electromagnetic fields off the potential barriers of black holes (BHs).

Extensions of Chandrasekhar's work using the Newman-Penrose formalism has continued into the present day, including investigations into quantizing the DE around various BHs \cite{Mukhopadhyay:1999,Jin:2000,Mukhopadhyay:2008}. The work closest in spirit of this present work, employing the \tit{tetrad} (vs Newman-Penrose) formalism discussed below, is that of Semiz \cite{Semiz:1992} for a magnetically charged BH. However, none of these works discuss, or are easily adaptable to, bound state solutions of the DE in SST, as is the focus of this work. It is this author's belief that the current ``pedagogical" approach presented in this work is a more accessible introduction for students into these topics, which could then be follow on investigations. Detailed comparison between this current work and the ones discussed above will be the subject of future research efforts.
 
In order to arrive to our desired destination, the DE in CST, we first need to stop once more at the discussion of the description of the observer's local laboratory. We need to distinguish between the use of various ``basis vectors" and corresponding ``1-forms" used to describe the observer's frame. The reason is that since the observer's local (frame) laboratory is described in terms of four basis vectors $\e_a$, we want to distinguish between coordinate bases (cryptically called \tit{holonomic} in the literature) and an orthonormal set of basis vectors (called, \tit{non-holonomic}). Why the two descriptions? In the spirit of GR, \tit{any} set of basis vectors are allowed, but these two are the one's most conveniently employed. 
Coordinate basis vectors are the ``easiest" to use and formally of the form $\e_\mu = \pd_\mu \defeq \pd/\pd x^\mu$ for coordinates $x^\mu$. (Here we have used $a\to\mu$ since the index $\mu$ reflects the surrounding CST). The defining property of a coordinate basis set is that their inner (dot) product defines the metric $\e_\mu \cdot \e_\mu = g_{\mu\nu}(x)$, and that the basis vectors commute, namely
$[\e_\mu, \e_\nu]=[\pd_\mu, \pd_\nu]= 0$ for $\mu\ne\nu$ reflecting the independence of the order of coordinate differentiation, i.e. $\pd_\mu\,\pd_\nu= \pd_\nu\,\pd_\mu$.

On the other hand, the orthonormal basis is more physical, and better suited to what the observer actually measures. It's defining property is that the inner product between the orthonormal basis vectors defines the \tit{local} metric, which in this case is the flat Minkowskian metric of SR.
The orthonormal basis vectors are of the form $\e_a = e_a^{\;\;\mu}(x)\,\pd_\mu$ where the 
\tit{tetrad components} $e_a^{\;\;\mu}(x)$ are spatially dependent. Thus, 
$[\e_a, \e_b]\ne 0$ in general. However, they do have the simplifying property that
$\e_a\cdot\e_b = \eta_{ab} = \trm{diagonal}(-1,1,1,1)$ the constant Minkowski SR metric - which is just a statement of the Equivalence Principle. 
(Note, the relativist's minus sign convention in front of the time component, vs the particle physicists convention of using  $\eta_{ab} = \trm{diagonal}(1,-1,-1,-1)$. If you pick up a random book an look at the ``sign of the times", you can instantly tell if the author is a particle physicist or a relativist, without even looking at the title. Try it!).
To recap this important distinction, we recap this once more below:
\bea{basis:vectors}
\trm{coordinate basis:}\qquad \e_\mu &=& \pd_\mu, \qquad\quad \e_\mu\cdot\e_\nu = g_{\mu\nu}(x), \\
\trm{orthonormal basis:}\qquad \e_a &=& e_a^\mu\pd_\mu, 
\qquad \e_a\cdot\e_a = \eta_{ab}(x).
\eea
(Note: to distinguish the two basis, some authors use a circumflex $\verb+^+$ over the index $\hat{a}$ so that $\e_0$ and $\e_{\hat{0}}$ denote the observer's 4-velocity 
in a coordinate and orthonormal basis, respectively.
 We will not have occasion to do this, since we will primarily use an orthonormal basis for the DE in CST).
 
 An example speaks a thousand words. Consider the simplest case of polar coordinates in two spatial dimensions, with coordinates $(r,\theta)$ such that $x=r\,\cos\theta$ and $y=r\,\sin\theta$ with line element
 $ds^2 = dr^2 + r^2 d\phi^2$.
 Then the coordinate basis vectors would be $\e_r = \pd_r$ and $\e_\phi = \pd_\phi$ and we have
 $\e_r\cdot\e_r = 1 = g_{rr}$ and $\e_\phi\cdot\e_\phi = r^2= g_{\phi\phi}$ with
 $\e_r\cdot\e_\phi=0$, leading to the metric 
 $g_{\mu\nu} = \tiny{\left(\begin{array}{cc} 1 & 0 \\0 & r^2\end{array}\right)}$. 
 Clearly $[\pd_r, \pd_\phi]=0$.
 
 In an orthonormal basis we would instead define $\e_r = \pd_r$ and $\e_\phi = \frac{1}{r}\,\pd_\phi$, with
 $\e_r\cdot\e_r = 1 = g_{rr}$ and $\e_\phi\cdot\e_\phi = 1= g_{\phi\phi}$ with
 $\e_r\cdot\e_\phi=0$, leading to the metric 
 $g_{ab} = \tiny{\left(\begin{array}{cc} 1 & 0 \\0 & 1\end{array}\right)} = \delta_{ab}$. 
 However, we now have 
 $[\e_r, \e_\phi]\,f(r,\phi) = [\pd_r, \frac{1}{r}\,\pd_\phi]\,f =-\frac{1}{r^2}\pd_\phi\,f = -\frac{1}{r}\e_\phi\,f$ which allows one to conclude (since $f$ was an arbitrary function) that 
 $[\e_r, \e_\phi]= -\frac{1}{r}\e_\phi \equiv \e_\phi\,C^{\phi}_{\;\;r \phi}(x)$.
 In the last step we have introduced the structure constants $C(x)$ defined by
 the commutators of the basis vector, $[\e_a, \e_b] =\e_c\, C^{ c}_{\;\;a b}(x)\,$ (with sum over the index $c$, see Hartle, p109), which just states that the commutator of the basis can be expanded in terms of a linear combination of the basis vectors.
 
% \blue{Move the following discussion of 1-forms to the SM.}
 
 % red 1417-1434
 {\color{black}
 A general vector (an, in general, tensor)  $\bs{v}$ is a geometric object, independent of the coordinates and basis vectors (user's laboratory frame) used to describe it. So we can write this as
 $\bs{v} = v^\mu(x)\,\e_\mu(x)$ using a coordinate basis vector description 
 (with \tit{contravariant} coordinate components $v^\mu$), or as 
 $\bs{v} = v^a(x)\,\e_a(x)$ using an orthonormal basis vector description (with physical orthonormal components $v^a(x)$).
 
 Note that the transition from coordinate basis to orthonormal basis (for this diagonal metric) can be performed by inspection by examining the metric and grouping terms as
 $ds^2 = (dr)^2 + (r d\phi)^2$ and somehow considering it's ``inverse," namely $(\pd_r, \frac{1}{r}\pd_\phi)$.
 Note that $r\,d\phi$ is not a total differential, so there is no coordinate associated with such an object.
 This intuition can be formalized by defining basis \tit{1-forms} which are \tit{dual} to the orthonormal basis vectors $\e_a$. We denote these 1-forms (generalized differentials) as $\bs{\theta}$ (Ryder's notation) which can be decomposed in terms of the coordinate (true) differentials $dx^\mu$ via 
 $\bs{\theta}^a = e^a_{\;\;\mu}(x) dx^\mu$. Note that we have purposely used the same symbol $e$ for the components of the one form, with the important distinction that the coordinate index $\mu$ is now on the bottom and the orthonormal index $a$ is on top (vs $\e_a^{\;\;\mu}$ associated with the basis vectors
 $\e_a = e_a^{\;\;\mu}(x)\,\pd_\mu$). Thus, even without a metric, we can define what we mean by dual by saying that a 1-form is an object that ``eats" vectors in the following sense (in both coordinate and orthonormal bases) $dx^\mu(\pd_\nu)\defeq \delta^\mu_{\;\;\nu}$, and 
 $\bs{\theta}^a(\e_b)\defeq \delta^a_{\;\;b}$.
 Thus, consider a general 1-form $\bs{w} = w_a(x)\bs{\theta}^a$ with (covariant) components $w_a(x)$.
 Then, we can have this 1-form act on a vector $\bs{v} = v^a(x)\,\e_a$ to give
 $\bs{w} (\bs{v}) = w_a\bs{\theta}^a (v^b\,\e_b) = w_a \,\bs{\theta}^a(\e_b)\,v^b =  w_a\,\delta^a_{\;\;b}\, v^b 
 =  w_a v^a \defeq \bs{w}\cdot\bs{v}$. Therefore, we can define and inner product without the need for a metric, and the metric merely serves to raise and lower components via $g_{\mu\nu} v^\nu = v_\mu$ and $\eta_{ab} v^a = v_b$.
} % red from 1417
 
%=========================================================
 \subsection{The DE in CST: covariant derivatives in GR}\label{sec:DE:CST:covar:deriv:GR}%========================================================= 
 With these preliminaries out of the way, we now get to the crux of the matter (and the part that is often an initial learning bottleneck for the beginning GR student). How does one define the derivative of basis vectors? Let's first work in the simpler coordinate basis vectors.
 Consider the following calculation for a posited, yet unknown, derivative which we denote as $\nabla_\mu$: 
 $\nabla_\mu \bs{v} = \nabla_\mu\big(v^\nu(x) \e_\nu(x)\big) = (\pd_\mu v^\nu)\,\e_\nu + v^\mu (\nabla_\mu\e_\nu)$.
 Here, we have made the (natural) assumption that $\nabla_\mu\to\pd_\mu$ on functions (of which $v^\nu$ are).
 Acting on basis vectors, we are yet unsure, so we just leave it as $\nabla_\mu$.
 This states that we must not only differentiate the components $v^\nu(x)$ of $\bs{v}$, but also its basis vectors $\e_\nu(x)$. But we are already familiar with this latter concept, since even in our simple 2D polar coordinate example above, the basis vector $\e_\phi = \pd_\phi$ is tangent to circles of constant radius $r$, and thus change direction (in the surrounding 2D Euclidean $\mathbb{R}^2$ space) as we vary the coordinate $\phi$.
As in the case of the 1-forms above, we assume that $\nabla_\mu\e_\nu$ can be expanded in terms of the basis vectors, and so we write $\nabla_\mu\e_\nu(x) = \G_{\mu\nu}^\lambda(x)\,\e_{\lambda}$ with proportionality functions $\G_{\mu\nu}^\lambda(x)$, the famous \tit{Levi-Civita connection}, which informs us as to how the basis vectors change as we move from point $x^\mu$ to point $x^\mu+u^\mu\,d\tau$ in the CST.
Inserting this into the full expression we have
$\nabla_\mu \bs{v} =  (\pd_\mu v^\nu)\,\e_\nu + v^\nu\,\G_{\mu\nu}^\lambda\e_\lambda$ which we can relabel dummy indices via $\lambda\leftrightarrow\nu$ to obtain
\be{cov:deriv}
\nabla_\mu \bs{v} = \left(\pd_\mu v^\nu +  \G_{\mu\lambda}^\nu v^{\lambda} \right) \e_\nu 
\defeq v^\nu_{;\mu}\,\e_\nu,
\ee
where the last expression defines the \tit{covariant derivative} (denoted conventionally by a semicolon vs a comma apropos for an ordinary coordinate derivative) of the components
$\nabla_\mu v^\nu \equiv v^\nu_{;\mu} = \pd_\mu v^\nu +  \G_{\mu\lambda}^\nu v^{\lambda}$.
The values of $\G_{\mu\lambda}^\nu$ are tied down 
(see \Eq{connection:coord})
by invoking the \tit{constancy of the metric} condition
$\nabla_\mu(g_{\alpha\beta})=0$, 
(which reduces to the identity $\pd_\mu(\eta_{\alpha\beta})=0$ in the observer's local laboratory). 
The covariant derivative is of fundamental importance in GR since the commutator of the covariant derivatives acting on a vector is proportional to the Riemann curvature tensor, from which Einstein's fundamental equations are derived: 
$[\nabla_\mu , \nabla_\nu]\,v^\alpha= v^\alpha_{;\mu;\nu} -v^\alpha_{;\nu;\mu}  = -R_{\beta\mu\nu}^\alpha v^\beta$. Note: the analogy in E\&M is the potential $A^\mu$ which acts as a $U(1)$ (\tit{gauge}) potential, so that the covariant derivative is $\nabla_\mu = \pd_\mu - A_\mu(x)$, and the analogue of the curvature is the Faraday tensor (containing the electric and magnetic fields as components) such that $[\nabla_\mu,\nabla_\nu] = \pd_\mu A_\nu - \pd_\nu A_\mu = F_{\mu\nu}$ (see Ryder, section 11.1).
The important point here is that the 4-potential $A^\mu(x)$ is spacetime dependent, and that changes in $A^\mu(x)$ (called \tit{local gauge transformations}) do not change the physical field $F_{\mu\nu}(x)$. This is called ``gauging" the E\&M field.

%=========================================================
 \subsection{The DE in CST: the spinor covariant derivative in GR}\label{sec:DE:CST:spinor:covar:deriv:GR}%=========================================================
%  \blue{This is probably the hardest concept for the novice student to grasp. One could just jump to the spinorial connection, but then it would probably appear to ``come out of nowhere." However, one could motivate how Weyl suggested the spinor should transform under world and local laboratory transformations and push  this background discussion to the SM.}
  
The new question to ask is: ``How does one gauge gravity?" This is the subject of Ryder\cite{Ryder:2009}, Chapter 11.3 (This chapter is titled ``Gauging Lorentz symmetry: torsion"). The main point is that up to now we've been discussing the covariant derivative for vectors (and tensors) which are associated with integer values of spin ($j = \{0, 1, 2,\ldots\}$, with $2 j+1$ components, i.e. scalars, vectors (e.g. photons), 2-tensor, (e.g. gravitons), etc\ldots).
But how does one define the covariant derivative for half-integer spin objects, specifically spin $1/2$ (e.g. electrons) with 2 components?

%\blue{Move this background discussion to the SM}

% red 1474-1502
{\color{black}
Recall that for the SR the total angular momentum is given by matrix 
$J_{\mu\nu} = -i\,(x_\mu\,\pd_\nu - x_\nu\,\pd_\mu)\,\mathbb{I} + \Sigma_{\mu\nu}$,
where $\Sigma_{\mu\nu}=\frac{i}{4}\,[\g_\mu, \g_\nu]$, and $\g_\mu = \eta_{\mu\nu}\,\g^\nu$ are the constant Dirac matrices of \Eq{DE:gamma:matrices} (since we are operating in the observer's local SR tangent plane/laboraotry).
As in SR, the underlying symmetry of GR (at least locally) is the Poincare group, which is the 10 parameter group of (3) rotations, (3) boosts, and (4) spacetime translations. These matrix operators satisfy as set of (involved) commutation relations 
$[J_{\mu\nu},J_{\alpha\beta}] = f_{\mu\nu,\alpha\beta}^{\rho\sigma}\,J_{\rho\sigma}$ (which reproduces the usual commutation relations for rotations if we set $\Sigma_{\mu\nu}\to 0$). The Poincare group admits both integer (vector, tensor) representation, as well as spinor (half-integer) representations. These representations are derived by consider small changes in the quantity under study. 

Following Ryder, Chapter 11.3, Herman Weyl proposed the following ansatz: the ($N$-dimensional) spinor $\psi$ transforms like a \tit{scalar} with respect to the ``world" transformations (i.e. with respect to the coordinate index $\mu$), but as a spinor \tit{with respect to local Lorentz transformations (LLT) in the local laboratory} (i.e. with respect to the index $a$ in the \tit{tangent space at $x$ in the CST where the observer's local laboratory instantaneously exists}).
These LLT transform the observer's instantaneous state of motion (in the flat Minkowski tangent plane) at $x$ from one type of motion to another, e.g. from a stationary observer at $x$, to an instantaneous freely falling observer at $x$, or to say an observer executing circular motion instantaneously at $x$, or to any kind of instantaneous motion. Thus, under infinitesimal changes in the surrounding CST small changes in the spinor $\delta\psi$ are given as $\delta\psi = -\xi^\mu\,\pd_\mu\psi$, i.e. with respect to the ordinary coordinate derivative 
$\pd_\mu$ apropos for a (world) scalar field. Here, $\xi^\mu = \omega^\mu_\nu x^\mu$ where for now, we consider $\xi^\mu$ as constants (since we are acting \tit{within} a given tangent plane at $x$). As an example, for an rotation infinitesimal in the $x-y$ plane by angle $\phi$ given by
$\tiny{
\left(\begin{array}{c}x'\\ y'\end{array}\right) = R(\phi) \left(\begin{array}{c}x\\ y\end{array}\right)
}$ 
(suppressing the $t$ and $z$ components for now, i.e. this should be embedded in a $4\times 4$ matrix)
where 
$\tiny{
R(\phi) = \left(\begin{array}{cc}\cos\phi & \sin\phi \\-\sin\phi & \cos\phi\end{array}\right)
}$ then 
$\omega^\mu_\nu$ is the anti-symmetric matrix given by 
$\tiny{
\frac{d R(\phi)}{d\phi}|_{\phi=0} = 
\left(\begin{array}{cc}0 & 1 \\-1 & 0\end{array}\right).
}$
However, for small changes in the spin, in the local observer's frame, the spinor changes are given by
$\delta\psi = -i\,\half\,\omega^{ab}\,\Sigma_{ab}\,\psi$. Where $\omega^{ab}$ are some constants describing the local Lorentz transformation (LLT) (as in SR).
} % red from 1474

Thus, for infinitesimal variations, the total change in the spinor is just the sum (to first order) of the two changes (variations) give by
$\delta\psi = -\xi^\mu\,\pd_\mu\,\psi -i\,\half\,\omega^{ab}\,\Sigma_{ab}\,\psi$.
We now ``gauge" this transformation by allowing both $\xi^\mu(x)$ and $\omega^{ab}(x)$
to be spacetime dependent (i.e. LLT varying at each point $x$)
in order to develop a spinor covariant derivative.

% red 1510-1517
{\color{black}
The derivation is not that hard but somewhat lengthy (detailed in Ryder, Chapter 11.3) yielding a form
$\psi_{|\mu} = \pd_\mu\,\psi + \half\,A^{ab}_\mu\,\Sigma_{ab}\,\psi$ such that the changes in the spinor 
$\delta\psi$ under LLT transform the same as $\psi$ itself, namely, $\delta(\psi_{|\mu}) = \half\omega^{ab}\,\Sigma_{ab}\,(\psi_{|\mu})$. (Note: the spinor covariant derivative is denoted by $\psi_{|\mu}$ to distinguish it from the covariant derivative acting on vectors and tensors, e.g. $v_{;\mu}$).
Most significantly, the commutator of the spinor covariant derivatives acting on $\psi$ are once again  proportional to (a spinor version of) the Riemann tensor (having both mixed tangent plane (Latin), and world (Greek) indices), namely
$\psi_{|\mu|\nu} - \psi_{|\nu|\mu} = -\half R^{ab}_{\mu\nu}\,\Sigma_{ab}\,\psi$.
} % red from 1510

With these preliminaries under one's belt, one can then derive the DE in CST, as detailed in Ryder, Chapter 11.4, pp416-418. The derivation is quite elegant, but here we indicate only the highlights.
We begin with the flat Minkowski SR DE (restoring factors of $\hbar$ and $c$) 
$i\,\hbar\,\g^\mu\pd_\mu\,\psi = - m\,c\, \psi$ (note the sign change on the mass, due to the local GR (e.g. Hartle, Ryder) metric, vs the particle physicist's (e.g. Greiner) Minkowski metric), noting that the Dirac matrices are the \tit{constant} ones discussed earlier for the SR DE \Eq{DE:gamma:matrices} (since we are in the observer's local laboratory, i.e. the instantaneous (locally flat, Minkowski) tangent space to the CST at the point $x$).
The net result of the gauging of gravity (really, gauging Lorentz symmetry) is that the covariant derivative for spinors boils down to
\be{Ryder:p416:128:cov:deriv}
\pd_\mu \to 
D_\mu \defeq \pd_\mu - \frac{i}{2}\,\G_{\alpha\beta\mu}\,\Sigma^{\alpha\beta} =
\pd_\mu + \frac{1}{8}\,\G_{\alpha\beta\mu}\,[ \g^\alpha, \g^\beta].  
\ee

Here, $\G_{\alpha\beta\mu} = g_{\alpha\rho}\,\G^{\rho}_{\beta\mu}$
where for coordinate basis vectors $\G^{\rho}_{\beta\mu}$ are the usual Levi-Civita connenction
given in terms of the metric by (derivable from the constancy of the metric condition $\mathbf{\nabla}^\mu\,g_{\alpha\beta}(x)=0$)
\be{connection:coord}
\trm{coordinate basis:}\quad
\G^{\mu}_{\alpha\beta} = 
\half\,g^{\mu\lambda}\,(\pd_\alpha\,g_{\lambda\beta} + \pd_\beta\,g_{\lambda\alpha}-\pd_\lambda\,g_{\alpha\beta}).
\ee
The $\g^\alpha$  appearing in \Eq{Ryder:p416:128:cov:deriv} are the \tit{constant} Dirac gamma matrices given prior in \Eq{DE:gamma:matrices}.

Thus,  we penultimately arrive at our desired goal, the DE in CST
\be{Ryder:p416:128}
i\,\hbar\,\g^\mu\,D_\mu\psi = 
i\,\hbar\,\,\g^\mu\,
\left(\pd_\mu + \frac{1}{8}\,\G_{\alpha\beta\mu}\,[ \g^\alpha, \g^\beta]\right)\,\psi = 
m\,c\,\psi.
\ee

We now introduce one more complication, namely, the translation of the above DE written in a coordinate basis, to a physical orthonormal set of basis vectors. This entails having an expression for the 
connection $\G_{\alpha\beta\mu}$ in an orthonormal basis. While straightforward, yet somewhat lengthy to derive (see Ryder, Chapter 3.13, pp107-110, Eq(3.259)) the results are a pleasing generalization, denoted by $\G_{abc}$, of the coordinate-based Levi-Civita connection 
$\G_{\mu\nu\lambda}$, given by
\bea{Ryder:p108:3.259}
\trm{orthonormal basis:}\quad 
\G_{abc}&=&
-\half\,
( 
C_{abc} + C_{bca} - C_{cab} 
)
 \\ 
C_{abc} &=& \eta_{ad}\,C^{d}_{\;\;bc}, \quad\trm{where}\quad [\e_a,\e_b] = \e_c\,C^{c}_{\;\;bc}.
\eea
(Note: many GR books, including Ryder (but not Hartle), use Greek indices on all basis vectors, coordinate and orthonormal, and the metric as well. One just has to be conscious of the context of the specific calculation to discern if the indices indicate global or local basis/metric, and hence which formula to utilize for the connection, \Eq{connection:coord} or \Eq{Ryder:p108:3.259}).

Finally, the DE in CST in an orthonormal basis is given by
\be{Ryder:p417:11.129} 
% with greek go to latin indices (PMA)
i\,\hbar\,\gamma^a (e_a + \G_a)\,\psi = m\,c\,\psi, \qquad \G_a\defeq\frac{1}{8}\,\G_{abc}[\g^b,\g^c].
\ee
Here, $e_a(\psi(x)) = e_a^\mu(x)\pd_\mu\psi(x)$ is the action of the orthonormal basis vector on the spinor (or any object), and we write $``\g^a\equiv\g^\mu"$ by which we mean (abusing notation)  that the 
$\g^a$ are numerically the \tit{same} constant Dirac gamma matrices as $\g^\mu$ (in SR, see \Eq{DE:gamma:matrices}).

For the Schwarzschild metric, one can straightforwardly work out the commutators of the orthonormal basis vectors 
(see Ryder, pp418-419, and \App{app:commutators:SST}) which are defined by (restoring again the boldface vector notation)
\be{Ryder:p417:11.131}
\e_0 = \frac{1}{c}\,\left(1- \frac{2 M_s}{r}\right)^{-1/2}\,\frac{\pd}{\pd t},\;\;
\e_1 = \left(1- \frac{2 M_s}{r}\right)^{1/2}\,\frac{\pd}{\pd r},\;\;
\e_2 = \frac{1}{r}\,\frac{\pd}{\pd \theta},\;\;
\e_3 = \frac{1}{r \sin\theta}\,\frac{\pd}{\pd \phi}
\ee
where $M_s\defeq G M/c^2 = \half\,r_s$, 
to finally arrive at our desired destination (see Ryder, Eq(11.139), p418)
\bea{Ryder:p418:11.139}
&{}&
i\,\hbar\,
\left\{
\left(1- \frac{2 M_s}{r}\right)^{-1/2}\,
\left[
\g^0\frac{1}{c}\frac{\pd\psi}{\pd t} - \frac{M_s}{2 r^2}\g^1\psi
\right]
+ \left(1- \frac{2 M_s}{r}\right)^{1/2}\g^1\,\frac{\pd \psi}{\pd r}
\right. \no
&{}& 
\left.
\hspace{0.20in} +\; 
\g^2\,\frac{1}{r}\frac{\pd \psi}{\pd \theta} +
\frac{M_s}{r^2}\,\left(1- \frac{2 M_s}{r}\right)^{1/2}\g^1\,\psi +
\g^3\,\frac{1}{r \sin\theta}\frac{\pd \psi}{\pd \phi} +
\frac{\cot\theta}{2 r}\,\g^2\,\psi 
\right\}
= m\,c\,\psi, \label{DE:CST}
\eea
 of a Dirac spin-1/2 particle of mass $m$ in the Schwarzschild spacetime.

 One thing we are struck by right away  is that nowhere in the above derivation has use been made of the classical GR conserved quantities $e$ and $\ell$ of \Eq{Hartle:p193:9.21} and \Eq{Hartle:p193:9.22}, respectively.
 However, upon further reflection, for a QM derivation this makes sense, since these
 constants of the motion involve $dt(\tau)/d\tau$ and $d\phi(\tau)/d\tau$ implying the classical notion of trajectories in spacetime, for which QM abandons, and replaces with the concept of stationary eigenstates over all space. Thus, even though tempting, it would make not make sense to replace quantities such as $\frac{\pd \psi}{\pd t}$ by 
 $\left(\frac{d t}{d\tau}\right)\,\frac{\pd \psi}{\pd \tau}\to e\,(1-\frac{2 M_s}{r(\tau)})^{-1}\,\frac{\pd \psi(\tau)}{\pd \tau}$, etc. since the the wave function $\psi$ would then be solely a function of $\tau$, and the QM concept of spatial eigenstates would not be possible.
 
%=========================================================
 \section{Bound states of the DE in SST}\label{sec:bound:states:DE:SST}%=========================================================
 We are now finally able to attempt to answer the student's original posed question: ``Does there exist bound states of the DE in SST, analogous to the bound states of the SR DE for either vector or scalar coupling?" Before we can answer this, it is helpful to massage \Eq{Ryder:p418:11.139} into a much more amenable, dimensionless form (especially for numerical calculations). By using the SR Dirac $\g^a$ matrices \Eq{DE:gamma:matrices} and multiplying through by $\rootF \defeq \sqrt{1-2 M_s/r}$, 
 and again letting $\psi = \tiny{\left(\begin{array}{c}\varphi \\ \chi\end{array}\right)}$, 
 \Eq{DE:CST} can be rearranged into the form % p1.15, p2.0 PMA 3Jun2022 notes
 \bea{DE:CST:pma:p5.2}
&{}&\left( i\,\hbar\,\frac{\pd}{\pd t} - m\,c^2\,\rootF\right)\,\varphi = c (\sigmahat\cdot\phat )\,\chi, \label{DE:CST:pma:p5.2:line1} \\
&{}&\left( i\,\hbar\,\frac{\pd}{\pd t} + m\,c^2\,\rootF\right)\,\chi = c (\sigmahat\cdot\phat )\,\varphi,\label{DE:CST:pma:p5.2:line2} \\
\trm{with}\quad \phat &=& 
 -i\,\hbar\,
 \left(
 \F\,\frac{\pd}{\pd r} + \F\, \frac{M_s}{r^2} - \frac{M_s}{2\,r^2},\; 
 \frac{\rootF}{r}\left(\frac{\pd}{\pd \theta} + \half\,\cot\theta \right),\;
 \rootF\,\frac{1}{r \sin\theta}\frac{\pd}{\pd \phi} 
 \right). \label{DE:CST:pma:p5.2:line3}
 \eea
 The first thing we note is that in going from SR to GR the rest mass goes from 
 $m\,c^2\to m\,c^2\,\rootF$, acting as a variable mass, that is the ordinary rest mass $m c^2$ at $r\to\infty$, and goes to zero at the Schwarzschild horizon $r\to 2 M_s$. This is one of the new features introduced by GR, the role of the horizon. Secondly, we see that \Eq{DE:CST:pma:p5.2:line1} and \Eq{DE:CST:pma:p5.2:line2} has the form of a SR free field DE, but with the radial potential terms buried within $\phat$, especially, 
$\hat{p}_r$ in \Eq{DE:CST:pma:p5.2:line3}.

If we now postulate the existence of stationary states, with each spinor having an $e^{-i\,E\,t/\hbar}$ temporal dependence, with E constant, we then have %PMA notes: 3Jun2022, p1.15, p2.0
\bea{DE:pma:p5.2:middle}
c (\sigmahat\cdot\phat )\,\chi &=& (E - m\,c^2\,\rootF)\,\varphi, \\
c (\sigmahat\cdot\phat )\,\varphi &=& (E + m\,c^2\,\rootF)\,\chi, 
\eea
 in strong analogy with the SR DEs \Eq{DE:RQMWE:p170:5:line:1} and \Eq{DE:RQMWE:p170:5:line:2}, but now with no explicit vector or  scalar coupling potential terms, $V_v(r)$ and $V_s(r)$ on the righthand side, and now additionally with a variable mass $m c^2\,\rootF$.
 Note that for $r\gg r_s=2 M_s$ we have
 $E + m\,c^2\,\rootF\approx E \pm m\,c^2\,(1-G M/ (r c^2)) = E \pm (mc^2 + V_G(r))$ where
 $V_G(r) = -G M m/r$ the Newtonian potential. Hence, we observe scalar (mass) coupling far from the horizon, as our intuition would expect.
 
 We now write the above equations in dimensionless form using natural units. 
 Defining the operator $\qhat$ via $\phat = -i\,\hbar\,\qhat$, and then
 dividing through by $\hbar c$ and recalling $\lambda_C = \frac{\hbar}{m c}$, we will define lengths as 
 $r = \lambda_C\,\rho$, and energies in terms of the rest mass via $\eps = \frac{E}{m c^2}$.
 We then obtain 
\bea{DE:pma:p5.3:top}
\hspace{-0.25in}
&{}& (\sigmahat\cdot\qhat )\,\chi = i\,(\eps - \rootF)\,\varphi, \label{DE:pma:p5.3:top:line1} \\
\hspace{-0.25in}
&{}& (\sigmahat\cdot\qhat )\,\varphi = i\,(\eps + \,\rootF)\,\chi, \label{DE:pma:p5.3:top:line2} \\
\hspace{-0.25in}
\qhat &=&
 \left(
 \F\,\frac{\pd}{\pd \rho} + \F\, \frac{m_s}{\rho^2} - \frac{m_s}{2 \rho^2},\; 
 \frac{\rootF}{\rho}\left(\frac{\pd}{\pd \theta} + \half\,\cot\theta \right),\;
\rootF\, \frac{1}{\rho \sin\theta}\frac{\pd}{\pd \phi} 
 \right), \;\; m_s \defeq \frac{M_s}{\lambda_C} = \frac{\half r_s}{\lambda_C}, \qquad \label{DE:pma:p5.3:top:line3} \\
 &\defeq&
 \left(
 \Qhat_1, \Qhat_2, \Qhat_3 \label{DE:pma:p5.3:top:line4}
 \right)
\eea 
Since our primary goal to find the simplest possible solution, not the most general solution, we will cut to the chase and look for only radial solutions along radial plunge orbits (hence avoiding all the complications of the orbital and spin angular momentum that arose in the SR DE) and define
\bea{PMA:ansatz}
\varphi &=& g(\rho)\,\tilde{\varphi}, \\
\chi &=& i\,f(\rho)\,\tilde{\chi},
\eea 
where  $\tilde{\varphi}$ and $\tilde{\chi}$ are constant 2-spinors. Recall in the SR DE, if 
$\ell = 0\Rightarrow m=0$ 
which implies $Y_{\ell=0,m=0}(\theta,\phi) = \frac{1}{\sqrt{4 \pi}}$ is a constant.
(Here $m$ is now the magnetic quantum number. The mass of the test particle, for which we have also used the same symbol $m$, is now subsumed in the Compton wavelength $\lambda_C$ and the dimensionless energy $\epsilon$, so there should be no confusion).
Let us now, with foresight, chose $\varphi = \tiny{\left(\begin{array}{c} a \\ b\end{array}\right)}$ and
$\chi = \tiny{\left(\begin{array}{c} a \\ -b\end{array}\right)}$ with $a, b$ constant such that $|a|^2 + |b|^2=1$. 
%We orient our axes so that $r$ and hence $\rho$ are along the $\hat{z}$ direction so that
%$\hat{\sigma}_1 = \hat{\sigma}_z = \tiny{\left(\begin{array}{cc} 1 & 0 \\ 0 &-1\end{array}\right)}$. 
Since the ``$1$" direction is along $r$ (see \Eq{Ryder:p417:11.131}) and hence $\rho$, 
we choose a definition of the Pauli spin matrices so that 
$\hat{\sigma}_1 \to \hat{\sigma}_z = \tiny{\left(\begin{array}{cc} 1 & 0 \\ 0 &-1\end{array}\right)}$.
%\blue{(Move to SM)\;}
% red 1683-
{\color{black} 
Then, substituting these definitions into  \Eq{DE:pma:p5.3:top:line1} and \Eq{DE:pma:p5.3:top:line2} we have
\bea{DE:pma:p5.3:bottom}
\hspace{-0.65in}
\left(\begin{array}{cc} \Qhat_1 & 0 \\ 0 &-\Qhat_1\end{array}\right)\,(i\,f)\,
\left(\begin{array}{c} a  \\ -b\end{array}\right) =
i\,(\eps -\rootF)\,g\, \left(\begin{array}{c} a  \\ b\end{array}\right) 
&\Rightarrow&
\begin{cases}
 a\,\Qhat_1\,f =  a (\eps -\rootF)\,g, \\ 
 b\,\Qhat_1\,f =  b (\eps -\rootF)\,g, 
\end{cases}
\Rightarrow \Qhat_1\,f =   (\eps -\rootF)\,g, \;\;\qquad \\
\hspace{-0.65in}
\left(\begin{array}{cc} \Qhat_1 & 0 \\ 0 &-\Qhat_1\end{array}\right)\,g\,
\left(\begin{array}{c} a  \\ b\end{array}\right) =
i\,(\eps +\rootF)\,(i\,f)\, \left(\begin{array}{c} a  \\ -b\end{array}\right) 
&\Rightarrow&
\begin{cases}
 a\,\Qhat_1\,g =  -a (\eps +\rootF)\,f, \\ 
 -b\,\Qhat_1\,g =  b (\eps +\rootF)\,f, 
\end{cases}
\hspace{-0.5em}
\Rightarrow \Qhat_1\,g =   -(\eps+\rootF)\,f. \;\;\qquad
\eea
Thus, for this choice of constant $\tilde{\varphi}$ and $\tilde{\chi}$, each spinor produces two equations, which are self-consistently the same. 
} %red from 1683
Therefore, our final dimensionless radial DE in SST equations are
\bea{DE:CST:p5.3:final}
\Qhat_1(\rho)\,f(\rho) &=&   \left(\eps -\sqrt{\mathcal{F}(\rho)}\right)\,g(\rho), \label{DE:CST:p5.3:final:line1} \\
\Qhat_1(\rho)\,g(\rho) &=&   -\left(\eps+\sqrt{\mathcal{F}(\rho)}\right)\,f(\rho), \label{DE:CST:p5.3:final:line2} \\
\trm{with}\quad 
\mathcal{F}(\rho) = 1 - \frac{2 m_s}{\rho},\quad 
\Qhat_1(\rho)&=&\F\,\frac{\pd}{\pd \rho} + \F\, \frac{m_s}{\rho^2} - \frac{m_s}{2\,\rho^2}, \;\;
\rho = r/\lambda_C, \;\; m_s = \frac{\half r_s}{\lambda_C}. \quad \label{DE:CST:p5.3:final:line3}
\eea

For a 2-spinor in the standard form 
$\tiny{\left(\begin{array}{c} \cos(\theta'/2) \\ \sin(\theta'/2)\,e^{i\phi'} \end{array}\right)}$
we can associate the point
$\hat{n} = (\sin\theta'\,\cos\phi',$ $\sin\theta'\,\sin\phi',\cos\theta')$
  with  polar and azimuthal angles $(\theta', \phi'$) on an ordinary 2-sphere 
  $S^2\in\mathbb{R}^3$, called the \tit{Bloch} sphere.
 (While we have oriented the spin $\hat{z}$-axis of the Bloch sphere with the world radial coordinate $r$ direction, 
 the internal spinor space angles $(\theta', \phi')$ should not be conflated with the Schwarzschild spacetime coordinates $(\theta, \phi)$).
 Spin up corresponds to $\theta'=0$ oriented along $\e_r$,
 and spin down with $\theta'=\pi$ oriented along $-\e_r$ (with $\phi'$ indeterminate at the poles, so it can be taken to be zero there).
 The general, constant spinor solutions we have used above are then of form, 
 $\tilde{\varphi} = \tiny{\left(\begin{array}{c} \cos(\theta'/2) \\ \sin(\theta'/2)\,e^{i\phi'} \end{array}\right)}$
 and 
 $\tilde{\chi} = \tiny{\left(\begin{array}{c} \cos(\theta'/2) \\ -\sin(\theta'/2)\,e^{i\phi'} \end{array}\right)}$
 with overlap $\IP{\tilde{\varphi}}{\tilde{\chi}} = \cos\theta'$, and thus are orthogonal at the Bloch sphere equator $\theta'=\pi/2$.
 For other possible choices of the spinors leading to a modified form of the radial equations \Eq{DE:CST:p5.3:final:line1} and \Eq{DE:CST:p5.3:final:line2}, see \App{app:other:spinor:solns}.

\subsection{Numerical solutions of the DE in SST}\label{subsec:Num:Solns:DE:SST}
The square root factor $\rootF$ in \Eq{DE:CST:p5.3:final:line1} and \Eq{DE:CST:p5.3:final:line2}
presents difficulties for the standard power series in $\rho$  solutions of these equations, as was employed in the NRQM SE and SR DE. Hence, here we make a \tit{conformal transformation} in order to map $\rho=\infty$ to a finite value, via the 
definition $\sin\Theta\defeq \sqrt{1 - 2 m_s/\rho}$ such that
$0\le\Theta\le\pi/2\leftrightarrow 2 m_s\le\rho\le\infty$.
Therefore, $\rho = \frac{2 m_s}{\cos^2\Theta}$ and 
$\frac{\pd }{\pd \rho} = \frac{1}{\pd \rho/\pd \Theta}\,\frac{\pd }{\pd \Theta} = \frac{\cos^3\Theta}{4 m_s \sin\Theta}\frac{\pd}{\pd \Theta}$ hence our DE in SST in a more numerically amenable form is given by
\bea{DE:CST:Qhat:Theta}
\Qhat_1(\Theta)\,f(\Theta) &=&   \left(\eps -\sin\Theta\right)\,g(\Theta), \label{DE:CST:Qhat:Theta:line1} \\
\Qhat_1(\Theta)\,g(\Theta) &=&   -\left(\eps+\sin\Theta\right)\,f(\Theta), \label{DE:CST:Qhat:Theta:line2}  \\ 
%%
%\Qhat_1(\rho)\to \Qhat_1(\Theta) &=& 
%\frac{\sin\Theta\,\cos^3\Theta}{4 m_s } \frac{\pd}{\pd \Theta} +
%\frac{\sin^2\Theta\,\cos^4\Theta}{4 m_s } -
%\frac{\cos^4\Theta}{8 m_s },  \label{DE:CST:Qhat:Theta:line3} \\
%%
\Qhat_1(\rho)\to \Qhat_1(\Theta)&=& 
\frac{\cos^4\Theta}{4 m_s }\,
\left(
\tan\Theta\,\frac{\pd}{\pd \Theta} + \sin^2\Theta -\half 
\right),  \\ \label{DE:CST:Qhat:Theta:line3}
\trm{where}\quad 0\le\Theta\le\pi/2 \; &\leftrightarrow&\; 2\,m_s\le\rho\le\infty \;\leftrightarrow\; 2\,M_s\le r\le\infty, \label{DE:CST:Qhat:Theta:line4}
\eea
for the coordinate region \tit{outside} the Schwarzschild horizon.
Note that the dimensionless constant $m_s=\frac{\half r_s}{\lambda_C}$ \tit{cannot} be absorbed into either $f$ or $g$ nor into the the definition of $\Theta$ itself 
(even if we were to have defined $\tilde{\rho} = \rho/2m_s$ with $\F\to 1-\frac{1}{\tilde{\rho}}$), 
and thus sets the scale for the problem. As shown in \App{app:Units}, for a solar mass BH, 
$r_s=1.48$ km $= 1.48\times 10^3$m and $\lambda_C = 2.246\times 10^{-12}$m, so that $m_s\sim 10^{15}$, an astronomically huge number. Nonetheless, with the courage of our intuition, but with warranted trepidation, we now seek numerical solutions of \Eq{DE:CST:Qhat:Theta:line1} and \Eq{DE:CST:Qhat:Theta:line2} for various values of $1\le m_s\le 10^{15}$.

%\blue{Move to SM}

% red 1776-
{\color{black}
A few other forms of the above equations are worthwhile to obtain a feel for part of the structure of the radial wave functions $f(\Theta)$ and $g(\Theta)$.
By using an integrating factor, we could simplify the above equations with the substitutions
\bea{DE:CST:Qhat:F:G}
f(\Theta)= \sqrt{\sin\Theta}\,e^{-\half\,\sin^2\Theta}\,F(\Theta) &\;\;\Rightarrow\;\;&
\frac{\sin\Theta\cos^3\Theta}{4 m_s }\,\frac{\pd F(\Theta)}{\pd \Theta} = \left(\eps -\sin\Theta\right)\,G(\Theta), \\\label{DE:CST:Qhat:F:G:line1}
g(\Theta)= \sqrt{\sin\Theta}\,e^{-\half\,\sin^2\Theta}\,G(\Theta) &\;\;\Rightarrow\;\;&
\frac{\sin\Theta\cos^3\Theta}{4 m_s }\,\frac{\pd G(\Theta)}{\pd \Theta} = -\left(\eps +\sin\Theta\right)\,F(\Theta). \label{DE:CST:Qhat:F:G:line2}
\eea
Lastly one could define a coordinate $x$ to the remove the prefactors in front of the above derivatives to yield
\bea{DE:CST:Qhat:F:G:x}
\frac{1}{4 m_s }\,\frac{\pd F(x)}{\pd x} &=& \big(\eps -\sin\Theta(x)\big)\,G(x), \label{DE:CST:Qhat:F:G:x:line1} \\
\frac{1}{4 m_s }\,\frac{\pd G(x)}{\pd x} &=& -\big(\eps +\sin\Theta(x)\big)\,G(x),   \label{DE:CST:Qhat:F:G:x:line2}\\
x(\Theta) = \int\frac{d \Theta}{\sin\Theta\cos^3\Theta} &=& \ln(\tan\theta) + \frac{1}{2\,\cos^2\theta}\equiv h(\Theta) \;\Rightarrow\;   \Theta(x) \defeq h^{-1}(x),  \label{DE:CST:Qhat:F:G:x:line3}
\eea
where $\Theta=\Theta(x)$ is now the inverse function of $x(\Theta)$ in \Eq{DE:CST:Qhat:F:G:x:line3}.
For numerical solutions we have found that all three of these formulations produce the same eigenvalues (as one would expect), and thus, in the following we will work directly with 
\Eq{DE:CST:Qhat:Theta:line1} and \Eq{DE:CST:Qhat:Theta:line2}.
} % red from 1776-1798

The numerical procedure to find both the eigenvalues $\eps$ and eigenfunctions 
$f(\Theta)$ and $g(\Theta)$ \Eq{DE:CST:Qhat:Theta:line1} and \Eq{DE:CST:Qhat:Theta:line2} is essentially a \tit{shooting method} \cite{NumRecinC:1992}. As in the SR DE we require the boundary conditions based on physical arguments, that $f(0) = f(\pi/2)=0$ and $g(0) = g(\pi/2)=0$. Since we are looking for bound states, we search for eigenvalues in the range $-1\le\eps\le 1$. Thus, by choosing a value of $\eps$ near zero and integrating inwards from $\Theta=\pi/2$ $\Theta=0$ to one obtains solutions for $f(0)$ and $g(0)$, which are in general non-zero. One then adjusts the value of $\eps$ (essentially performing a line search in $\eps$) and re-integrates again from $\Theta=\pi/2$ to $\Theta=0$ until the desired boundary conditions on $f$ and $g$ are met at $\Theta=0$.
In this fashion one can find the lowest eigenvalue of $\eps_1$ that yields both $f(0)=0$ and $g(0)=0$. 
One then repeats this procedure by searching for $\eps_2$ in the range
$|\eps_1|\le\eps_2\le 1$ and $-1\le\eps_2\le-|\eps_1|$, and similarly for larger eigenvalues.

\begin{figure}[h]
%\begin{tabular}{cc}
%\includegraphics[width=7.0in,height=2.5in]{fig_DE_SST_evals_no_diffusion_ms_5_10_100_1000_1e15_18Jun2022} %&
\includegraphics[width=7.0in,height=2.5in]{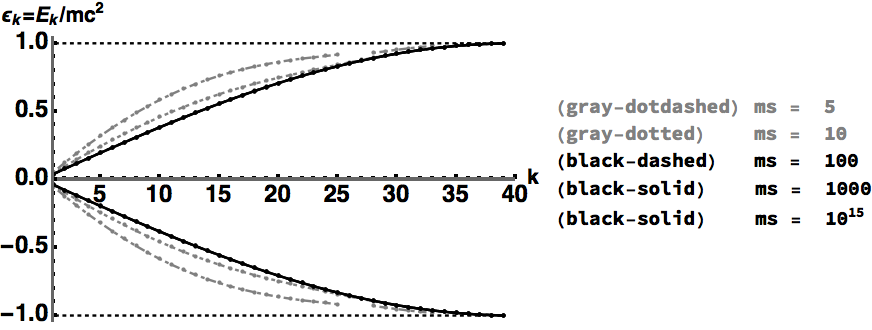} %&
%\includegraphics[width=3.0in,height=1.5in]{alphaplus_wfdivw0_2_beta_0p1}
%\end{tabular}
\caption{Eigenvalues of \Eq{FG:eval:oprs:line1} and \Eq{FG:eval:oprs:line2} for $m_s\in\{5, 10, 10^2, 10^3, 10^{15}\}$, from outside to inside, with the dashed black lines at $\pm 1$
as a guide lines for the  boundary of the bound state energy region.
}\label{fig:DE:SST:evals:ms:5:10:100:1000:1e15} 
\end{figure}

\tit{Mathematica}\cite{Mathematica} has such a routine called \ttt{NDEigensystem} which performs this numerical procedure for coupled differential operators 
$\mathcal{L}_1\big(u_i(x,y,\ldots), v_i(x,y,\ldots),\ldots\big) = \lambda_i\,u_i(x,y,\ldots)$, 
$\mathcal{L}_2\big(u_i(x,y,\ldots), v_i(x,y,\ldots),\ldots\big) = \lambda_i\,v_i(x,y,\ldots)$, \ldots, 
returning the eigenvalue and eigenfunction pair $\{\lambda_i, \{u_i, v_i\}\}$ for $i=\{1,2,\ldots,78\}$ (the maximumn number of eigenvalues that \ttt{NDEigensystem} can return; see \App{app:NumericalMethods} for illustrative examples).
We therefore write our coupled pair of radial wave equations 
\Eq{DE:CST:Qhat:Theta:line1} and \Eq{DE:CST:Qhat:Theta:line2}
as the coupled eigenvalue equations
\bea{FG:eval:oprs}
\mathcal{L}_1(f,g) &=& \hspace{1.5em}\Qhat(\Theta) f(\Theta) + \sin\Theta\,g(\Theta) 
\quad\Rightarrow\quad \mathcal{L}_1(f,g) = \eps\,g(\Theta), \label{FG:eval:oprs:line1} \\
\mathcal{L}_2(f,g) &=& -\left( \Qhat(\Theta) g(\Theta) + \sin\Theta\,f(\Theta) \right)
\;\Rightarrow\quad \mathcal{L}_2(f,g) = \eps\,f(\Theta). \label{FG:eval:oprs:line2}
\eea
For now we simply treat $m_s$ as a variable parameter and plot the eigenvalues from \ttt{NDEigensystem}
for $m_s\in\{5, 10, 10^2, 10^3, 10^{15}\}$ as shown in \Fig{fig:DE:SST:evals:ms:5:10:100:1000:1e15}.
%============================
%============================
The dashed black lines at $\pm 1$ denote the upper bounds for the bound state energy region for 
$|\eps_k = E_k/m\,c^2|<1$. The gray-dashed curve (working outside to inside) for $m_s=5$ shows missing values, since the these eigenvalues had imaginary components. But as the curves for 
$m_s=10$ (gray-dotted), 
$m_s=100$ (black-dashed),
$m_s=1000$ (overlapping black-solid) 
show, the eigenvalues quickly settle down to the black-solid curve for $m_s>10^2-10^3$. In fact, this latter curve is also valid when we used $m_s=10^{15}$ appropriate for a solar mass BH.
The value of the first and last five eigenvalues are found to be
$\eps_k=\pm\,\{0.0379246, 0.0781011, 0.117483, 0.156412, 0.195074,\ldots,$ $0.980712, 0.987621, 0.992999, 0.996871, 0.999206\}$.
To validate the numerical procedure, we also numerically computed the eigenvalues of the SR DE  \Eq{Greiner:p184:2:line1} and \Eq{Greiner:p184:2:line2} using \ttt{NDEigensystem} employed above for the DE in SST, and compared the results to the analytic formula \Eq{Greiner:p186:case:1} for vector coupling, with very good numerical agreement (for further details see\, \cite{SR:DE:Numerical:Note}).
%\medskip

%\blue{The following Subsection B on numerical diffusion was included because of the non-smoothness of the eigen wave functions computed without numerical diffusion. Note however that the eigenvalues used in the following results are all quoted WITHOUT using numerical diffusion. So this whole subsection, and smoothness issue, including the Fig(3), Fig(4) and Fig(5), could be moved to the SM.}

% red 1845-1965
{\color{black} 
\subsection{Adding numerical diffusion}\label{subsec:numerical:diffusion}
\Eq{FG:eval:oprs:line1} and \Eq{FG:eval:oprs:line2} are \tit{convection dominated} equations, which \ttt{NDEigensystem} warns, and suggests that one add (artificial) numerical diffusion.
In fact, an examination of the numerical wave function solutions reveal cusp-like behaviors in places, typical of such convection dominated problems.
There is a standard numerical remedy to this problem (resulting from numerical integration instability) described in many computational physics books, but particularly lucidly (as are other numerous other topics treated similarly)  in \tit{Numerical Recipes in C} \cite{NumRecinC:1992} (or in Fortran, or Fortran90, or C++, if you prefer), written by a group of numerical relativists, see Chapter 19.1, pp834-839), by Press, Teukolsky, Vetterling and Flannery. It entails adding 
to the righthand side of the following 1D example problem
$\pd_t f(x) = v(x) \pd_x f(x)$, a second order term of the form $\alpha_d\,\pd^2_{xx}$.
As discussed in Press \tit{et al}., this added diffusive term stabilizes the equation, when one performs von Neumann stability analysis, for the growth of small fluctuations (in this example, with $v(x)$ and $\alpha_d$ treated as constants). In our problem, the stability analysis is bit more complicated. 
As such, we have found that adding a non-constant diffusion term
$\frac{\cos\Theta}{m_s}\, \pd^2_{\Theta\Theta} \,g$ to $\mathcal{L}_1$ (since it contains $\pd_{\Theta}\,g$), 
and similarly, a term
$\frac{\cos\Theta}{m_s}\, \pd^2_{\Theta\Theta} \,f$ to $\mathcal{L}_2$ (since it contains $\pd_{\Theta} \,f$),
smooths the wave functions $f(\Theta)$ and $g(\Theta)$ somewhat, without severely changing the eigenvalue spectrum. This is shown in \Fig{fig:DE:SST:evals:with:without:diffusion}.
%============================
\begin{figure}[h]
%\begin{tabular}{cc}
%\includegraphics[width=5.0in,height=2.5in]{fig_DE_SST_with_25K_and_without_diffusion_1e15_18Jun2022} %&
\includegraphics[width=5.5in,height=2.5in]{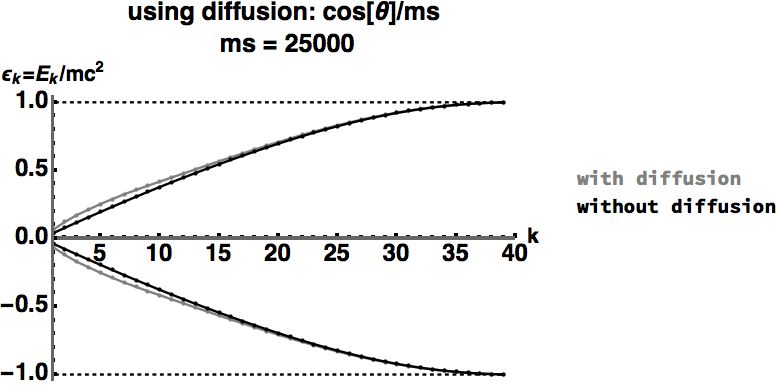}
%\includegraphics[width=3.0in,height=1.5in]{alphaplus_wfdivw0_2_beta_0p1}
%\end{tabular}
\caption{Eigenvalues of \Eq{FG:eval:oprs:line1} and \Eq{FG:eval:oprs:line2} for $m_s=25,000$, 
(black curve) without numerical diffusion, and
(gray curve) with numerical diffusion.
}\label{fig:DE:SST:evals:with:without:diffusion} 
\end{figure}
%============================
The reason for the use of $\alpha_d = \cos\Theta/m_s$ is that the added diffusion term of the same order of magnitude 
$1/m_s$ as the coefficients in $\mathcal{L}_1$ and $\mathcal{L}_2$, and the function $\cos\Theta$ ``turns on" the diffusion 
as we proceed from $\Theta=\pi/2$ (radial infinity) inwards to $\Theta=0$ (the horizon) where the spikiness of the wave functions without numerical diffusion are observed.
The role of the diffusion term on the wave functions can be seen in \Fig{DE:SST:evecs:diff:nodiff:f1:g1:f2:g2}
and \Fig{DE:SST:evecs:diff:nodiff:f3:g3:f4:g4},
for the lowest four eigenvalues $\eps_{k=\{1,2,3,4\}}$.
%============================
\begin{figure}[h]
\begin{tabular}{cc}
\includegraphics[width=3.0in,height=1.5in]{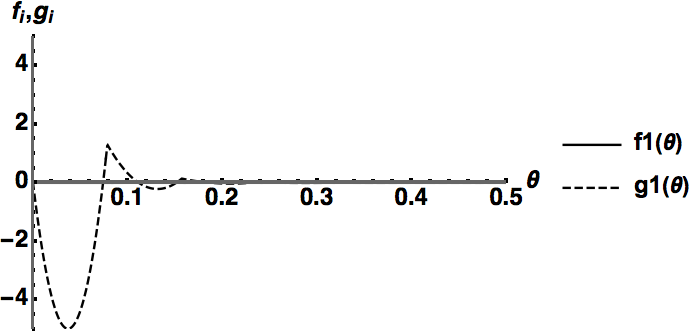} &
\includegraphics[width=3.0in,height=1.5in]{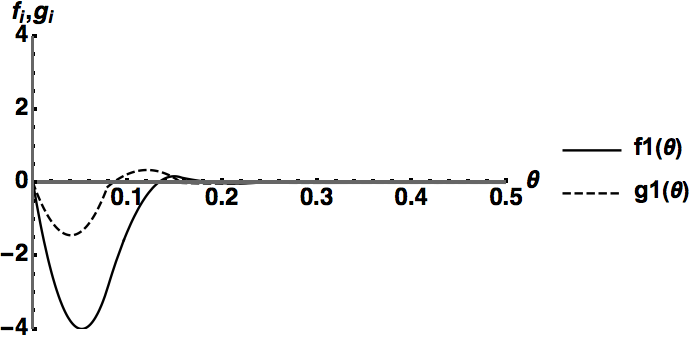} \\
\includegraphics[width=3.0in,height=1.25in]{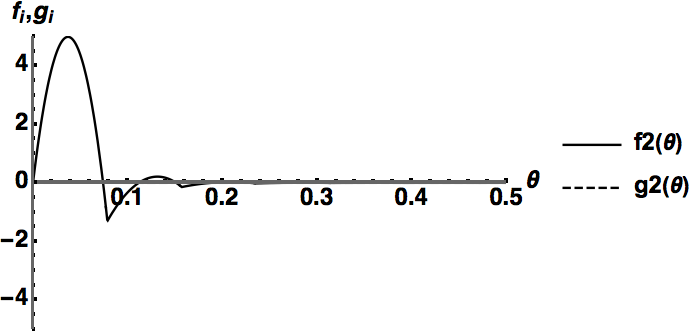} &
\includegraphics[width=3.0in,height=1.25in]{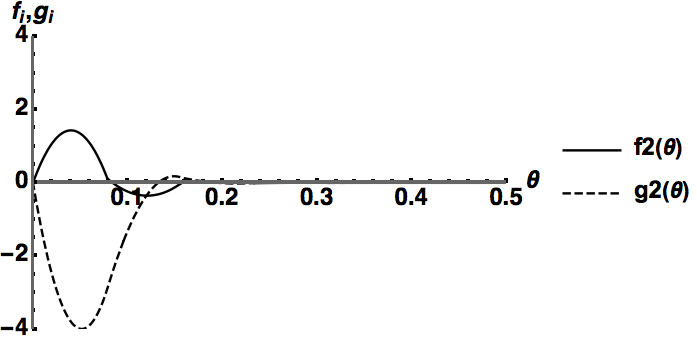}
\end{tabular}
\caption{Wave functions of \Eq{FG:eval:oprs:line1} and \Eq{FG:eval:oprs:line2} for $m_s=25,000$, 
(left column) without numerical diffusion, and
(right column) with numerical diffusion,
for the lowest two eigenvalues $\eps_{k=\{1,2\}}$.
}\label{DE:SST:evecs:diff:nodiff:f1:g1:f2:g2} 
\end{figure}

%============================
% fig_5_top_left_DE_SST_AJP_26Jun2022
%============================
\begin{figure}[h]
\begin{tabular}{cc}
\includegraphics[width=3.0in,height=1.5in]{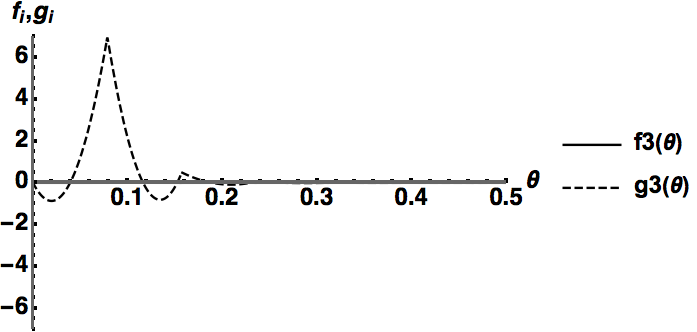} &
\includegraphics[width=3.0in,height=1.5in]{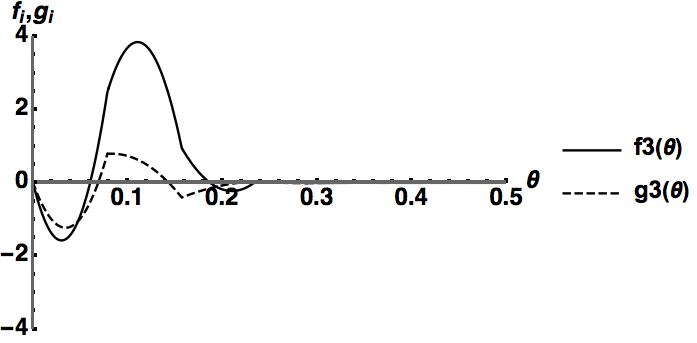} \\
\includegraphics[width=3.0in,height=1.5in]{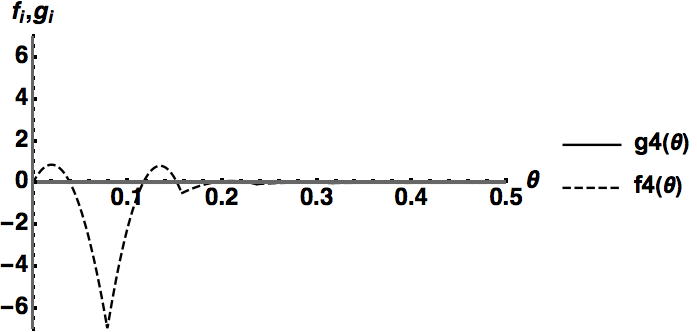} &
\includegraphics[width=3.0in,height=1.5in]{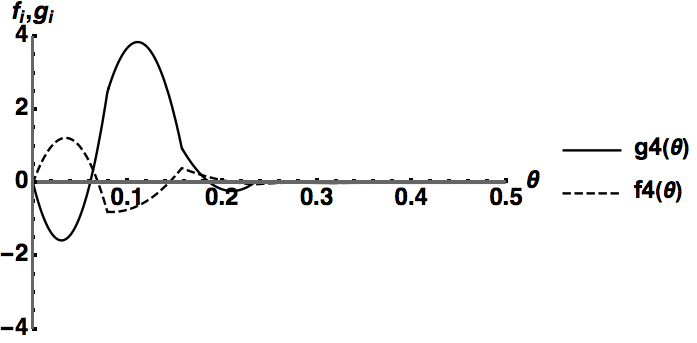}
\end{tabular}
\caption{Wave functions of \Eq{FG:eval:oprs:line1} and \Eq{FG:eval:oprs:line2} for $m_s=25,000$, 
(left column) without numerical diffusion, and
(right column) with numerical diffusion,
for the eigenvalues $\eps_{k=\{3,4\}}$.
}\label{DE:SST:evecs:diff:nodiff:f3:g3:f4:g4} 
\end{figure}
%============================
%\clearpage
%\newpage
%============================

The left column are plots without numerical diffusion, while the right column are with numerical diffusion.
While in \Fig{DE:SST:evecs:diff:nodiff:f1:g1:f2:g2}(left) the plot of $f_1(\Theta)$ appears to be suppressed, it is actually non-zero, but so small that is does not register on the plot.  In \Fig{DE:SST:evecs:diff:nodiff:f1:g1:f2:g2}(right), with diffusion, $f_1(\Theta)$ is brought out (black curve) and the sharp cusps of $g_1(\Theta)$ on the left (numerical non-smoothness due to lack of diffusion) is smoothed out (as is $f_1(\Theta)$) on the right. 
The smoothing for wave functions for larger (magnitude) eigenvalues is still uneven in places as can be seen 
 \Fig{DE:SST:evecs:diff:nodiff:f1:g1:f2:g2} and  \Fig{DE:SST:evecs:diff:nodiff:f3:g3:f4:g4}, but this resulted from our ``best guess" diffusion coefficient taken to be $\alpha_d = \cos\Theta/m_s$, with $m_s=25,000$. These plots indicate that a more sophisticated 
 $\Theta$-dependent diffusion coefficient is most likely warranted. But for our purpose of exploring the student's intuition, our chose form is sufficient for illustration. There is a balancing act involved here since the choice of a  $\Theta$-dependent diffusion coefficient can alter the values of the eigenvalues. For the choice of diffusion we employed, the altered eigenvalues were 
$\eps^{(\trm{diffusion})}_k=\pm\,\{0.0642606, 0.120916, 0.171706, 0.212431, 0.25394,\ldots,$ 
$0.981319, 0.988035, 0.993213, 0.996967, 0.999226\}$, for $m_s=25,000$, which are not too different from the eigenvalues obtained without diffusion, given by 
$\eps_k=\pm\,\{0.0379246, 0.0781011,$ $0.117483, 0.156412, 0.195074,\ldots,$ $0.980712, 0.987621, 0.992999, 0.996871, 0.999206\}$.
We consider the latter eigenvalues without diffusion (with $m_s = 10^{15}$ apropos for a solar mass BH) as the true numerical eigenvalues.
For a reasonable, desired degree of smoothness in the  corresponding wave functions, we allow for the numerical diffusion we have chosen (after several ``numerical experiments") which does not significantly alter the eigenspectrum, as shown in \Fig{fig:DE:SST:evals:with:without:diffusion}.

} % red from 1776-1798

\subsection{Comparison of eigenvalues of DE in SST with SR DE and SE+SR}
\begin{figure}[h]
\begin{tabular}{cc}
\hspace{-0.75in}
\includegraphics[width=4.0in,height=2.0in]{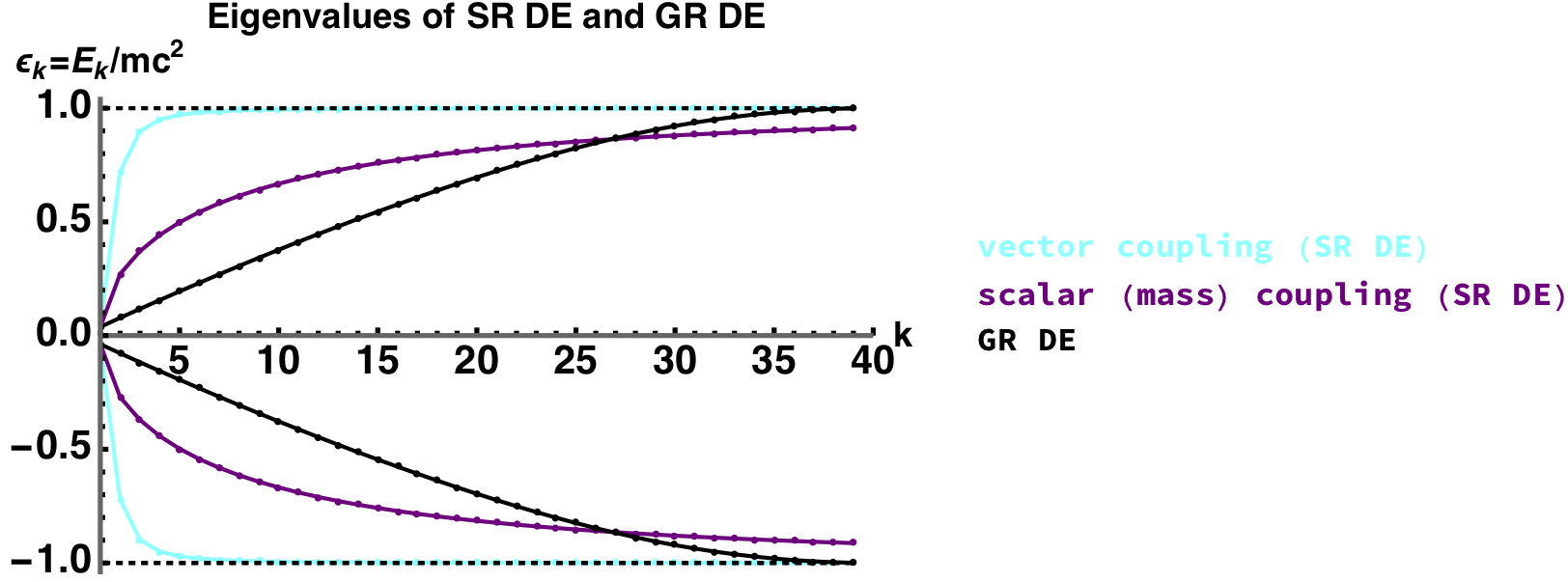} &
\hspace{-0.1in}
\includegraphics[width=4.0in,height=2.0in]{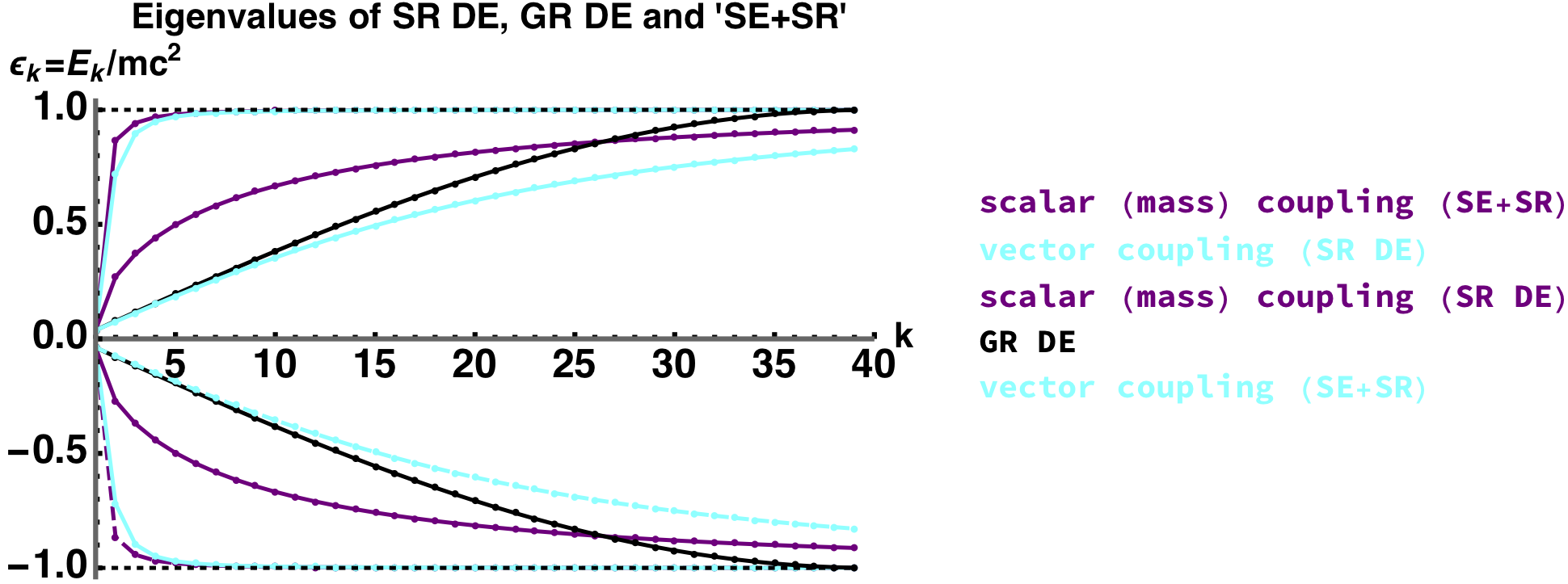}
\end{tabular}
\caption{Eigenvalues of 
(Left: top to bottom)
(cyan curve) SR DE formula with vector coupling coefficient 
$\alpha_v=0.999281$ fitted to  $\eps_1$ of DE in SST,
(purple curve) SR DE formula with scalar coupling coefficient 
$\alpha_s=26.25$ fitted to $\eps_1$ of DE in SST,
(black curve) GR DE .
(Right: top to bottom)
Same figures as on left plot, but with added
(outermost purple curve) ``SE+SR" scalar coupling formula with  $E_I=0.998562$ fitted to $\eps_1$ of DE in SST, and
(innermost cyan curve) ``SE+SR" vector coupling formula with   $E_I=694.278$ fitted to $\eps_1$ of DE in SST.
%(Right: top to bottom)
%(purple curve) ``SE+SR" scalar coupling formula with  $E_I=0.998562$ fitted to $\eps_1$ of DE in SST,
%(cyan curve) SR DE formula with vector coupling coefficient 
%$\alpha_v=0.999281$ fitted to  $\eps_1$ of DE in SST (same as left plot),
%(purple curve) SR DE formula with scalar coupling coefficient 
%$\alpha_s=26.25$ fitted to $\eps_1$ of DE in SST (same as left plot),
%(black curve) GR DE without diffusion (same as left plot).
%(cyan curve) ``SE+SR" vector coupling formula with   $E_I=694.278$ fitted to $\eps_1$ of DE in SST.
}\label{fig:DE:SST:evals:compare:SR:DE:and:SE:DEscalar:vector:coupling} 
\end{figure}

Lastly, in this section 
we compare the eigenvalues
$\eps_k$ of the DE in SST (gray-curve: with diffusion; black curve, without diffusion) 
in \Fig{fig:DE:SST:evals:with:without:diffusion} 
against previous eigenvalue formulas.
In \Fig{fig:DE:SST:evals:compare:SR:DE:and:SE:DEscalar:vector:coupling}(left) 
the outer cyan curve is the SR-DE for vector coupling \Eq{Greiner:p186:case:2} with $\alpha\to~\alpha_v=0.999281$ fitted to $E_1^{(vector)}/mc^2=\eps_1$ using $n=1$ and $j=\half$.
The inner purple curve is the the SR-DE for 
scalar coupling \Eq{Greiner:p186:case:1} with $\alpha'\to\alpha_s=26.25$ fitted to $E_1^{(scalar)}/mc^2=\eps_1$ 
using $n=1$ and $j=\half$.
As our intuition would suspect, the (purple) scalar coupling formula for the SR DE in a sense more closely approximates (qualitatively) the DE in SST (black curve).
%============================
%\begin{figure}[h]
%\begin{tabular}{cc}
%\hspace{-0.75in}
%%\includegraphics[width=4.0in,height=2.25in]{fig_DE_SST_and_SR_DE_formaula_19Jun2022} &
%\includegraphics[width=4.0in,height=2.25in]{fig_6_left_DE_SST_AJP_26Jun2022} &
%\hspace{-0.1in}
%%\includegraphics[width=4.0in,height=2.25in]{fig_DE_SST_and_SE+SR_formaula_19Jun2022}
%\includegraphics[width=4.0in,height=2.25in]{fig_6_right_DE_SST_AJP_26Jun2022}
%\end{tabular}
%\caption{Eigenvalues of 
%(Left:, top to bottom)
%(cyan curve) SR DE formula with vector coupling coefficient 
%$\alpha_v=0.999281$ fitted to  $\eps_1$ of DE in SST (no diffusion),
%(purple curve) SR DE formula with scalar coupling coefficient 
%$\alpha_s=26.25$ fitted to $\eps_1$ of DE in SST (no diffusion),
%(gray curve) GR DE with diffusion $m_s=25,000$,
%(black curve) GR DE without diffusion.
%%
%(Right: top to bottom)
%(purple curve) SE+SR scalar coupling formula with  $E_I=0.998562$ fitted to $\eps_1$ of DE in SST (no diffusion),
%(gray curve) GR DE with diffusion $m_s=25,000$,
%(black curve) GR DE without diffusion.
%(cyan curve) SE+SR vector coupling formula with   $E_I=694.278$ fitted to $\eps_1$ of DE in SST (no diffusion),
%}\label{fig:DE:SST:evals:compare:SR:DE:and:SE:DEscalar:vector:coupling} 
%\end{figure}
%============================

%In \Fig{fig:DE:SST:evals:compare:SR:DE:and:SE:DEscalar:vector:coupling}(right) 
%we compare the DE in SST with the SE+SR formulas \Eq{E_s} and \Eq{E_v} which we had termed 
%for the Schr\"{o}dinger Equation (SE) 
%``scalar coupling" and ``vector coupling" respectively, since the position of the square root being either in the numerator or in the denominator, respectively, mimics that of the SR DE scalar and vector coupling formulas
%\Eq{Greiner:p186:case:1} and \Eq{Greiner:p186:case:2}, respectively. 
%Here ``SE+SR" (Schr\"{o}dinger Equation plus Special Relativity) means that we ``add" $m c^2$ to the NRQM resulting SE energies $E_k$, and interpret the result as the first order approximation to some unknown (from a non-relativistic point of view) SR DE formula, involving square roots (either in the numerator or denominator).
In \Fig{fig:DE:SST:evals:compare:SR:DE:and:SE:DEscalar:vector:coupling}(right) 
we compare the DE in SST with the SE+SR formulas \Eq{E_s} and \Eq{E_v} which we had termed
for the Schr\"{o}dinger Equation (SE)
 ``scalar coupling" and  ``vector coupling" respectively, since the position of the square root being either in the numerator (denominator) mimics that of the SR DE scalar (vector) coupling formula
\Eq{Greiner:p186:case:1} (\Eq{Greiner:p186:case:2}). 
Here ``SE+SR" (Schr\"{o}dinger Equation plus Special Relativity) means that we ``add" $m c^2$ to the NRQM resulting SE energies $E_k$, and interpret the result as the first order approximation to some unknown (from a non-relativistic point of view) SR DE formula, involving square roots (either in the numerator or denominator).
For the 
``scalar coupling" case we fit $\eps_1 =\sqrt{1-\frac{E_I}{n^2}}$ for $n=1$ 
to find  $E^{(vector)}_I=0.998562$, 
while for the 
``vector coupling" case we fit $\eps_1 =\frac{1}{\sqrt{1+\frac{E_I}{n^2}}}$ for $n=1$ 
to find  $E^{(scalar)}_I=694.278$.
The are plotted in \Fig{fig:DE:SST:evals:compare:SR:DE:and:SE:DEscalar:vector:coupling}(right) using the same colors as the (left) plot, namely, cyan for vector coupling and purple for scalar coupling.
Counter to our intuition, now it is the SE ``vector coupling" formula that most closely mimics the eigenvalues of the DE in SST (black curve). This flipping of our intuitive interpretation of the SE+SR formulas \Eq{E_s} and \Eq{E_v} is most likely due to the fact that at the values of $E^{(scalar)}_I=0.998562$ and $E^{(vector)}_I=694.278$, a first order expansion of the square root (as we assumed by adding $m\,c^2$ to $E_I$) is not warranted at these large values of $E_I$.
Nonetheless, it is curious that the roles of the SE ``scalar" and ``vector" coupling flip, and nearly approximate (qualitatively) their SR DE opposites, vector and scalar coupling, respectively,
%\Fig{fig:DE:SST:evals:compare:SR:DE:SE:SR:scalar:vector:coupling:All}.
\Fig{fig:DE:SST:evals:compare:SR:DE:and:SE:DEscalar:vector:coupling}(right). %============================
%\begin{figure}[h]
%%\begin{tabular}{cc}
%%\includegraphics[width=5.0in,height=2.5in]{fig_DE_SST_and_SR_DE_and_SE+SR_formaulas_ALL_19Jun2022} %&
%\includegraphics[width=5.0in,height=2.5in]{fig_7_DE_SST_AJP_26Jun2022} %&
%%\includegraphics[width=3.0in,height=1.5in]{alphaplus_wfdivw0_2_beta_0p1}
%%\end{tabular}
%\caption{Composite of the left and right plots \Fig{fig:DE:SST:evals:compare:SR:DE:and:SE:DEscalar:vector:coupling}. 
%}\label{fig:DE:SST:evals:compare:SR:DE:SE:SR:scalar:vector:coupling:All} 
%\end{figure}
%============================

%==================================================================
\section{Solutions of the DE in SST inside the horizon?}\label{sec:DE:SST:inside:horizon}
%==================================================================
For the solutions of the DE in SST in the previous section we defined the angle $\Theta$ by
$\sin\Theta~=~\sqrt{1-\frac{2 M_s}{r}}$, for $0\le\Theta\le\pi/2\leftrightarrow 2 M_s\le r \le \infty$, the region outside the horizon. Following our intuition, suppose we \tit{anlaytically continue} $\Theta\to i\,x$ so that 
$\sin\Theta\to i\,\sinh x$ and $\sqrt{1-\frac{2 M_s}{r}}\to i\,\sqrt{\frac{2 M_s}{r}-1}$, i.e. defining
$\sinh x \defeq \sqrt{\frac{2 M_s}{r}-1}$, for $0\le r\le 2 M_s \leftrightarrow \infty\ge x\ge 0$. To keep the resulting 
radial equations real (since the above procedure introduces explicit factor of $i$), we are led to consider redefining the ``energies" as $\eps\to i\,\eta$. 

\begin{figure}[h]
%\begin{tabular}{cc}
%\includegraphics[width=6.5in,height=2.5in]{fig_DE_SST_inside_BH_19Jun2022} %&
%\includegraphics[width=6.5in,height=2.5in]{fig_8_DE_SST_AJP_26Jun2022} %&
\includegraphics[width=4.0in,height=2.5in]{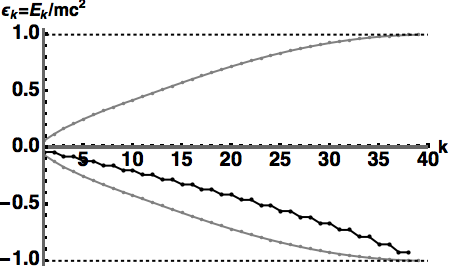} %&
%\includegraphics[width=3.0in,height=1.5in]{alphaplus_wfdivw0_2_beta_0p1}
%\end{tabular}
\caption{Eigenvalues of 
(black curve) DE in SST inside the horizon, \Eq{L1:opr:inside:horizon} and \Eq{L2:opr:inside:horizon}, \\
(gray curve) DE in SST outside the horizon,  \Eq{FG:eval:oprs:line1} and \Eq{FG:eval:oprs:line2},
both with $m_s=10^{15}$.
}
\label{fig:DE:SST:evals:inside:outside:horizon} 
\end{figure}

Holding off on interpretation for now, and boldly (but not blindly) pushing forward, this leads to the following set of eigenvalue equations that we can once again use \ttt{NDEigensystems} to solve 
 \Eq{FG:eval:oprs:line1} and \Eq{FG:eval:oprs:line2}, now with $f(\Theta)\to F(x)$ and $g(\Theta)\to i\,G(x)$
\bea{L1:L2:oprs:inside:horizon}
\hspace{-0.25in}
\mathcal{L}_1 &=& 
\frac{\cosh^4 x}{4 m_s}
\left(
-\tanh x \,\frac{\pd}{\pd x}   + \sinh^2 x + \half
\right)\,G(x) - \sinh x\,F(x), \;\;
\Rightarrow\; \mathcal{L}_1(F,G) = \eta\,F(x), \qquad  \label{L1:opr:inside:horizon} \\
\hspace{-0.25in}
\mathcal{L}_2 &=& 
-\frac{\cosh^4 x}{4 m_s}
\left(
\tanh x \,\frac{\pd}{\pd x}   + \sinh^2 x + \half
\right)\,F(x) - \sinh x\,G(x), \;\;
\Rightarrow\; \mathcal{L}_2(F,G) = \eta\,G(x). \qquad \label{L2:opr:inside:horizon}
\eea
These equations produce the ``eigenvalues" $\eta_k$ (black curve) in \Fig{fig:DE:SST:evals:inside:outside:horizon} with only negative values inside the horizon.
Again, we show the eigenvalues (gray curve) $\eps_k$ outside the horizon, computed in the previous section, for comparison.
%============================
%============================

We must now reconcile imposing an interpretation on the results shown in  \Fig{fig:DE:SST:evals:inside:outside:horizon}, and for the meaning of the values of $\eta_k<0$. 
At first glance setting $\eps = E/m c^2 \to i\,\eta$ seems to imply either imaginary mass, or imaginary energies, and therefore appears non-sensical.
Upon further reflection, note that the temporal and radial portion of the Schwarzschild metric
$ds^2_{outside} = -\left(1-\frac{2 M_s}{r}\right)\,c^2\,dt^2 +  \left(1-\frac{2 M_s}{r}\right)^{-1}\,dr^2$ for $r>2 M_s$ switches signs to
$ds^2_{inside} = \left(\frac{2 M_s}{r}-1\right)\,c^2\,dt^2 -  \left(\frac{2 M_s}{r}-1\right)^{-1}\,dr^2$ inside for $r<2 M_s$.
The interpretation is that the timelike variable is associated with the term with the $-1$ (times a positive factor), and thus becomes $r$ inside the horizon. Thus, the variables $t$ and $r$ switch roles and become spacelike and timelike respectively, inside the horizon. Consequently, moving to the the singularity $r=0$ is inevitable, since ``time" $r$ always moves forward for inward falling particles inside the horizon. 
In fact, $\eps$ now has the interpretation of a momentum (vs an energy, see Hartle's\cite{Hartle:2009} discussion of the Penrose process for extracting energy from a rotating BH, Kerr metric).).
Further, the stationary states, which outside had a temporal dependence of
$e^{-i \eps_k\,t}$ now become upon replacement $\eps_k\to i\,\eta_k$, factors of $e^{\eta_k\,t}$, which for $\eta_k<0$ become evanescent like wave functions decaying to the singularity as $t$ increases. Therefore, the fact the numerics actually produces solutions with only $\eta_k<0$ seems to fit with our intuition.
(Note that if we had instead chosen  $\eps_k\to -i\,\eta_k$ then $e^{-i\eps_k\,t} \to e^{-\eta_k\,t}$ for which $\eta>0$ solutions would have been produced for $t>0$).
Recall that outside the horizon $e^{\pm i k r}$ are waves moving radially outward ($+$) 
and inward ($-$), with $\hbar k$ being the magnitude of the momentum $p_k$, with energy
$p_k c = \hbar k c$. Thus it is conceivable (or at least plausibly interpretable) that $|\eta| \to \frac{(E=\hbar k c)}{m c^2}$ measures a quantized momentum inside the BH.

%=========================================================
\section{Summary and Conclusion}\label{sec:Summary:Conclusion}
%=========================================================
It has been a somewhat lengthy journey to attempt to answer a seeming simple question posed by an astute student (in 15 words or less):
``In analogy with the Coulomb problem, can one talk about bounds states in Schwarzschild spacetime?"
To answer this question, we had to assemble theoretical background concepts, and analytic and numerical solution techniques for 
(i) the Schr\"{o}dinger Equation (SE) in a central potential (the Coulomb problem),
(ii) the Special Relativity (SR) generalization to the Dirac Equation (DE) in flat (Minkowski spacetime, MST), and finally
(iii) the General Relativity (GR) generalization of the DE to curved spacetime (CST), specifically, the central symmetric Schwarzschild spacetime (SST). Driven by intuition, and bolstered by the courage of our convictions, we attempted to put some theoretical, analytic, and numerical terra firma beneath feet, though knowing at times we might be pushing an analogy (or two) beyond the limits of credulity.  The results in \Sec{sec:bound:states:DE:SST}, and in particular, the bound state eigenvalues 
$|E_k|< m c^2$ numerically found in \Fig{fig:DE:SST:evals:with:without:diffusion} seem to indicate that the answer to the student's question appears to be affirmative. 
Recall that the SR DE admits eigenvalues for scalar coupling for any value of $\alpha' >0$, small, or large \Eq{Greiner:p186:case:1}, so it is not surprising that the GR DE also emits quantized eigenvalues for $\alpha'$ very large.
Further, the comparison of the the eigenvalues for the DE in SST with formulas from both the SR DE and the SE+SR, as shown in 
\Fig{fig:DE:SST:evals:compare:SR:DE:and:SE:DEscalar:vector:coupling}(right),
%\Fig{fig:DE:SST:evals:compare:SR:DE:SE:SR:scalar:vector:coupling:All}, 
shows that the student's intuition, and question, had merit. 

The intent of this investigation was to highlight a research strategy that could be employed by the student, drawing upon knowledge obtained in upper undergraduate to first year graduate physics education, supplemented by a host of well written and highly useful expository QM and GR textbooks currently available to the student, with enough worked exercises that this whole endeavor could be the seed of an independent research project.

Much more remains that could be investigated, as we have only scratched the surface here, and peeked under a few lids, on a host container of fascinating NR and SR QM, and GR topics worthy of deeper investigation (e.g. angular momentum and spin details in the SR and GR Dirac equation, solutions of massless particles in Schwarzschild and other spacetimes often used to model photons, and the role of spinor transformations under the Poincare group of rotations, boost and translations, to name just a few).  It is hoped that this investigation has served as a vehicle that might pique the research interest of the inquisitive and curious student, and spur them to further guided or self study, complemented by a host available instructive textbooks, and hopefully beyond, to the research literature itself.
\medskip

%\blue{This ``test your comprehension" of the material extra credit problem could be moved to the SM if space is required.}

% red 2116-2139
{\color{black}
\flushleft{\underline{\bf Extra credit (research) project:}}
\newline
For the ambitious student (or professor!), you can verify your understanding of this material by redoing the DE in CST calculation presented in the main text for the \tit{Reissner-N{\o}rdstrom} metric (see Ryder\cite{Ryder:2009}, p265) for a star with mass $M$ and electric charge $Q$ given by spherically symmetric metric
\be{RN:metric}
\hspace{-0.25in}
ds^2 = -c^2 \left(1- \frac{2 G M }{c^2 r} + \frac{G Q^2}{c^4 r^2}\right) dt^2 
+ \left(1- \frac{2 G M }{c^2 r} + \frac{G Q^2}{c^4 r^2}\right)^{-1}\,dr^2
+ r^2\,\left(d\theta^2 + \sin^2\theta\,d\phi^2\right).
\ee
As before, $\frac{2 G M }{c^2 r} = \frac{r_s}{r}$, while
$\frac{G Q^2}{c^4 r^2} = 
\left(\frac{e^2}{\hbar c}\right) 
\left(\frac{Q^2}{e^2}\right) 
\left(\frac{G \hbar}{c^3 r^2}\right)
=
\alpha_c\,
\left( \frac{Q}{e} \right)^2\,
\left( \frac{\ell_p}{r} \right)^2,
$
where $\ell_p$ is the Planck length and $\alpha_C$ is the fine structure constant (see \App{app:Units}).
In the spirit of John Wheeler (who coined the term black hole), what might one anticipate about any possible solutions even before embarking on the lengthy calculation (to back up one's intuition)?
Hint: compare the magnitude and length scales of the two $r$-dependent terms in \Eq{RN:metric}.
Note: Astronomically speaking, the Reissner-N{\o}rdstrom is not considered particularly physical, since any charge on a BH will most likely be rapidly neutralized by surrounding in-falling matter. However, the metric is an intuitively satisfying extension of the Schwarzschild metric, extending the sole extensive properties of the BH from mass, to mass plus charge - and it is amenable to the analysis presented in this work.
} % red from 2116

%=========================================================
% Appendices
%=========================================================
%\clearpage
%\newpage
%=========================================================
\appendix
%=========================================================
\section{Units}\label{app:Units}
In this appendix we detail the natural scale lengths (in SI units) used to write both the SE and DE (SR and GR) in dimensionless form. This will aid in the comparison of the equations and eigenvalues of the bound states.

\subsection{The Coulomb problem}\label{app:subsec:Coulomb}
Let $e>0$ be the magnitude of the charge on the electron, 
$\hbar= 1.054\times 10^{-34}$ J s Planck's constant and 
$c\approx 3 \times 10^8$ m/s the speed of light in vacuum. The fundamental dimensional constant that sets the scale for all electromagnetic phenomena is the fine structure constant $\alpha$ defined by
\be{alpha}
\alpha = \frac{e^2}{\hbar c}\sim \frac{1}{137}.
\ee
(see the inside cover of CTQMv1).

Let $m_e=9.11 \times 10^{-31}$ kg  be the mass of the electron, with rest mass
$m_e c^2 \sim 0.5$ MeV where $1eV = 1.6 \times 10^{-19}$ J.
Then, the natural length scale for the electron (or any particle of mass $m$) is its Compton wavelength
$\lambda_C$ given by
\be{lambda:c}
\lambda_C = \frac{\hbar}{m c} \overset{m\to m_e}{=} 2.426 \times 10^{-12}\, \trm{m}.
\ee
This length scale depends on the mass of the particle, and can be interpreted as the distance a classical electron must move in order to be $(Rayleigh)$ resolved in position to be considered as having moved a small distance.
For electromagnetic phenomena, we can define a natural length scale called the ``classical" radius of the electron $r_e$, depending $e$ defined as 
\be{r:e}
r_e = \frac{e^2}{m_e c^2} = \left(\frac{e^2}{\hbar c}\right)\,\left(\frac{\hbar c}{m_e c^2}\right) 
= \alpha\,\lambda_C = 2.82 \times 10^{-15}\,\trm{m}.
\ee

\subsection{The Bohr Atomic Model}\label{app:subsec:Bohr:Model} 
The semi-classical Bohr atomic model  was the first successful attempt to describe the eigenenergies of the hydrogen atom. It is based on the hypothesis that the electron describes a circular orbit of radius $r$ about the proton obeying the following equations (CTQMv1, p791, using $\mu\to m_e\to m$)
\bea{Bohr:Model}
E &=& \frac{1}{2} m v^2 - \frac{e^2}{r}, \quad \trm{-classical kinetic/potential energy balance equation},\\
\frac{m v^2}{r} &=& \frac{e^2}{r^2}, \hspace{0.55in}\quad \trm{-Coulomb force equals mass times radial acceleration}, \\
m v r &=& n \hbar, \hspace{0.6in}\quad\trm{-postulated quantization condition; integer values of action}.\quad
\eea
The simple to derive solutions of these equations leads to
\bea{Bohr:Model:solns}
E_n &=& -\frac{E_I}{n^2}, \quad\trm{with ionization energy}\quad E_I = \frac{m e^4}{2 \hbar^2} = \frac{1}{2} \alpha^2\,m c^2,\\ 
r_n &=& n^2\,a_0,  \quad\trm{with Bohr radius}\quad a_0 = \frac{\hbar^2}{m e^2} = \frac{\lambda_C}{\alpha}\sim 137\,\lambda_C, \\
v_n &=& \frac{v_0}{n}, \quad\quad\trm{lowest circular orbit velocity}\quad v_0 = \frac{e^2}{\hbar} \sim \alpha\,c=\frac{1}{137}\, c.
\eea

\subsection{Gravitation}\label{app:sec:Gravity}
Gravity introduces a  new fundamental constant, the gravitational constant $G$, that sets the scale for gravitational interactions. 
From the Newtonian potential 
$V_G=-\frac{G M m}{r}$ we see that $G$ has units of energy-length/mass$^2$. 
Therefore, by dimensional arguments alone, there should exist a new mass defined solely in terms the fundamental constants. Since the above indicates that mass$^2\sim$ energy-length/$G$ and 
$\hbar$ has units of energy-time, and so $\hbar c$ has units of energy-length, we have that
\be{Mp}
M_p = \sqrt{\frac{\hbar c}{G}} \approx 2.2\times 10^{-5} \trm{g} = 10^{-8} \trm{kg},
\ee
defining the Planck mass.
A new gravitational length, the Planck length $\ell_p$ is then defined as the Compton wavelength of the Planck mass
\be{lp}
\ell_p = \frac{\hbar}{M_p c} =  \sqrt{\frac{\hbar G}{c^3}}\approx 4.5 \times 10^{-34} \trm{m}.
\ee
The Planck length is thought to be the scale at which spacetime breaks down as a classical continuum concept and quantum effects become paramount (warranting a full blown quantum theory of gravity).
Finally, the Planck time $t_p$ is defined as
\be{lp}
t_p = \frac{\ell_p}{c} =  \sqrt{\frac{\hbar G}{c^5}}\approx 1.5 \times 10^{-42} \trm{s}.
\ee

Lastly, from a classical Newtonian argument of the largest possible escape velocity for a mass $M$, one has
$E= \frac{1}{2} m v^2 - \frac{G M m}{r} =0$ with $v\to c$ yielding the Schwarzschild radius $r_s$ of the mass $M$
\be{rs}
r_s = \frac{2 G M}{c^2}.
\ee

The naive analogue of the E\&M fine structure constant is
\be{alpha:G}
\alpha_C = \frac{e^2}{\hbar c} \to \alpha_G = \frac{G M m}{\hbar c} 
= \left(\frac{M}{M_p}\right)\,\left(\frac{m}{M_p}\right)
\ee
For a solar mass black hole (BH) $M=M_{\bigodot}=2 \times 10^{30}$ kg, and the mass of the electron
$m=m_e = 9.11\times 10^{-31}$ kg, we have $\alpha_G\sim 4 \times 10^{15}\gg 1$.
Thus, we should be careful (suspicious) of such a naive translation from E\&M to gravity when pushing the analogy of a ``gravitational" Bohr model, since $E^{C}_I = \frac{1}{2} \alpha_C m c^2$ would naively suggest 
$E^{G}_I\overset{?}{\to} \frac{1}{2} \alpha_G m c^2\gg m c^2$, which might suggest that no  gravitational bound states would exist. One of the primary goals in the main text is to investigate the ``go/no-go" validity of this point.

%%=========================================================
%\section{Solutions of the NR Schr\"{o}dinger Equation for the Coulomb potential}\label{app:NRSE:Coulomb}
%
%%=========================================================
%\section{Separation of variables for the DE in a central potential}\label{app:SRDE:sep:var}
%
%%=========================================================
%\section{Solutions of the SR Dirac Equation for vector (Coulomb) and scalar coupling}\label{app:SRDE:central:potential}
\medskip

%\blue{The following two Appendices B and C could be moved to the SM to save space if required.}

% red 2337-2571
{\color{black} 
%=======================================================================
\section{The commutators, spinor connection and DE using orthonormal basis vectors in the SST}\label{app:commutators:SST}
%========================================================================
In a coordinate basis with 
\be{app:Schw:basis:vectors}
\e_t =\frac{\pd}{\pd t},\;\;
\e_r = \frac{\pd}{\pd r},\;\;
\e_\theta =\frac{\pd}{\pd \theta},\;\;
\e_\phi =\frac{\pd}{\pd \phi}, \quad
\e_\mu\cdot\e_\nu = g_{\mu\nu}
\ee
The connection $\G^{\mu}_{\alpha\beta}$ (Christoffel symbols) are given by the standard formula
\be{app:Levita:Civita:coord:basis}
\trm{coordinate basis:}\quad
\G^{\mu}_{\alpha\beta} = 
\half\,g^{\mu\lambda}\,(\pd_\alpha\,g_{\lambda\beta} + \pd_\beta\,g_{\lambda\alpha}-\pd_\lambda\,g_{\alpha\beta}).
\ee

In the following (see Ryder, Chapter 11.4, pp416-418, also p109), we show how the commutators 
of the orthonormal basis vector $\e_a$ for the Schwarzschild metric, 
given by
\bea{app:Ryder:p417:11.131}
\hspace{-0.5in}
\e_0 = \frac{1}{c}\,\left(1- \frac{2 M_s}{r}\right)^{-1/2}\,\frac{\pd}{\pd t},\;\;
\e_1 &=& \left(1- \frac{2 M_s}{r}\right)^{1/2}\,\frac{\pd}{\pd r},\;\;
\e_2 = \frac{1}{r}\,\frac{\pd}{\pd \theta},\;\;
\e_3 = \frac{1}{r \sin\theta}\,\frac{\pd}{\pd \phi}, \\
\hspace{-0.5in}
\e_a\cdot\e_b &=& \eta_{ab} \defeq \trm{diagonal}(-1,1,1,1,), \quad \trm{(GR convention)},
\eea
%see Ryder, pp418-419
with $M_s = \frac{G M}{c^2} = \half\,r_s$,
 produce the spinor connection coefficients $\G_{abc}$ defined by
\bea{app:Ryder:p108:3.259}
\trm{orthonormal basis:}\quad 
\G_{abc}&=&
-\half\,
( 
C_{abc} + C_{bca} - C_{cab} 
)
 \\ 
C_{abc} &=& C_{ab}^{\;\;\;d}\,\eta_{dc}\,, \quad\trm{where}\quad [\e_a,\e_b] =C_{ab}^{\;\;\;c}\, \e_c.
\eea
Note: the connection can be defined by the \tit{sum} of the two formulas 
\Eq{app:Levita:Civita:coord:basis} and \Eq{app:Ryder:p108:3.259}, since for the case of coordinate basis vectors the commutators are zero, and for the case of an orthonormal basis, the metric is constant, so their partial derivatives are zero.
(Note: Ryder switches which index on $C$ is raised/lowered in going from (3.262) p109 (first index), to 
(11.134) p 417 (last index). In this appendix, we follow the latter convention).

Consider now the generic commutator acting on a arbitrary function $f$
\bea{commutators}
[\e_a, \e_b]\,f &=& \Big(e_a^\mu(x)\pd_\mu\Big)\, \Big(e_b^\nu(x)\pd_\nu\,f(x)\Big) - 
                              \Big(e_b^\nu(x)\pd_\nu\Big)\, \Big(e_a^\mu(x)\pd_\mu\,f(x)\Big), \no
&=& 
\Big(e_a^\mu(x)  \Big(\pd_\mu\,e_b^\nu(x)\Big) \Big)\,\pd_\nu\,f(x) - 
\Big(e_b^\nu(x) \Big(\pd_\nu\, e_a^\mu(x)  \Big)\Big)\,\pd_\mu\,f(x), \no
&=& 
\left(
\e_a\Big(e_b^\nu(x)\Big)\,\pd_\nu\, - \e_b\Big(e_a^\mu(x)\Big)\,\pd_\mu\,
\right)\,f, \no
\Rightarrow
[\e_a, \e_b] &=& 
\left(\e_a\Big(e_b^\nu(x)\Big) - \e_b\Big(e_a^\nu(x)\Big)\right)\,\pd_\nu\, \no
&\equiv& [\e_a, \e_b]^\nu\,\pd_\nu, \no %=  [\e_a, \e_b]^\nu\, e_\nu^{\;\; c}\,\e_c, \no
&=&  [\e_a, \e_b]^\nu\, e_\nu^{\;\; c}\,\e_c, \no
 {\Rightarrow [\e_a, \e_b]} &\equiv& C_{a b}^{\;\;\; c}\,\e_c.
\eea
where in the fourth line we dropped the $f(x)$ since it was arbitrary, and relabeled the dummy indices
$\mu\to\nu$ in the second term. We also used the inverse of the tetrad $\e_a^{\;\;\mu}(x)$ defined from
$\e_a = e_a^{\;\;\nu}\,\pd_\nu \Rightarrow \pd_\nu = (e_a^{\;\;\nu})^{-1}\,\e_a \defeq e_\nu^{\;\;a}\,\e_a$.
(Note the switch in the position of the indices $a$ and $\nu$, indicating the inverse tetrad).
Since the metric is diagonal, the inverse tetrad components are just the one over the tetrad components, e.g.
for $e_0^t = \frac{1}{c}\,\left(1-\frac{2 M_s}{r}\right)^{-1/2}$ we simply have
$e_t^0 = c \left(1-\frac{2 M_s}{r}\right)^{1/2}$, etc. and are easy to compute by hand.
The general formula \Eq{commutators} is useful for systematic calculations of the commutators by symbolic manipulation programs such as Mathematica\cite{Mathematica}, especially when the metric is non-diagonal.

As a representative example, we have
\bea{}
[\e_0, \e_1] 
&=&
\left[
\left(1-\frac{2 M_s}{r}\right)^{-1/2}\,\frac{1}{c}\,\frac{\pd}{\pd t},
\left(1-\frac{2 M_s}{r}\right)^{1/2}\,\frac{\pd}{\pd r}
\right], \no
&=& -\left(1-\frac{2 M_s}{r}\right)^{1/2}\, 
\left(\frac{\pd}{\pd r}\,\left(1-\frac{2 M_s}{r}\right)^{-1/2}\right)\,\frac{1}{c}\,\frac{\pd }{\pd t}, \no
&=& \frac{M_s}{r^2}\,\left(1-\frac{2 M_s}{r}\right)^{-1}\,\frac{1}{c}\,\frac{\pd}{\pd t}, \no
&=& \frac{M_s}{r^2}\,\left(1-\frac{2 M_s}{r}\right)^{-1/2}\,\e_0, \\
\Rightarrow C_{01}^{\;\;\; 0} &=& \frac{M_s}{r^2}\,\left(1-\frac{2 M_s}{r}\right)^{-1/2} = -C_{010},
\eea
where in the last line we have used $C_{010}=C_{01}^{\;\;\; a}\,\eta_{a0} = (-1)\,C_{01}^{\;\;\; 0}$.
Thus, when lowering terms with the temporal index $a=0$ we pick up a minus sign (with the GR local SR metric convention $\eta_{ab}=\{-1, 1, 1, 1\}$), while we do not for terms with spatial indices $a=(1,2,3)$.

The rest of the commutators can be worked out in a similar fashion, and the non-zero ones are given by
(see Rydyer, p417)
\bea{}
C_{010} &=& -\frac{M_s}{r^2}\,\left(1-\frac{2 M_s}{r}\right)^{-1/2} = -C_{100},\\
C_{122}=-C_{212} &=& -\frac{1}{r}\,\left(1-\frac{2 M_s}{r}\right)^{1/2} = C_{133}=  -C_{313},\\ 
C_{233} &=& -C_{323} = -\frac{\cot\theta}{r}.
\eea

Substituting these results in \Eq{app:Ryder:p108:3.259} yields the non-zero
Christoffel symbols for the orthonormal basis: (see Ryder, p418)
%\bea{}
%\G_{010} &=& -\G_{100} =  \frac{M_s}{r^2}\,\left(1-\frac{2 M_s}{r}\right)^{-1/2}, 
%\hspace{1.0in} \G_{001} = 0,\\
%%
%\G_{122} &=& \G_{212} =\G_{133} = -\G_{313} =  \frac{M_s}{r^2}\,\left(1-\frac{2 M_s}{r}\right)^{1/2}, \quad
%\G_{221} = \G{331} = 0, \\
%%
%\G_{233} &=& -\G_{323} = \frac{\cot\theta}{r}, 
%\hspace{2.25in} \G_{332}=0.
%\eea

\begin{alignat}{3}
\G_{010} &= -\G_{100} &&=  \frac{M_s}{r^2}\,\left(1-\frac{2 M_s}{r}\right)^{-1/2}, 
\qquad &&\G_{001} = 0,\\
\G_{122} = -\G_{212} =\G_{133} &= -\G_{313} &&=  \frac{M_s}{r^2}\,\left(1-\frac{2 M_s}{r}\right)^{1/2}, 
&&\G_{221} = \G_{331} = 0,\\
\G_{233} &= -\G_{323} &&= \frac{\cot\theta}{r}, 
&&\G_{332}=0.
%  &[x \mapsto s]x                       &&= s && \\
%  &[x \mapsto s]y                       &&= y \qquad                                          &&\text{als } y\neq x \\
%  &[x \mapsto s](\lambda(y)t_1) &&= \lambda(y)[x \mapsto s]t_1 \qquad &&\text{als } y \neq x \text{ en   } y \not \in FV(s) \\
%  &[x \mapsto s](t_1 \; t_2)&&= ([x \mapsto s]t_1)\;([x \mapsto s]t_2)       &&
\end{alignat}

Finally, for the DE in CST
\be{app:DE:CST}
i\,\hbar\,\g^\a\left(e_a + \G_a\right)\,\psi = m c\,\psi \qquad (\trm{writing, conventionally}\; \e_a\to e_a)
\ee
the spinor connection coefficients 
\be{app:spinor:connection}
\G_a = \frac{1}{8}\,\G_{abc}\,[\g^a, \g^b],
\ee 
are given by (see Ryder, p418)
\bea{app:spinor:conn:coeffs}
\G_0 &=&  \frac{M_s}{2 r^2}\,\left(1-\frac{2 M_s}{r}\right)^{-1/2}\,\hat{\alpha}^1, \quad \G_1 = 0, \\
%\G_1 &=& 0, \\
\G_2 &=& -\frac{i M_s}{2 r^2}\,\left(1-\frac{2 M_s}{r}\right)^{1/2}\,\hat{\Sigma}^3, \\
\G_3 &=& \frac{i M_s}{2 r^2}\,\left(1-\frac{2 M_s}{r}\right)^{1/2}\,\hat{\Sigma}^2 
-\frac{i\,\cot\theta}{2 r}\, \hat{\Sigma}^1,
\eea
where
\be{}
\hat{\alpha}^i = 
\left(\begin{array}{cc}0 & \hat{\sigma}^i \\ \hat{\sigma}^i  & 0\end{array}\right), \qquad
\hat{\Sigma}^i = 
\left(\begin{array}{cc}\hat{\sigma}^i  & 0\\ 0  & \hat{\sigma}^i\end{array}\right).
\ee
One then uses these relations
\be{}
\g^0\,\hat{\alpha}^1 = -\g^1, \;\;
\g^2\,\hat{\Sigma}^3 = -\g^3\,\hat{\Sigma}^2 = i\,\g^1, \;\;
\g^3\,\hat{\Sigma}^1 = i\,\g^2, 
\ee
where the (constant) Dirac matrices are given by
\be{}
\g^0 = \left(\begin{array}{cc}\mathbb{I}  & 0 \\ 0  &-\mathbb{I}\end{array}\right), \qquad
\g^i =\left(\begin{array}{cc} 0  & \hat{\sigma}^i \\ -\hat{\sigma}^i  &0\end{array}\right),
\ee
and inserts all this into \Eq{app:spinor:connection}, and then into 
%\Eq{app:DE:CST} 
 $i\,\hbar\,\gamma^a (e_a + \G_a)\,\psi$ \Eq{Ryder:p417:11.129},
to arrive at the DE in SST as given by \Eq{Ryder:p418:11.139} in the main text
(repeated below, see Ryder, p418)
\bea{app:DE:SST}
&{}&
i\,\hbar\,
\left\{
\left(1- \frac{2 M_s}{r}\right)^{-1/2}\,
\left[
\g^0\frac{1}{c}\frac{\pd\psi}{\pd t} - \frac{M_s}{2 r^2}\g^1\psi
\right]
+ \left(1- \frac{2 M_s}{r}\right)^{1/2}\g^1\,\frac{\pd \psi}{\pd r}
\right. \no
&{}& 
\left.
\hspace{0.20in} +\; 
\g^2\,\frac{1}{r}\frac{\pd \psi}{\pd \theta} +
\frac{M_s}{r^2}\,\left(1- \frac{2 M_s}{r}\right)^{1/2}\g^1\,\psi +
\g^3\,\frac{1}{r \sin\theta}\frac{\pd \psi}{\pd \phi} +
\frac{\cot\theta}{2 r}\,\g^2\,\psi 
\right\}
= m\,c\,\psi, \label{app:DE:CST}
\eea
 for a Dirac spin-1/2 particle of mass $m$ in a Schwarzschild field.

%=========================================================
\section{Other spinor solutions to \Eq{DE:CST:p5.3:final:line1} and \Eq{DE:CST:p5.3:final:line2}}\label{app:other:spinor:solns}
%=========================================================
Let us consider a more general ansatz for the spinors $\varphi$ and $\chi$:
\be{}
\varphi = g(\rho)\,
\left(\begin{array}{c}
\g\,Y_{\ell_1,m_1} \\
\delta\,Y_{\ell_2,m_2} 
\end{array}\right)\,
\qquad
\chi = i\,f(\rho)\,
\left(\begin{array}{c}
\g\,Y_{\ell_3,m_3} \\
\delta\,Y_{\ell_4,m_4} 
\end{array}\right),
\ee
where $Y_{\ell_i, m_i}(\theta,\phi)$ are the usual spherical harmonics\cite{Arfken:2012}

If we retain the angular contributions in the DE in SST \Eq{DE:pma:p5.3:top:line1}-\Eq{DE:pma:p5.3:top:line3}, 
the full set of equations are
\bea{DE:general}
\left(\begin{array}{cc}
\Qhat_\rho & \frac{\rootF}{\rho}\,\Ohat^+ \\
\frac{\rootF}{\rho}\,\Ohat^- & -\Qhat_\rho
\end{array}\right)\, (i\,f)
\left(\begin{array}{c}
\gamma\,Y_{\ell_3,m_3} \\
\delta\,Y_{\ell_4,m_4} 
\end{array}\right)
&=&
i\,(\eps -\rootF)\,g\,
\left(\begin{array}{c}
\alpha\,Y_{\ell_1,m_1} \\
\beta\,  Y_{\ell_2,m_2} 
\end{array}\right), \label{DE:general:line1} \\
&{}&\no
%
%\trm{and}\quad  
\left(\begin{array}{cc}
\Qhat_\rho & \frac{\rootF}{\rho}\,\Ohat^+ \\
\frac{\rootF}{\rho}\,\Ohat^- & -\Qhat_\rho
\end{array}\right)\, g
\left(\begin{array}{c}
\alpha\,Y_{\ell_1,m_1} \\
\beta\,Y_{\ell_2,m_2} 
\end{array}\right)
&=&
i\,(\eps +\rootF)\,(i\,f)\,
\left(\begin{array}{c}
\gamma\,Y_{\ell_3,m_3} \\
\delta\,  Y_{\ell_4,m_4} 
\end{array}\right), \label{DE:general:line2}
\eea
where 
\bea{Q:O:pm}
\Qhat_1(\rho)&=&\F\,\frac{\pd}{\pd \rho} + \F\, \frac{m_s}{\rho^2} - \frac{m_s}{2\,\rho^2}, \label{app:Q:opr} \\
\Ohat^\pm &=& \left(\frac{\pd }{\pd \theta} + \half\cot\theta\right) \pm \left( -i\,\frac{\pd }{\pd \phi}\right). \label{app:O:opr}
\eea
The spherical harmonics have the properties (see Arfken\cite{Arfken:2012}, Chapter 11)
\bea{Ylm:properties}
%\hspace{-0.5in}
\left( -i\,\frac{\pd }{\pd \phi}\right)\, Y_{\ell, m} &=&  m\,Y_{\ell, m}, \label{Ylm:properties:phi} \\
%
%\hspace{-0.5in}
\left(
\frac{\pd }{\pd \theta} + \half\,\cot\theta
\right) 
Y_{\ell, m} &=& 
\cot\theta
\left(
\ell\,e^{i\,m \phi} + \half
\right)\, Y_{\ell, m}
-\frac{(-1)^m}{\sin\theta}\,\sqrt{\frac{2 \ell+1}{2 \ell-1}}\,\sqrt{\ell^2-m^2}\, Y_{\ell-1,m}. \qquad
\label{Ylm:properties:theta}
\eea
For equatorial orbits $\theta=\pi/2$ we have $\cot\theta\overset{\theta\to\pi/2}{\longrightarrow} 0$.
Further, we note that for  $\theta=\pi/2$ we have\cite{Arfken:2012}
\be{Ylm:theta:pidiv2}
Y_{\ell, m}(\theta=\pi/2, \phi) = 
\begin{cases}
\sqrt{\frac{2 \ell +1}{4 \pi}}\, \frac{\sqrt{\ell-m)!\,(\ell+m)!}}{(\ell-m)!\,!(\ell+m)!!}\, e^{i\,m \phi}, & 
\ell+m\;\;\trm{even, \;or } \; 
(\ell,m) \in \left\{\trm{(even, even),  (odd, odd)}\right\} \\
0 &  \ell+m\;\;\trm{\;odd, \;or}\; (\ell,m) \in \left\{\trm{(even, odd),  (odd, even)}\right\},
\end{cases}
\ee
so that on the righthand side of \Eq{Ylm:properties:theta}
$Y_{\ell-1,m}(\theta,\phi)\overset{\theta\to\pi/2}{\longrightarrow} 0$ for $\ell-1+m$ odd, or $\ell+m$ even.
Therefore, for $(\ell, m)\in \{\trm{(even, even), (odd, odd)}\}$, the righthand side of \Eq{Ylm:properties:theta} 
is zero. (Note: Zeroing out these terms will lead to one possible solution that will let us avoid the $Y_{\ell-1, m}$ term when trying to cancel the spherical harmonics from both sides of all equations).

Continuing, as in the main text, let us take
$\gamma\to\alpha$ and $\delta\to-\beta$ and further require that $\alpha/\beta = \beta/\alpha$ or 
$\alpha^2=\beta^2$. Then, with the normalization $|\alpha|^2 + |\beta|^2=1$  this implies 
$\beta = \pm\alpha = \frac{1}{\sqrt{2}}$.
Substituting in these results, \Eq{DE:general:line1} and \Eq{DE:general:line2} become
\bea{DE:general:theta:pidiv2}
(\Qhat_\rho f) Y_{\ell_3, m_3} - f\, \frac{\rootF}{\rho}\, m_4\,Y_{\ell_4, m_4} &=& (\eps-\rootF)\,g\,Y_{\ell_1, m_1}, \label{DE:general:theta:pidiv2:line1} \\ 
(\Qhat_\rho f) Y_{\ell_4, m_4} - f\, \frac{\rootF}{\rho}\, m_3\,Y_{\ell_3, m_3} &=& (\eps-\rootF)\,g\,Y_{\ell_2, m_2}, \label{DE:general:theta:pidiv2:line2} \\
&{}&\no
(\Qhat_\rho g) Y_{\ell_1, m_1} + g\, \frac{\rootF}{\rho}\, m_2\,Y_{\ell_2, m_2} &=& -(\eps+\rootF)\,f\,Y_{\ell_3, m_3}, \label{DE:general:theta:pidiv2:line3} \\
(\Qhat_\rho g) Y_{\ell_2, m_2} + g\, \frac{\rootF}{\rho}\, m_1\,Y_{\ell_1, m_1} &=& -(\eps+\rootF)\,f\,Y_{\ell_4, m_4}. \label{DE:general:theta:pidiv2:line4}
\eea
Now, \Eq{DE:general:theta:pidiv2:line1} is the same equation as \Eq{DE:general:theta:pidiv2:line2},
and \Eq{DE:general:theta:pidiv2:line3} is the same equation as \Eq{DE:general:theta:pidiv2:line4}, if we have
$(\ell_i, m_i) = (\ell, m)$ for all $i\in\{1,2,3,4\}$. This choice then results in the spherical harmonics canceling from both sides of all four equations and we end up with just two radial equations:
\bea{DE:radial:eqns:m}
\Qhat_\rho f  - f\, \frac{\rootF}{\rho}\, m &=& (\eps-\rootF)\,g, \label{DE:radial:eqns:m:line1} \\
\Qhat_\rho g  + g\, \frac{\rootF}{\rho}\, m &=& -(\eps+\rootF)\,f, \label{DE:radial:eqns:m:line2}
\eea
for $m\in\{-\ell, -\ell+1,\ldots, \ell-1, \ell\}$, for a given $\ell$.
For $m=0$, \Eq{DE:radial:eqns:m:line1} and \Eq{DE:radial:eqns:m:line2}
reduce to \Eq{DE:CST:p5.3:final:line1} and \Eq{DE:CST:p5.3:final:line2}, respectively in the main text 
independent of the value of $\ell$ (though from \Eq{Ylm:theta:pidiv2} $Y_{\ell, m}(\theta=\pi/2, \phi)\ne 0$ only for even $\ell$ for $m=0$).
} % red from 2337
%=========================================================

\section{A discussion of the numerical methods used to find the eigenvalues of the SR DE and DE in SST}\label{app:NumericalMethods}
In this appendix we discuss the essentials of the numerical methods that were used to compute the eigenvalues of the SR DE and DE in SST in the main text. The \textit{Mathematica}\cite{Mathematica} routine \ttt{NDEigensystem} used a finite element method (FEM)  to computing both the eigenvalues and eigenfunctions. Here we outline the essentials of the FEM  for computing eigenvalues/eigenfunctions and discuss issues concerned with their accuracy and convergence. One finds that the eigenfunctions are much more sensitive to the numerical procedure than the eigenvalues.
The primary source we reference is the informative article by Boffi\cite{Boffi:2010}, and slide presentation by van der Vegt\cite{vanderVegt:2021}. We illustrate these methods using \ttt{NDEigensystem} on some simple 1D systems where the analytic solutions are well known. 

We then discuss the added complications that arise in coupled first order eigenvalue/eigenfunction equations that arise in the radial DE (which is first order space). These are advection (``flow-like") equations which can be unstable unless one introduces numerical dissipation (diffusion) as discussed in \tit{Numerical Recipes in C}\cite{NumRecinC:1992}, and outlined here. The addition of numerical dissipation is mainly required to smooth (damp) out the high (spatial) frequencies of the eigenfunctions, and has a lesser effect on the values of the computed eigenvalues. In the main text we are primarily concerned with the eigenvalues of the DE (both SE, SR and GR) and their comparison, and all eigenvalues are quoted without the use of numerical dissipation. Thus, this discussion is mainly relevant only to \Sec{subsec:numerical:diffusion}, and does not effect the primary eigenvalue results of the other subsections of  \Sec{sec:bound:states:DE:SST} and \Sec{sec:DE:SST:inside:horizon}. However, it is an important enough topic in general, that the reader ought to be aware of the issues involved.

\subsection{The Finite Element Method 1D Eigenvalue Problems}\label{subsec:FE}
The finite element method (FEM) is a ubiquitous efficient solution method for solving coupled partial differential equations (PDEs) in complicated domains. Crudely, it involves the use of a mesh to grid the domain of interest and function expansions such that the PDEs are reduced to an algebraic set of equations that can be solved by a variety of efficient means. Here we tackle only the most elementary of problems to illustrate the very basics of the FEM (for further details, see Boffi\cite{Boffi:2010} and van der Vegt\cite{vanderVegt:2021}).

Consider the eigenvalue problem for the one-dimensional (1D) Laplace operator
\bea{1D:Laplace:Eq}
-u''(x) &=& \lambda \, u(x), \qquad \trm{in}\; \Omega = [0,\pi],\label{1D:Laplace:Eq:line1}\\
u(0) &=& u(\pi) = 0 \quad \trm{on}\; \pd\Omega = \{0\}\cup\{1\} \label{1D:Laplace:Eq:line2},
\eea
where prime denotes differentiation by $x$.
The analytic solution to \Eq{1D:Laplace:Eq:line1} with boundary conditions (BC) \Eq{1D:Laplace:Eq:line2} is elementary and is given by $u(x) = \sin(k\,x)$ for $k=\{1,2,3,\ldots\}$ so that $\lambda=k^2$.

We now convert this problem into its \tit{weak form} by considering a (variational) set of functions 
$v(x)\in V = H_0^1(x)$ with the property that $v(0)=v(\pi)=0$. Then multiplying both sides of 
\Eq{1D:Laplace:Eq:line1}  by $v(x)$ and integrating by parts, one obtains 
\bea{}
\int_0^\pi u'(x)\,v'(x)\,dx = \lambda\,\int_0^\pi u(x)\,v(x)\,dx, \quad \forall\; v(x)\in V,
\eea{}
where the boundary term $u'(x)\,v(x)\big|_0^\pi=0$ since $v(x)$ vanishes on the boundary $\pd\Omega$.

A \tit{Galerkin} approximation is employed using a finite-dimensional subspace 
$V_h=\trm{span}\{\varphi_1,\ldots,\varphi_N\}\subset~V$ and consists of looking for 
discrete eigenvalues $\lambda_h\in \mathbb{R}$, and non-vanishing eigenfunctions $u_h(x)\in V_h$ such that
\be{FEM:P1}
\int_0^\pi u'_h(x)\,v'_h(x)\,dx = \lambda\,\int_0^\pi u_h(x)\,v_h(x)\,dx, \quad \forall\; v(x)\in V.
\ee
One discretizes the interval $[0,\pi]$ by points $x_i$ and 
uses the (standard) 1D piecewise linear basis $\mathcal{P}_1$ given by
\be{P1:basis}
\varphi_i(x) = 
 \begin{cases}
  \frac{x-x_{i-1}}{x_h-x_{i-1}}  & \text{if } x\in[x_{i-1},x_i] \\
  \frac{x_{i+1}-x}{x_{i+1}-x_{i-1}}  & \text{if } x\in[x_i,x_{i+1}]\\
  0 & \text{otherwise}.
\end{cases}
\ee
Note that these nearest-neighbor overlapping  ``tent-functions" $\varphi_i(x)$ have the properties
$\varphi_i(x_{i-1}) = \varphi_i(x_{i+1})=0$ on the boundaries of their domain $\varphi_i(x)\in [x_{i-1},x_{i+1}]$,
while $\varphi_i(x_{i})=1$, and is zero everywhere else. This converts \Eq{FEM:P1} into the matrix equation
\be{FEM:matrix:eqn}
A\, x = \lambda\, M\,x,
\ee
where the ``stiffness" matrix $A=\{a_{ij}\}_{i,j=1}^{N}$ is given by 
\be{A:ME}
a_{ij} = \int_0^\pi \varphi'_i(x)\,\varphi'_j(x)\,dx=
\frac{1}{h}
\begin{cases}
  2  & \text{for } i=j \\
  -1  & \text{for } |i-j|=1\\
  0 & \text{otherwise},
\end{cases}
\ee
for interval grid size $\Delta x=h$, and the ``mass" matrix is given by
\be{M:ME}
M_{ij} = \int_0^\pi \varphi_i(x)\,\varphi_j(x)\,dx=
\frac{1}{h}
\begin{cases}
  4/6  & \text{for } i=j \\
  1/6  & \text{for } |i-j|=1\\
  0 & \text{otherwise},
\end{cases}
\ee
with $i,j\in\{1,2,\ldots,N\}$, where $N$ is the number of internal nodes in the interval $[0,\pi]$.
Standard techniques such as the conjugate gradient method are used to solve \Eq{FEM:matrix:eqn}, vs costly and often unstable direct inversion methods (since the matrices are sparse, since the functions $\varphi_i$ have small support). For problems that are not too large, sparse LU decompositions and Cholesky decompositions still work well.

In this simple case one is able to compute the eigenfunctions and eigenvalues exactly:
given $k\in \mathbb{N}$, the $k$th eigenspace is generated by the interpolant of the continuous solution
\be{u:hk}
u_h^{(k)}(i\,h) = \sin(k\,i\,h), \quad i=1,\ldots,N,
\ee
with corresponding eigenvalue
\be{lambda:hk}
\lambda_h^{(k)} = (6/h^2)\,\frac{1-\cos kh}{2+\cos kh} 
\approx k^2 + (k^4/12)\,h^2 + \mathcal{O}(k^6 h^4)\;\; \trm{as } h\to 0.
\ee
Thus, as the grid size goes to zero $h\to 0$, one can deduce the optimal estimates
\be{scaling}
|| u^{(k)}- u_h^{(k)} ||_V = \mathcal{O}(h), \qquad || \lambda^{(k)}- \lambda_h^{(k)} ||_V = \mathcal{O}(h^2),
\ee
with $u_h^{(k)}(x) = \sin(k x)$ and $\lambda_h^{(k)} = k^2$.
Note that the eigenvalues converge at one higher order of the grid size $h$ than the eigenfunctions.

The function call in \tit{Mathematica}
\begin{center}
\small{
\begin{lstlisting}[mathescape,language=Mathematica,
caption={NDEigensystem for 1D Laplacian: default settings},label=Mathematica:Code:1]
		Block[{$\$$MaxExtraPrecision = $\infty$},
 		      NDEigensystem[-u''[x], u[x], {x, 0, $\pi$}, 6
		]
\end{lstlisting}
}
\end{center}
produces the eigenvalues $\{0., 1., 4.00005, 9.00061, 16.0034, 25.0128\}$.
Note that the righthand side of the eigenvalue equation $\lambda\,u[x]$ is implied in 
\ttt{NDEigensystem} and therefore not explicitly stated.
Greater accuracy can be obtain by choosing different options to compute the eigenvalues and eigenfunctions as in \Lst{Mathematica:Code:2} below
\newline
\newline
\begin{center}
\small{
\begin{lstlisting}[mathescape,language=Mathematica,
caption={NDEigensystem for 1D Laplacian: FEM},label=Mathematica:Code:2]
Block[{$\$$MaxExtraPrecision = $\infty$},
      NDEigensystem[{-u''[x], DirichletCondition[u[x] == 0, True]}, 
                      u, {x, 0, $\pi$}, 6,
         Method -> {"PDEDiscretization" -> {"FiniteElement", 
                    "MeshOptions" -> {"MaxCellMeasure" -> 0.0001}},
                    "Eigensystem" -> {"Arnoldi", "MaxIterations" -> 10000}}
      ]
 ]
\end{lstlisting}
}
\end{center}
which produces the eigenvalues $\{1., 4., 9., 16., 25., 36.\}$.
\begin{figure}[h]
%\begin{tabular}{cc}
\includegraphics[width=4.5in,height=2.5in]{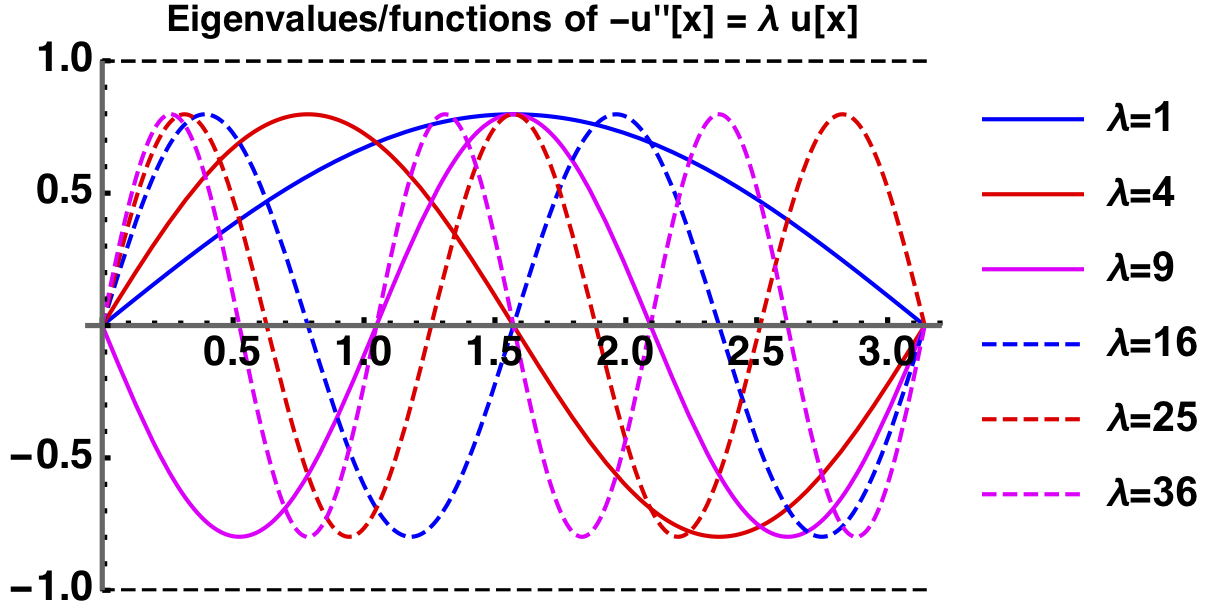} %&
%\end{tabular}
\caption{Eigenvalues and eigenfunctions of 1D Laplacian from \Lst{Mathematica:Code:2}.
}
\label{fig:evals:evecs:1D:Laplacian} 
\end{figure}

The statement \verb+DirichletCondition[u[x] == 0, True]+ enforces Dirichlet BCs on $\pd\Omega$.
The eigenfunctions for the eigenvalues generated from \Lst{Mathematica:Code:2} are shown in \Fig{fig:evals:evecs:1D:Laplacian}. Note that the eigenfunctions do not scale between $[-1,1]$ (a consequence of the numerical method), but this can be remedied by numerically normalizing the square of the eigenfunctions to unity on the interval $\Omega=[0,\pi]$. While, the eigenvalues can be computed fairly accurately for large $k$, the eigenfunctions do not fair as well as $k$ increases. This is a well known effect and arises from the highly oscillatory nature of the eigenfunctions, requiring an increasingly fine mesh to keep the approximation error within the same accuracy of the eigenvalues. Too fine a mesh can also lead to an accumulation of roundoff errors.
In general, the method employed in \verb+NDEigensystem+ can have an effect on the evaluation of  both the eigenvalues and eigenfunctions (see Mathematica documentation of \verb+NDEigensystem+ \cite{Mathematica}).

Coupled eigenvalue equations can be handled by \verb+NDEigensystem+, such as equations of the form
$\mathcal{L}_u[u,v] = \lambda\,u$ and $\mathcal{L}_v[u,v] = \lambda\,v$ (which is the form of the SR DE and GR DE), and only the operators $\mathcal{L}_u$ and $\mathcal{L}_v$ need be supplied to \verb+NDEigensystem+. Both Dirichlet and homogeneous Neumann (derivative zero on $\pd\Omega)$ BCs can also be handled.
 
 \subsection{Advective equations and the issue of numerical dissipation}
 The SR DE and GR DE dealt with in this paper are by design first order in both time and space, to treat all dimensions on an equal footing. This can lead to numerical issues, as will be shortly discussed. The following is based on the clear discussion given in \tit{Numerical Recipes in C}\cite{NumRecinC:1992} Section 19.1 on flux conservation initial value problems (pp 834-839). While the radial DE equations we discuss in the work are actually coupled first order equations, it is best to first consider the simple 1D scalar problem
 \be{NRC:19.1.6}
 \frac{\pd u}{\pd t} = -v\,\frac{\pd u}{\pd x},
 \ee
 where for simplicity and illustrative purposes we take $v>0$ as constant.
 The analytic solution is well known and is any arbitrary function $f(x-vt)$. The solution is pinned down by that particular function which satisfies the BCs. \Eq{NRC:19.1.6} is called an ``advective" equation since the quantity $u$ is transported by a ``fluid flow" with velocity $v$. 
 
 Let us consider a simple finite difference scheme (vs a FEM) with equally spaced points along both the time and spatial axes, given by $x_j = x_0 + j\,\Delta x$ for $j=0,1,\ldots,J$ and 
 $t_n = t_0 + n\,\Delta t$ for $n=0,1,\ldots,N$. Let $u_{j}^{n} = u(t_n, x_j)$ and choose the straight forward forward time, centered space (FTCS) finite differencing scheme which yields 
 \be{NRC:19.1.11}
 \frac{u_{j}^{n+1} - u_{j}^{n}}{\Delta t} = -v\,\left(\frac{u_{j+1}^{n} - u_{j-1}^{n}}{2\Delta x}  \right). 
 \ee
 Here the time derivative is accurate to $\mathcal{O}(\Delta t)$, while the 
 spatial derivative is $\mathcal{O}(\Delta x^2)$.
 
 We can use \tit{von Neumann stability analysis} to determine if this finite differencing scheme is stable.
 Since the equation and differencing scheme is linear, the later also holds true for small fluctuations 
 $\delta u_j^n \equiv \xi^n\,e^{i k j \Delta x}$ where ($i=\sqrt{-1}$, while $j,k$ are integers). Here we imagine that the coefficients of the equations vary so slowly during the fluctuation that we can consider them constant.
 Substituting  $\delta u_j^n$ into \Eq{NRC:19.1.11} and factoring out $\delta u_j^n$ from both sides of the equation yields, after simple algebra, the expression for the amplitude $\xi$
 \be{NRC:19.1.13}
 \xi(k) = 1 -i\,\frac{v\,\Delta t}{\Delta x}\,\sin k\Delta x \quad\Rightarrow\quad 
 |\xi|^2 = 1 + \sin^2 k\Delta x >1, \; \forall\; k,\, \Delta x>0.
 \ee
 Thus, this FTCS scheme is \tit{unconditionally unstable} since small fluctuations grow   $|\xi|>1$ exponentially in time. (For nonlinear equations one linearizes the ODE or PDE and solves a similar linear equation to first order in
  $\delta u_j^n$ for the amplitude $\xi$).
  
 A simple modification of \Eq{NRC:19.1.11} (associated with Lax\cite{NumRecinC:1992}) fixes the stability problem.
 The \tit{Lax Method} is to substitute $u_j^n\to\thalf (u_{j+1}^{n}+u_{j-1}^{n})$ (its average value at neighboring spatial grid points) on the lefthand side of \Eq{NRC:19.1.11} (just in the term $-u_j^n$). This leads to the temporal update scheme for $u_{j}^{n}$ given by
 \be{NRC:19.1.15}
u_{j}^{n+1} = \half\,(u_{j+1}^{n}+u_{j-1}^{n})
 -\frac{v\,\Delta t}{2\Delta x}\,\left(u_{j+1}^{n} - u_{j-1}^{n} \right). 
 \ee
 Performing a similar stability analysis on \Eq{NRC:19.1.15} leads instead now to the amplitude equation
 \be{NRC:19.1.16}
 \xi(k) = \cos k\Delta x -i\,\frac{v\,\Delta t}{\Delta x}\,\sin k\Delta x \;\Rightarrow\; 
 |\xi|^2 = \cos^2 k\Delta x + \left(\frac{v\,\Delta t}{\Delta x}\right)^2\,\sin^2 k\Delta x < 1, 
 \trm{for } \frac{v\,\Delta t}{\Delta x}< 1.
 \ee
 The last inequality in \Eq{NRC:19.1.16} is the famous \tit{Courant stability condition}, which states
 that the spatial grid size $\Delta x$ must be greater than the distance $v\,\Delta t$ for neighboring points
 $u_{j\pm1}^n$ at time $t_n$ to propagate information to $u_j^{n+1}$ at time $t_{n+1}$. If this condition is not met, the numerical scheme is missing information that can influence the evolution of $u_j^n$, creating instabilities that can grow in time.
 
 Further insight into \Eq{NRC:19.1.15} can be obtained by adding $-u_j^n$ 
 to both sides of the equation, rearranging the constant factors, and then taking the continuum limit
 \bea{NRC19.1.18:19.1.19}
u_{j}^{n+1} - u_j^n&=& 
 -\frac{v\,\Delta t}{2\Delta x}\,\left(u_{j+1}^{n} - u_{j-1}^{n} \right) 
 +\half\,(u_{j+1}^{n}-2\,u_j^n+u_{j-1}^{n}), \no
 \Rightarrow\;
 \frac{u_{j}^{n+1} - u_j^n}{\Delta t} &=& 
 -v\,\left(\frac{u_{j+1}^{n} - u_{j-1}^{n}}{2\Delta x} \right) 
 +\frac{(\Delta x)^2}{2 \Delta t}\,\left(\frac{u_{j+1}^{n}-2\,u_j^n+u_{j-1}^{n}}{\Delta x^2}\right),\no
 \Rightarrow\;
 \frac{\pd u}{\pd t} &=& -v\,\frac{\pd u}{\pd x}  +\frac{(\Delta x)^2}{2 \Delta t}\,\nabla^2 u.
  \eea
  The last term in \Eq{NRC19.1.18:19.1.19} shows that a \tit{numerical (spatial) dissipation} or
  \tit{numerical  diffusion} term 
  $\tfrac{(\Delta x)^2}{2 \Delta t}\,\nabla^2 u$ has been added to the original advective equation \Eq{NRC:19.1.6} in order to render the latter stable. The net effect of this diffusion term is smooth out the shock-like structures that can appear in the eigenfunctions for these equations by damping out high spatial frequencies $k>1/\Delta x$, which the finite differencing does not resolve 
  when using a grid with smallest spatial scale $\Delta x$.

\begin{figure}[h]
\begin{tabular}{ccc}
\includegraphics[width=2.25in,height=1.5in]{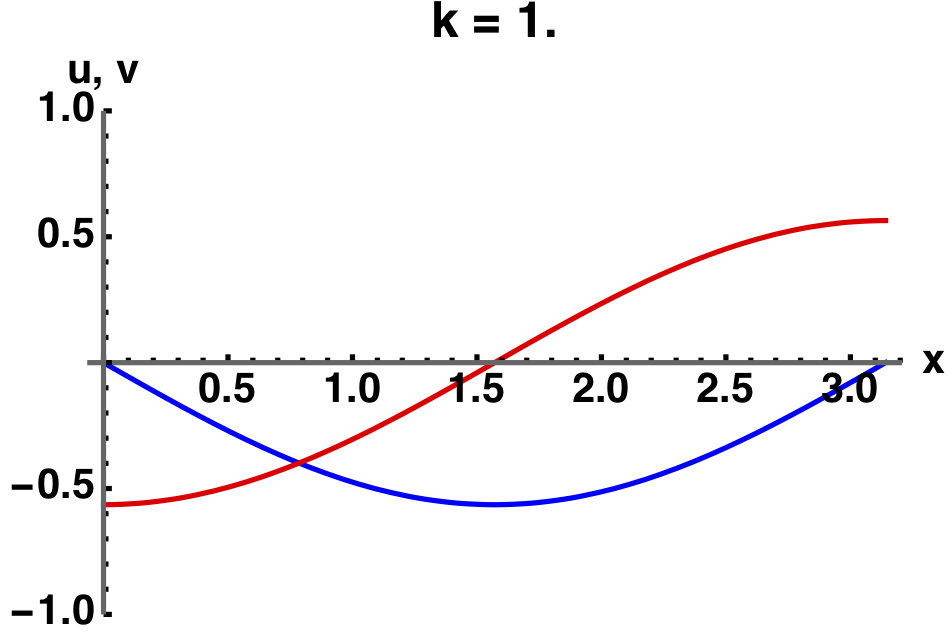} & \hspace{0.5in} & 
\includegraphics[width=2.25in,height=1.5in]{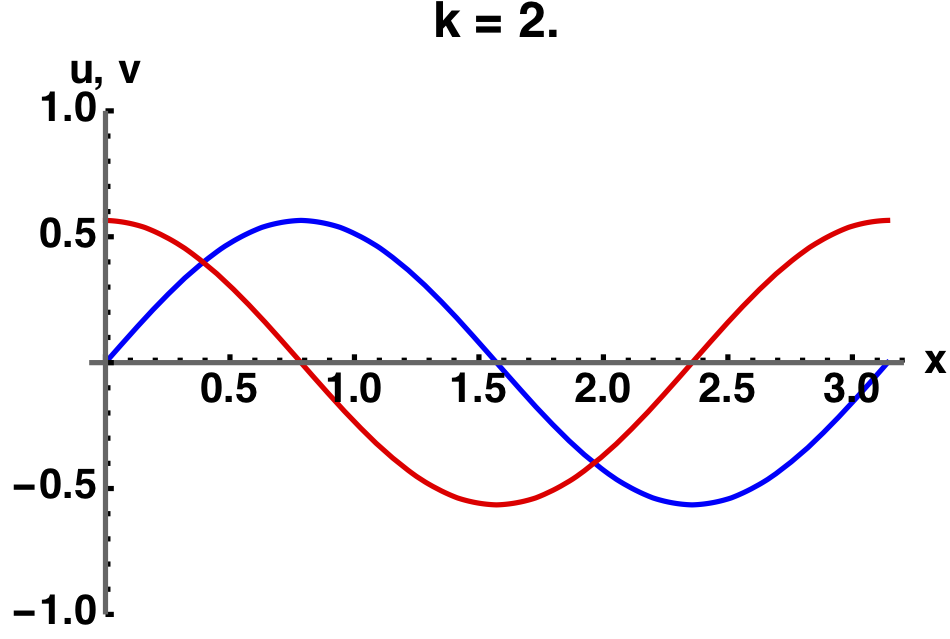} \\
\includegraphics[width=2.25in,height=1.5in]{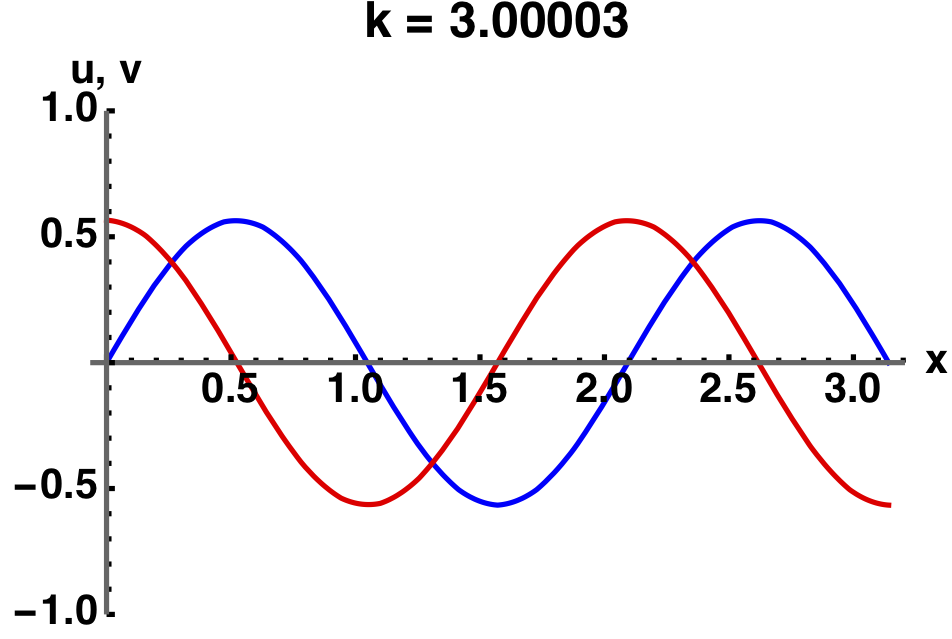} & \hspace{0.5in} & 
\includegraphics[width=2.25in,height=1.5in]{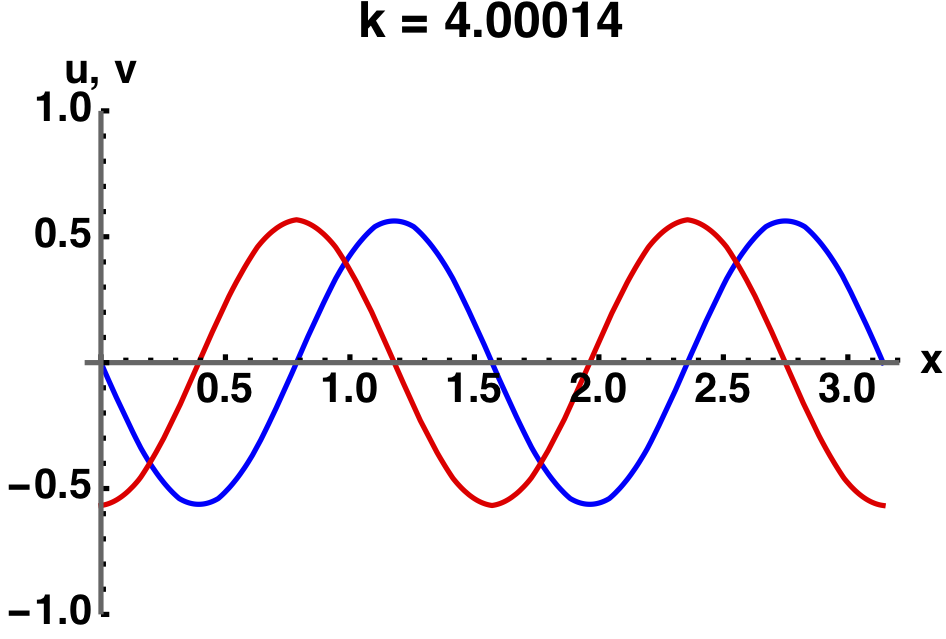} 
\end{tabular}
\caption{$u$(blue), $v$(red) eigenfunctions for the first four lowest eigenvalues of
the coupled first order eigenvalue equations 
$\mathcal{L}_u(u,v)\equiv -v'(x) = k\,u(x)$ and 
$\mathcal{L}_v(u,v)\equiv u'(x) = k\,v(x)$ .
}\label{fig:coupled:u:v:x:evals:evecs}
\end{figure}

Let us consider writing the original 1D Laplacian eigenvalue equation as a pair of first order coupled equations by defining $u'(x) = k\,v(x)$, with $\lambda=k^2$. Then,  $u''(x) = -k^2\,u(x) =  k\,v'(x)$ or $-v'(x) = k\,u(x)$. Thus, 
$\mathcal{L}_u=-v'(x)$ (since  $k\,u(x)$ appears on the righthand side of the latter equation) and
$\mathcal{L}_v=u'(x)$ (since  $k\,v(x)$ appears on the righthand side of the first equation).
The straightforward call to $\verb+NDEigensystem+$ is then
\begin{center}
\small{
\begin{lstlisting}[mathescape,language=Mathematica,
caption={NDEigensystem for coupled first order equations: no diffusion},label=Mathematica:Code:3]
Block[{$\$$MaxExtraPrecision = $\infty$},
    NDEigensystem[{-v'[x], u'[x] + NeumannValue[0, x == $\pi$], 
    DirichletCondition[u[x] == 0, True]}, {u, v}, {x, 0, $\pi$}, 12]
]
\end{lstlisting}
}
\end{center}
Since we know the eigenfunction solutions of  coupled first order equations should be
$u_k(x) = \sin(k x)$ and $v_k(x) = \cos(k x)$ we have included a homogeneous von Neumann BC on 
the second $\mathcal{L}_v$ equation (since $v'_k(\pi) = -k\,\sin(k \pi)=0$).
The above code produces the eigenvalues
$k =\{0., 0., -1., 1., -2., 2., -3.00003, 3.00003, 4.00014, -4.00014, -4.9592, 4.9592\}$.
The $u,v$ eigenfunctions for the first four lowest eigenvalues are shown in \Fig{fig:coupled:u:v:x:evals:evecs}.

\begin{figure}[h]
\begin{tabular}{ccc}
\includegraphics[width=2.25in,height=1.5in]{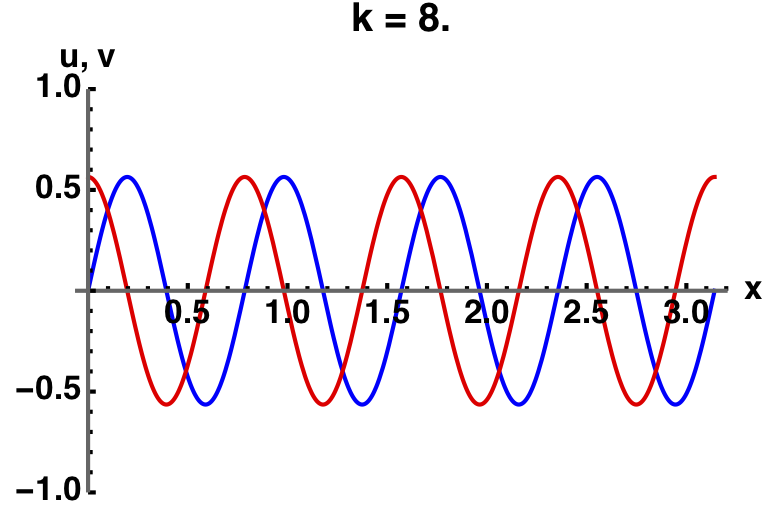} & \hspace{0.5in} & 
\includegraphics[width=2.25in,height=1.5in]{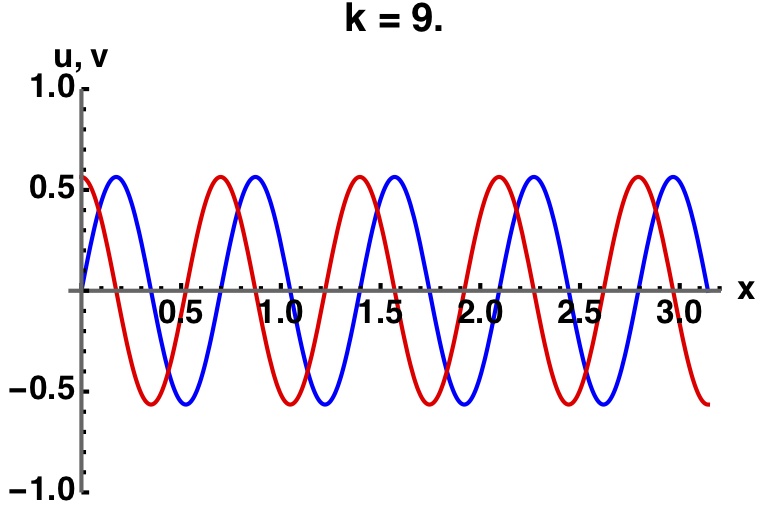} 
\end{tabular}
\caption{$u$(blue), $v$(red) eigenfunctions for the eigenvalues $k=\{8,9\}$ of
the coupled first order eigenvalue equations 
$\mathcal{L}_u(u,v)\equiv -v'(x) = k\,u(x) + \alpha\,u''(x)$ and 
$\mathcal{L}_v(u,v)\equiv u'(x) = k\,v(x) + \alpha\,v''(x)$ 
for $\tfrac{v \Delta t}{\Delta x}=\alpha=10^{-8}$.
}\label{fig:coupled:u:v:x:evals:evecs:k:8:9}
\end{figure}

If we now include numerical dissipation with $\tfrac{v \Delta t}{\Delta x}=\alpha=10^{-8}$
as in \Lst{Mathematica:Code:4} 
\begin{center}
\small{
\begin{lstlisting}[mathescape,language=Mathematica,
caption={NDEigensystem for coupled first order equations: with numerical diffusion $\alpha=10^{-8}$},label=Mathematica:Code:4]
Block[{$\$$MaxExtraPrecision =$\infty$},
   NDEigensystem[{-v'[x] + $\alpha$ u''[x], u'[x] + $\alpha$ v''[x] + NeumannValue[0, x == $\pi$],
      DirichletCondition[u[x] == 0, True]}, {u[x], v[x]}, {x, 0, $\pi$}, 12,
     Method -> {"PDEDiscretization" -> {"FiniteElement", 
         "MeshOptions" -> {"MaxCellMeasure" -> 0.01}},
       "Eigensystem" -> {"Arnoldi", "Criteria" -> "RealPart", 
         "MaxIterations" -> 10000}}
     ] // Chop;
 ]
\end{lstlisting}
}
\end{center}
we instead obtain the eigenvalues
$k=\{1., 2., 3., 4., 4.9938, 5., 6., 7., 8., 9., 9.99264, 10.\}$.
The eigenfunctions for $k=\{8,9\}$ are shown in \Fig{fig:coupled:u:v:x:evals:evecs:k:8:9}.
Note that the last pair of eigenvalues are numerically repeated and the associated eigenfunctions for 
$k=\{9.99264,10\}$ are not well behaved. To achieve better eigenfunctions a combination of a finer grid size and the appropriate dissipation coefficient $\alpha=\tfrac{v \Delta t}{\Delta x}$ is required (which is somewhat of an art).

The lesson of this appendix is that for both the SR DE and the DE in SST considered in this paper, it is far easier  to obtain the eigenvalues (and they are more accurate) than the eigenfunctions themselves. In the main text we focused primarily on the computation of the eigenvalues, except in \Sec{subsec:numerical:diffusion} where we briefly tackled the issue of the GR DE, using the numerical dissipation methods (though FEM) discussed here. Since the GR DE involves a coupled pair of highly nonlinear first order equations, the computation of the eigenfunctions is much more difficult, and an attempt at a variable $\alpha$ was employed (that perturbed the lowest eigenvalues only slightly) that helped smooth the lowest eigenfunctions. As such, all eigenvalues quoted in the main text were those obtained without the use of numerical dissipation, and further effort would be required in order to obtain smoother eigenfunctions for larger eigenvalues.

%=========================================================
%\clearpage
%\newpage
%=========================================================
\begin{acknowledgments}
The author would like to thank Shannon Ray for useful discussions, 
and a thorough reading and re-derivation of all the main equations and the numerical eigenvalue solutions.
The author would like to acknowledge support of this work from
the Air Force Office of Scientific Research (AFOSR).
%%%The author would like to thank C.C. Alsing for useful discussions.
%%Any opinions, findings and conclusions or recommendations
%%expressed in this material are those of the author(s) and do not
%%necessarily reflect the views of Air Force Research Laboratory.
%%%
%%The appearance of external hyperlinks does not constitute endorsement by the United States Department of Defense  of the linked  websites, or the information, products, or services contained therein. The Department of Defense  do not exercise any editorial, security, or other control over the information you may  find at these locations.
\smallskip
\newline
The author has no conflicts to disclose.
\end{acknowledgments}


\begin{thebibliography}{99}
\bibitem{SR:DE:Numerical:Note} To check the validity of the numerical scheme employed using \ttt{NDEigensystem}  we also solved the SR DE  for the Coulomb problem in the same fashion. Namely, in \Eq{Greiner:p184:2:line1} and \Eq{Greiner:p184:2:line2} we let $r=\lambda_C\,\rho = \lambda_C\,\tan\Theta$,
where $\lambda_C = \frac{\hbar}{m c}$, so that 
$r=0\leftrightarrow \Theta=0$ and $r=\infty\leftrightarrow \Theta=\pi/2$, and write these as the following eigenvalue equations: 
$
\cos^2\Theta\frac{dG(\Theta)}{d\Theta} + \k\,\cot\Theta\,G(\Theta) - 
\big(1+ (\alpha-\alpha')\,\cot\Theta\big)\,F(\Theta) = \eps\,F(\Theta)
$,
and 
$
-\cos^2\Theta\frac{dF(\Theta)}{d\Theta} + \k\,\cot\Theta\,F(\Theta) +
\big(1- (\alpha+\alpha')\,\cot\Theta\big)\,G(\Theta) = \eps\,G(\Theta)
$,
where $\eps = E/m c^2$ and $\k = \mp(j+1/2)$ if $j=\ell\pm1/2$.
Recall that the exact eigenvalues to first order are given by 
$\epsilon_n\approx 1-\delta_n = 1-\half\frac{\alpha_C^2}{n^2}\sim \mathcal{O}(1)$, since
$\delta_n \ll 1$ ($E_I =13.6$eV and $m_e\,c^2 = 0.5\times 10^6$ eV).
%
Solving these coupled eigenvalues equations with \ttt{NDEigensystems} 
for $j=\half,\, \alpha=1/137,\, \alpha'=0$ (Coulomb potential, vector coupling, spin $\half$) produces the 
eigenvalues  corrections (relative to the rest mass) 
%
$\delta_n^{(numerical)} = \{2.664 10^{-5}, 6.660 10^{-6}, 2.960 10^{-6}, 1.665 10^{-6}, 
1.063 10^{-6}, 8.408 10^{-7}, 7.019 10^{-7}\}$,
%
to be compared to the exact values from the analytical formula \Eq{Greiner:p186:case:2}
$\delta_n^{(exact)} = \{2.664 10^{-5}, 6.660 10^{-6}, 2.960 10^{-6}, 1.665 10^{-6}, 
1.066 10^{-6}, 7.400 10^{-7}, 5.437 10^{-7}\}$,
using a grid-size discretization of $\Delta\Theta = 10^{-4}$.
Numerical differences arise from the smallness of the  atomic scale energies (eV) relative to the larger rest mass energy (MeV).
Greater numerical precision can be obtained, for example by refining the grid size $\Delta\Theta$.
\end{thebibliography}
\end{document}